\documentclass[aps,prl,reprint,groupedaddress,nofootinbib]{revtex4-2}

\usepackage[utf8]{inputenc}
\usepackage{amsmath}
\usepackage{amssymb}
\usepackage{graphicx}
\usepackage{bm}
\usepackage{xcolor}

\usepackage[
  colorlinks=true,
  citecolor=blue,
  urlcolor=blue,
  linkcolor=blue,
  hypertexnames=false
]{hyperref}

\newcommand{\new}[1]{\textcolor{black}{#1}}
\newcommand{\New}[1]{\textcolor{black}{#1}}

\begin{document}

% Main text
\title{Broken Ergodicity and the Violation of the Fluctuation–Dissipation Theorem Lead to Generalization Beyond Overfitting in Machine Learning}


\author{Chan Li$^1$}
\author{Nigel Goldenfeld$^{1,2}$}%
\email{nigelg@ucsd.edu}
\affiliation{$^1$Department of Physics, University of California San Diego, 9500 Gilman Drive, La Jolla, California 92093, USA.
\\
$^2$Hal\i c\i o\u{g}lu Data Science Institute, University of California San Diego, 9500 Gilman Drive, La Jolla, California 92093, USA
}

% \author{Chan Li}
% \affiliation{Department of Physics, University of California San Diego, 9500 Gilman Drive, La Jolla, California 92093, USA}

% \author{Nigel Goldenfeld}
% \email{nigelg@ucsd.edu}
% \affiliation{Department of Physics, University of California San Diego, 9500 Gilman Drive, La Jolla, California 92093, USA}
% \affiliation{Hal\i c\i o\u{g}lu Data Science Institute, University of California San Diego, 9500 Gilman Drive, La Jolla, California 92093, USA}


%\date{\today}

\begin{abstract}
%Double descent reveals a non-monotonic learning curve, yet its theoretical origin remains poorly understood from the perspective of critical phenomena. We formulate double descent as a critical transition by developing a dynamical mean-field theory that captures the long-time behavior of a teacher-student linear regression model trained with stochastic gradient descent. Our theory reveals critical exponents and scaling collapse, attributes the divergence of generalization error to a diverging susceptibility, identifies the interpolation threshold as a dynamical transition point from equilibrium to non-equilibrium learning, and shows that regularization acts as a finite-size effect that smooths this transition.

The remarkable ability of modern neural networks to generalize improves with increasing network capacity, even when the number of model parameters or effective degrees of freedom exceeds the number of training data points.  This phenomenon is all the more surprising given that generalization error diverges when the number of model parameters approaches a critical value from below.  Here we use dynamical mean field theory to show that this so-called \lq\lq double descent" behavior is the outcome of a phase transition in the stochastic field theory describing the training process. We calculate the critical exponents and scaling function of the double descent phase transition, and show that it is marked by a breakdown of the fluctuation-dissipation theorem associated with broken ergodicity. The corresponding response function has the same functional form as the simple London model of the superconducting transition, with the rigidity of the wave function corresponding to the neural network's ability to generalize accurately.

\end{abstract}
%\keywords{}

\maketitle

%\textit{Introduction.}---
%%%%%%%%%%%%%%%%%%%%%%%%%%%%%%%%%%%%%%%%%%%%%%%%%%%%%%%%%%%%%%%%%
The success of modern neural networks in generalization might naively be attributed to having enough parameters or degrees of freedom (so-called \lq\lq network capacity") to represent the training data.  From this perspective, having too few parameters to represent a complex data set is analogous to trying to fit a high order polynomial with a quadratic or linear function, and so would not be expected to exhibit good generalization. Increasing the order of the fitting polynomial might be considered a strategy to improve the situation, but once the order is too large (ie. there are too many fitting parameters), the representation of the training data is contaminated by its inherent noise, and the ability to generalize will be impaired.  A sweet spot between these two extremes would be the natural operating point for a neural network, balancing what is referred to as \lq\lq bias" (underfitting the signal) and \lq\lq variance" (overfitting the noise), whilst still remaining in a regime of underparameterization compared to the size of the training data \cite{hastie2009elements,geman1992neural,mehta2019high}.  

The surprising feature of modern neural networks is that they operate far above this sweet spot, in an over-parameterized regime, with a huge number of parameters compared to the training data, and thus would be expected to have poor generalization ability.  Modern neural networks defy this expectation by interpolating the training data with no error while still generalizing well \cite{belkin_2019}.  In particular, the generalization error $\varepsilon_g$ (defined precisely below) as a function of model capacity reveals a remarkable phenomenon known as \lq\lq double descent": $\varepsilon_g$ initially decreases at small capacity towards a local minimum at the sweet spot, then rises sharply near a threshold of network capacity, and then falls again in the highly overparameterized regime, where modern neural networks achieve unexpectedly strong performance. This threshold, known as the \lq\lq interpolation threshold", marks the transition from the classical bias–variance tradeoff to the modern interpolating regime of overparameterization.  In the underparameterized regime, training error $\epsilon_t$ (defined precisely below) is nonzero and generalization error follows the classical bias–variance tradeoff. At the interpolation threshold and beyond, the training error vanishes: models interpolate the training data perfectly yet the generalization error decreases from infinity as a function of increasing network capacity! Thus, the network can still achieve low generalization error, a phenomenon often described as benign overfitting \cite{bartlett2020benign}.

Double descent is observed universally across diverse architectures and tasks {in an offline setting where a finite set of training data is repeatedly used} \cite{belkin_2019}, including such important applications as large language models \cite{cao2023instruction} and protein folding prediction \cite{ahdritz2024openfold}. To account for this universality, it is natural to interpret double descent as a phase transition in statistical physics \cite{goldenfeld2018lectures_m}, reflecting two qualitatively distinct learning regimes \cite{liao2020random,d2020double,spigler2018jamming,geiger2019jamming,sclocchi2024different}.  A useful analogy is the jamming transition in granular matter \cite{liu2010jamming}.  At the interpolation threshold, networks exhibit a sharp change from nonzero to vanishing training error \cite{spigler2018jamming,geiger2019jamming,sclocchi2024different} with the satisfaction of every physical constraint in a jammed material being the analogy of fitting all the training data.  Random matrix theory \cite{marcenko1967distribution} has been used to analyze double descent in the static thermodynamic limit of tractable high-dimensional models \cite{liao2020random,d2020double,hastie2022surprises,mei2022generalization,rocks2022memorizing,rocks2022bias} and in dynamics \cite{advani2020high,mignacco2020dynamical}.
%
%while dynamical mean-field theory (DMFT) provides a self-consistent description of stochastic gradient descent that tracks both training and test errors throughout learning \cite{mignacco2020dynamical}.
%
Overall, none of these approaches fully explain the mechanism of the transition, the critical behavior near the interpolation threshold, and what happens within the overparameterized regime where the training error is identically zero but the generalization error decreases with network capacity.

The purpose of this Letter is to provide a theoretical explanation of the mechanism underlying the double descent transition, and to identify the emergent phenomena which give rise to the benign overfitting phase of machine learning. {Unlike previous analyses that focus on the static least-norm solution obtained from zero initialization, we study the full training dynamics with arbitrary initialization, and thus discovered a novel phase transition which we describe below.} We formulate double descent in linear regression trained by stochastic gradient descent {in an offline setting}, and analyze its thermodynamic limit using dynamical mean-field theory. We show that the generalization error is governed by the susceptibility of the training dynamics, explaining its divergence at the interpolation threshold.  We compute the critical exponents and scaling function near the critical point. 
%The training error is a filtered version of the generalization error; in the overparameterized regime the filter’s gain on the relevant modes vanishes, driving the training error to zero. 
We identify a second fundamental aspect of double descent: it is a dynamical phase transition, where learning transitions from equilibrium to nonequilibrium dynamics with increasing overparameterization, as shown by a breakdown of the relevant fluctuation-dissipation theorem (FDT) \cite{yaida2018fluctuation,han2021fluctuation,kubo1966fluctuation_m,crisanti2003topical} {(see Supplemental Material \cite{SM} for technical details (Sec. I-V) and background information (Sec. VI)).}  
%
%Proceeding further into the overparameterized regime, the training error remains zero, but the degree of ergodicity breaking increases, with concomitant decrease in generalization error. 
%
The breakdown of the FDT arises through the behavior of an appropriate response function, which has the same form and analytic properties as the response function in the London model for superconductivity \cite{london1935electromagnetic_m,tinkham2004introduction_m}. A persistent current emerges below the superconducting transition temperature, whose stability is quantified by wave function rigidity; its magnitude is dependent on the fraction of superconducting electrons. {Similarly, in the overparameterized regime of machine learning, the response function reveals an emergent rigidity of the neural network, dependent on the fraction of degrees of freedom contributing to effective generalization, and associated with a persistent memory. In both cases, these emergent phenomena arise from ergodicity breaking, where the dynamics is restricted to a subspace of the full phase space of the system, which is associated with its initial condition. The nature of these two transitions is mathematically identical, providing a novel explanation for benign overfitting beyond the transition into the over-parameterized regime. Consequently,} a neural network is able to generalize well, because its response to a perturbation (prompt) is localized within the context of its initial condition.

\textit{Dynamical mean-field theory for high-dimensional stochastic gradient descent.}--- To understand how double descent emerges dynamically and in the long-time limit from the collective behavior during training, we employ dynamical mean-field theory (DMFT) to model the learning dynamics of neural networks trained with stochastic gradient descent \cite{cugliandolo2023recent,mignacco2020dynamical,agoritsas2018out,gerbelot2024rigorous}. While the framework is broadly applicable, we focus on a minimal teacher-student setup amenable to exact analysis. The teacher is a single-layer linear model that receives $N$ input vectors $\mathbf{x}^\mu \in \mathbb{R}^d$ and generates scalar labels $y^\mu=\frac{1}{\sqrt{d}} \sum_{i=1}^d \beta_i x_i^\mu+\epsilon^\mu$ for $\mu=1, \ldots, N$, where $\beta_i$ denotes the ground-truth weight connecting input unit $i$ to the output, and $\epsilon^\mu$ is sample-specific Gaussian noise with zero mean and variance $D$. For simplicity, we consider isotropic training data drawn i.i.d. from a standard normal distribution, $x_i^\mu \sim \mathcal{N}(0, 1)$. The scaling factor $1/\sqrt{d}$ ensures that the weighted sum is independent of the input width. The student has the same architecture as the teacher, with trainable weights $\boldsymbol{\hat{\beta}} \in \mathbb{R}^d$, and maps each input $\mathbf{x}^\mu$ to a prediction $\hat{y}^\mu=\frac{1}{\sqrt{d}}\sum_{i=1}^d \hat{\beta}_i x_i^\mu$. Training minimizes the {empirical squared loss} between {model} predictions and teacher-generated labels, augmented by an $\ell_2$ regularization term with strength $\lambda$ :
% Training minimizes the mean squared error (MSE) between predictions and teacher-generated labels, augmented by an $\ell_2$ regularization term with strength $\lambda$ :
{
\begin{equation}
\mathcal{L}=\frac{1}{2} \sum_{\mu=1}^N \left(\hat{y}^\mu-y^\mu\right)^2+\frac{\lambda}{2}\|\hat{\boldsymbol{\beta}}\|_2^2.\\
\label{eq1}
\end{equation}
}
Optimization is performed via stochastic gradient descent: 

\begin{equation}
% \hat{\beta}_i(t+1)=\hat{\beta}_i(t)-\zeta \frac{\partial \mathcal{L}}{\partial \hat{\beta}_i(t)},
{\hat{\beta}_i(t+1)=\hat{\beta}_i(t)-{\zeta} \frac{\partial\left[ \frac{1}{2} \sum_{\mu=1}^N s^{\mu}\left(\hat{y}^\mu-y^\mu\right)^2+\frac{\lambda}{2}\|\hat{\boldsymbol{\beta}}\|_2^2\right]}{\partial \hat{\beta}_i(t)}},
\label{eq2}
\end{equation}
where $s^\mu \in\{0,1\}$ is a binary random variable drawn independently at each iteration, with $\mathbb{P}\left(s^\mu=\right.$ $1)=\gamma$, indicating whether sample $\mu$ is selected for training at step $t$. The parameter {$\zeta$} denotes the learning rate, and $t$ indexes the iteration step. Considering the ${\zeta}\to 0$ limit, the dynamics can be approximated by a continuous-time equation in the form of a stochastic gradient flow (SGF):

\begin{equation}
%\dot{\hat{\beta}}_i(t)=- \frac{\partial \mathcal{L}}{\partial \hat{\beta}_i(t)},
% \partial_t \hat{\beta}_i(t)=- \frac{\partial \mathcal{L}}{\partial \hat{\beta}_i(t)},
{
\partial_t \hat{\beta}_i(t)=- \frac{\partial \left[ \frac{1}{2} \sum_{\mu=1}^N s^{\mu}\left(\hat{y}^\mu-y^\mu\right)^2+\frac{\lambda}{2}\|\hat{\boldsymbol{\beta}}\|_2^2\right]}{\partial \hat{\beta}_i(t)}},
\label{eq3}
\end{equation}
which reduces to the standard gradient flow when $\gamma=1$.  The performance of the model is evaluated via the training error $\varepsilon_t$ and generalization error $\varepsilon_g$, with $\varepsilon_t$ computed on the entire training set $\mathbf{x}$ and $\varepsilon_g$ on unseen samples $\mathbf{x}^0$ from the same distribution.
%The performance of the model is evaluated via the training error $\varepsilon_t$ and generalization error $\varepsilon_g$, both in MSE form, with $\varepsilon_t$ computed on the entire training set $\mathbf{x}$ and $\varepsilon_g$ on unseen samples $\mathbf{x}^0$ from the same distribution.

We now turn to the analysis of the stochastic gradient flow in the 
%infinite-size 
{thermodynamic} limit $N \rightarrow \infty, d \rightarrow \infty$ with the complexity parameter $\alpha=d / N$ held fixed. The regime $\alpha<1$ corresponds to underparameterization but lacks the first descent of the classical bias–variance tradeoff, as the model is well specified and the estimator remains unbiased yet variance-dominated \cite{hastie2022surprises}. The interpolation threshold at $\alpha=1$ marks a singular transition to the overparameterized phase ($\alpha>1$), where the number of parameters exceeds the data constraints and the model exactly interpolates the training set. 
% In this thermodynamic limit, DMFT provides an exact macroscopic description by mapping the coupled high-dimensional dynamics onto an effective stochastic process for a single representative weight $\hat{\beta}(t)$:
{In this thermodynamic limit, after averaging over the quenched disorder and invoking self-averaging, a saddle-point (mean-field) approximation of the dynamical generating functional (see Sec.~I of the Supplemental Material \cite{SM}) leads to a decoupling of the high-dimensional learning dynamics into statistically equivalent and independent representative components. Dynamical mean-field theory (DMFT) then yields an exact macroscopic description in terms of an effective stochastic process for a single typical (representative) weight $\hat{\beta}(t)$}:
\begin{equation}
\begin{aligned}
\partial_t \hat{\beta}(t)
& =-\frac{ \gamma }{\alpha} \int_0^\infty dt^\prime \left(\gamma R +\delta \right)^{-1}(t, t^\prime)\left(\hat{\beta}(t^\prime)-\beta\right)\\&-\lambda \hat{\beta}(t)+\eta(t),\\
\label{eq4}
\end{aligned}
\end{equation}
where $\delta(t)$ is the Dirac delta function and $\beta$ denotes the representative ground-truth weight.
$R\left(t, t^{\prime}\right) $ is the response function, measuring the change in the model state at time $t$ due to a perturbation $F$ applied at $t^{\prime}$. 
The Gaussian noise $\eta(t)$ has zero mean and covariance $\hat{D}(t_1,t_2)$, 
\begin{equation}
\hat{D}\left(t, t^{\prime}\right)=\frac{\gamma^2}{\alpha}(\gamma R+\delta)^{-1} *(D+C) *\left(\gamma R^{\top}+\delta\right)^{-1}\left(t, t^{\prime}\right).
\end{equation}
where the operator $*$ denotes convolution over intermediate times, defined as $\left(A * B\right)\left(t, t^{\prime}\right)=\iint A\left(t, t^{\prime \prime}\right) B\left(t^{\prime \prime}, t^{\prime}\right) \mathrm{d} t^{\prime \prime}$. The correlation function $C(t,t^\prime)$  characterizes temporal correlations of the deviation of the student weights from the teacher.  The dynamics is then fully specified by a closed set of self-consistent equations for $R(t_1, t_2)$ and $C(t_1, t_2)$ (see Sec.~I of the Supplemental Material \cite{SM}).

Compared to the original high-dimensional dynamics, the effective dynamics offers a clearer physical picture. 
The memory kernel $(\gamma R+\delta)^{-1}$ guides the weights toward the ground truth by integrating past responses and encoding the accumulation of information over time. 
Stochastic fluctuations $\eta(t)$ continually perturb this motion, with a variance that, as we show below, is determined by the training error and reflects the intrinsic uncertainty of the learning process.
Learning thus emerges from the competition between signal extraction through the kernel and noise injection through the fluctuations, which together set the ultimate precision of generalization.

The regularization strength $\lambda$ acts as a relevant control parameter for the double-descent transition. Here we analyze the unregularized case $\lambda=0$, where the generalization error diverges at the interpolation threshold, in order to expose the existence of the phase transition and the emergent features of machine learning in the overparameterized regime.  In the End Matter, we calculate how non-zero $\lambda$ modifies the phase transition, and compute the scaling laws and crossover exponents.

\medskip
\textit{Susceptibility as a physical origin of generalization divergence.}--- The generalization error is defined in {mean squared error (MSE)} form as $\varepsilon_g(t) = \langle \left(\frac{1}{\sqrt{d}} \sum_{i=1}^d x_i^0[\hat{\beta}_i(t)-\beta_i]-\epsilon^0\right)^2 \rangle$, where the average is over test inputs $\vec{x}^0\in\mathbb{R}^d$ and label noise $\epsilon^0$, drawn independently from the same distribution as the training data. In the thermodynamic limit, one obtains $\varepsilon_g(t) = C(t, t)+D$, with $C(t,t)$ the equal-time correlation function from the self-consistent DMFT equations and $D$ the label-noise variance (see Sec.~II.A of the Supplemental Material \cite{SM}). When $\lambda=0$, the frequency–domain analysis shows that for $\alpha<1$, the asymptotic error is (see Sec.~V.B.3 of the Supplemental Material \cite{SM}):
\begin{equation}
\varepsilon_g(t\to\infty)=D(\gamma \chi+1) = \frac{D}{1-\alpha},
\label{eq:alpha<1ge}
\end{equation} 
where $\chi=\int_0^{\infty}R(\tau)d\tau$ denotes the susceptibility of the learning dynamics, well defined only in equilibrium. 
% The generalization error thus depends linearly on $\chi$, which diverges as $\chi = \frac{\alpha}{\gamma \left(1-\alpha\right)}$ when $\alpha\to1^{-}$. This interesting interpretation suggests that generalization error can be viewed as the system’s response to a perturbation by the test data. 
{Eq.~\eqref{eq:alpha<1ge} shows that the generalization error depends linearly on $\chi$, which diverges as $\chi = \frac{\alpha}{\gamma \left(1-\alpha\right)}$ when $\alpha\to1^{-}$. This relation, which follows from linear response theory, suggests that the generalization error can be viewed in a formal sense as the response of the trained system to a hypothetical perturbation by an unseen test sample, even though the test data do not affect the training of the weights. }
%Its linear dependence shows that the divergence of generalization error at the interpolation threshold arises from the divergence of susceptibility, a hallmark of phase transitions in statistical physics.  

However, susceptibility is not defined when $\alpha>1$, because the dynamics is out of equilibrium as we will show later. The system enters a non-ergodic phase with persistent memory: the response fails to decay and instead saturates at $R(\tau \to \infty)=\frac{\alpha-1}{\alpha}$, which we denote as the \textit{ergodicity-breaking degree}, quantifying the persistent memory of an initial perturbation. 
In this regime, the asymptotic generalization error depends explicitly on this quantity, reflecting the retention of the initial condition $C(0,0)$ and showing a behavior analogous to the $\alpha<1$ regime near criticality (see Sec.~V.B.3 of the Supplemental Material \cite{SM}):
\begin{equation}
\varepsilon_g(t\to\infty)=\frac{(\alpha-1)\,C(0,0)}{\alpha}+\frac{D\alpha}{\alpha-1}, 
\label{eq:alpha>1ge}
\end{equation}
where $C(0,0) = \langle \hat{\beta}^2(0)\rangle + r$, with $\langle \hat{\beta}^2(0)\rangle$ the variance of the estimator’s initialization and $r$ the mean-squared norm of the ground-truth parameters per dimension. 
%
%Note that the mean-squared generalization error (an absolute comparison between the machine learning output and the ground-truth label) increases with $r$, because a stronger ground-truth signal contributes proportionally to the total error. This counterintuitive behavior arises from using an absolute metric; a relative measure such as $\varepsilon_g / r$ provides a more appropriate characterization, decreasing with signal strength and signal-to-noise ratio (SNR), $r/D$.
%
The {\it relative} generalization error at steady state $\varepsilon_g/r \sim \langle \hat{\beta}^2(0)\rangle /r +  \frac{(\alpha-1)}{\alpha}+\frac{D\alpha}{r(\alpha-1)}$ decreases as the ground-truth signal increases (i.e. large $r$) towards the value ${(\alpha-1)}/{\alpha}$, which is the rigidity characterizing the degree of ergodicity-breaking. This satisfies the intuition that performance should improve with increasing ground-truth signal and decreasing signal-to-noise ratio ($r/D$).

%Since the mean-squared generalization error scales with $r$, a stronger signal produces a larger error in Eq.~\ref{eq:alpha>1ge}, which appears counterintuitive until one realizes that the dimensionless generalization error is normalized by the variance of the label noise $D$, which does reduce for larger signal strength. The most appropriate way to evaluate the quality of generalization is to use a normalized measure of the scale-free cosine similarity between the estimator and the ground truth; this increases with signal strength and signal-to-noise ratio (SNR), $r/D$ (see sec.~V.B.3 of the Supplemental Material \cite{SM}).

The training error is also defined in MSE form $\varepsilon_t(t)=\Big\langle \Big(\tfrac{1}{\sqrt{d}}\sum_{i=1}^d x_i^1[\hat{\beta}_i(t)-\beta_i]-\epsilon^1\Big)^2 \Big\rangle$,
where a single training sample $\vec{x}^1$ is drawn from the dataset for evaluation. The average is taken over both the training dynamics and the dataset, so the dynamical measure is correlated with the evaluated data.  In the thermodynamic limit, the training error can be expressed in terms of the observables introduced above and is proportional to the noise variance (see Sec.~II.B of the Supplemental Material \cite{SM}),
\begin{equation}
\varepsilon_{t}(t)=\big[(\gamma R+\delta)^{-1}*(D+C)*(\gamma R^{\top}+\delta)^{-1}\big](t,t).
\end{equation}
While the generalization error measures the equal-time correlation, the training error is a kernel-modified version of the same quantity, reshaped by the memory operator $(\gamma R+\delta)^{-1}$.  
In the overparameterized regime this kernel vanishes, and the transformation suppresses the signal entirely, causing the training error to approach zero as a dynamical consequence of overparameterization.

Similarly, the training error can be expressed in terms of the susceptibility (see sec.~V.B.3 of the Supplemental Material \cite{SM}),
\begin{equation}
\lim_{t \to \infty}\varepsilon_{t}(t)=(\gamma \chi+1)^{-2}\,\varepsilon_g(t\to\infty),
\end{equation}
which evaluates to $D(1-\alpha)$ for $\alpha<1$.  For $\alpha>1$, the equilibrium description breaks down and the quantity $\chi=\int_0^{\infty}R(\tau)d\tau$ diverges, driving the training error to zero. 

\begin{figure}
\centering
\includegraphics[width=0.5\textwidth]{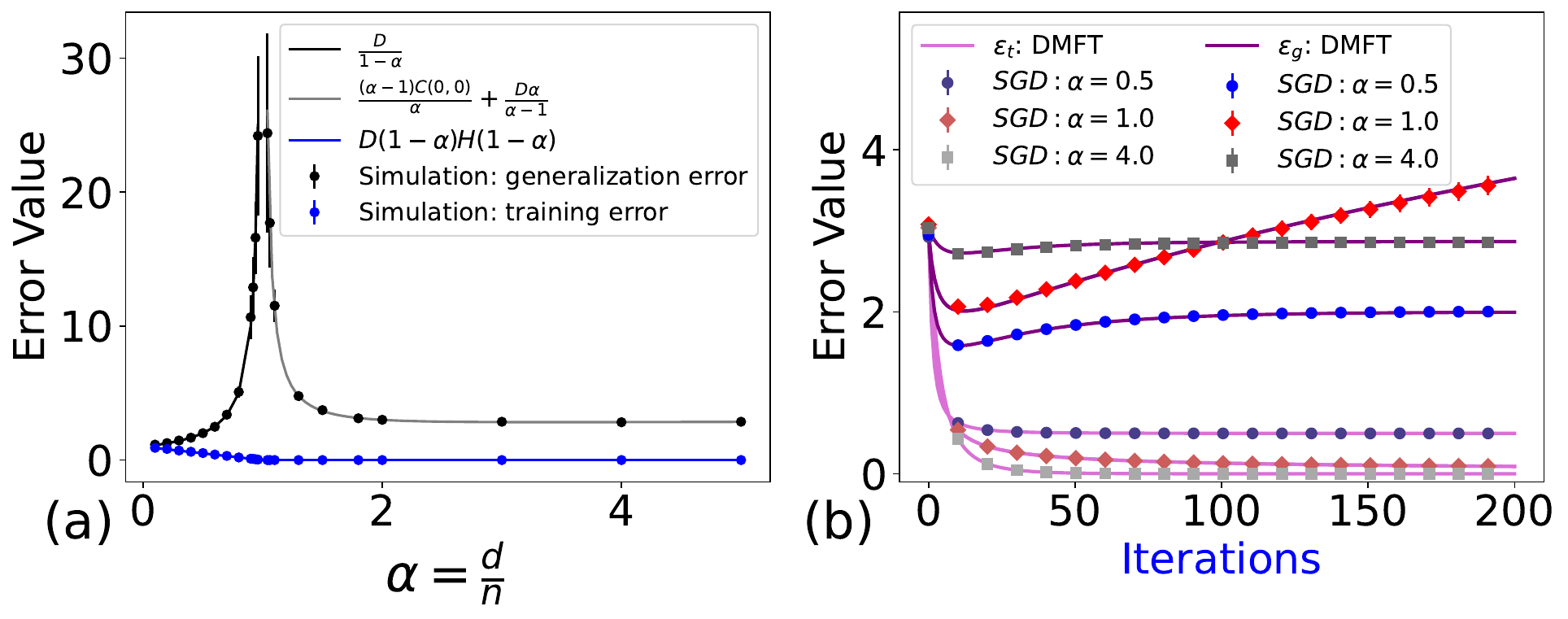}
\caption{Comparison between numerical simulations and theoretical predictions for the generalization error $\varepsilon_{g}(t)$ and training error $\varepsilon_{t}(t)$ in an unregularized ($\lambda=0$) single-layer linear neural network ($N=1,000$, $d=\alpha N$, $D=1$, $r=1$, $\gamma=0.5$, and ${\zeta}=0.1$). 
(a) Asymptotic values (training time $T = 10^4$) of $\varepsilon_{g}$ and $\varepsilon_{t}$ as functions of model capacity $\alpha$. 
Markers denote simulations (averaged over ten independent runs), while solid lines show analytical predictions. $H(x)$ is the Heaviside step function, which equals $1$ for $x>0$ and $0$ for $x\leq0$.
(b) Time evolution of $\varepsilon_{g}(t)$ and $\varepsilon_{t}(t)$ for different $\alpha$ ($T=20$). The number of {iterations} is ${\frac{T}{\zeta}}=200$.
Markers denote simulations (averaged over ten independent runs), and solid lines indicate DMFT results. }
\label{error}
\end{figure}

Numerical simulations confirm that the dynamics of both training and test errors are accurately captured by DMFT, and their asymptotic values agree precisely with the analytical predictions in Eqs.~(\ref{eq:alpha<1ge}) and~(\ref{eq:alpha>1ge}), as shown in Fig.~\ref{error}.

\textit{Universal scaling and data collapse of the generalization error.}---
From the asymptotic behavior of the generalization error near criticality in Eqs. (\ref{eq:alpha<1ge}) and (\ref{eq:alpha>1ge}), we see that 
$\varepsilon_g(t\to\infty)$ diverges as the interpolation threshold is approached when $\alpha\to 1^{\pm}$, suggesting the scaling $\varepsilon_g(t\to\infty)\sim |1-\alpha|^{-\sigma}$ with an exponent $\sigma = 1$ in DMFT.  During training, both training and generalization error are time-dependent, but eventually saturate at long enough times to a steady state value.  The saturation time is $\alpha$-dependent, and so we anticipate from general phase transition principles \cite{goldenfeld2018lectures_m} that a dynamic scaling data collapse should be present near the transition point $\alpha=1$.  This scaling ansatz takes the form 
\begin{equation}
\varepsilon_g
\approx \frac{D}{|1-\alpha|^\sigma}\,
F\!\left(\frac{|1-\alpha|^\phi (\gamma t)}{4}\right),
\label{eq:udc}
\end{equation}
where we have introduced a new exponent $\phi$ and a universal scaling function $F(z)$.  As $\alpha \rightarrow \pm 1$ but $t$ remains arbitrary, the singularity in front of $F$ must be absorbed by the scaling function, implying that for  values of its argument $z \rightarrow 0$, it must take the form $F(z)\sim z^{\sigma/\phi}$. At $\alpha=1$, the generalization error never saturates but grows in time as $\varepsilon_g(t)\sim (\gamma t)^{\sigma/\phi}$, where the exponent $\phi$ characterizes the growth with training time and $\gamma$ is the probability of a sample being included in the stochastic training process. The form of Eq.~(\ref{eq:udc}) implies that there is a characteristic time scale $\tau(\alpha)$ for the convergence to steady scaling, with a form $\tau = (\gamma |1-\alpha|^\phi)^{-1}$ that diverges as $\alpha\rightarrow 1$, which is a critical slowing down \cite{goldenfeld2018lectures_m}.  Calculation of the critical exponents and universal scaling function are presented in the End Matter.

\textit{Double descent and dynamical transitions at the interpolation threshold.}—
For $\lambda=0$, the interpolation threshold at $\alpha=1$ marks not only the phase transition between under- and over-parameterization, but also a dynamical transition from equilibrium to non-equilibrium, closely analogous to the superconducting transition \cite{london1935electromagnetic_m} and the gelation transition \cite{goldenfeld1992dynamic}, which exhibit counterparts of the scaling laws given above. For example, in a simple conductor, the conductivity $\tilde\sigma(\omega)$ as a function of frequency $\omega$ quantifies the current response $\tilde J(\omega)$ to an external field $\tilde E(\omega)$: $\tilde J(\omega) = \tilde\sigma (\omega) \tilde E(\omega)$ (Ohm's Law). In machine learning, the response function in frequency space $\tilde{R}(\omega)$ plays the analogous role to $\tilde\sigma(\omega)$, describing how the average deviation {$\Delta\hat{\beta}(t)\equiv\langle\hat{\beta}(t)-\beta\rangle$} responds to a source $\tilde F(\omega)$ (see sec.~VI.F of the Supplemental Material \cite{SM}): $
\Delta \hat{\beta}(\omega)=\tilde{R}(\omega)\, \tilde F(\omega)$.
%This equation is a linear response relation between the deviation and an external frequency-dependent source $\tilde F(\omega)$, analogous to Ohm's law in a simple conductor.

For $\alpha<1$, the learning dynamics of the neural network obeys the fluctuation–dissipation theorem (FDT) (see sec.~VI.D of the Supplemental Material \cite{SM} and \cite{yaida2018fluctuation}), consistent with equilibrium dynamics.  The response function $\tilde R(\omega)$ is causal and analytic with no poles in the lower half of the complex frequency domain, as is the case for the conductivity $\tilde\sigma(\omega)$ of a simple conductor. For $\alpha>1$, the imaginary part of $\tilde{R}(\omega)$ develops an additional pole at $\omega=0$, leading to an equation of motion for the average deviation in the long time limit (see sec.~VI.A, VI.B and VI.F of the Supplemental Material \cite{SM}):
\begin{equation}
\partial_t\Delta \hat{\beta}(t)=\frac{\alpha-1}{\alpha}\, F(t).
\label{eq:sc_ml}
\end{equation}
This equation tells us that the external source governs not the value of the average deviation, but its rate of change.  In particular even when the source $F(t)$ is zero, $\Delta \hat{\beta}(t) = \Delta \hat{\beta}(0)$, showing that the system has a persistent memory of its initial condition at arbitrarily long times, with strength given by the rigidity ${(\alpha-1)}/{\alpha}$. {This ergodicity breaking (see Sec. VI.E of the Supplemental Material \cite{SM}) confines the training dynamics to a restricted subspace around the initialization, and} as $\alpha$ increases further into the over-parameterized regime, this persistent contribution strengthens, leading to benign overfitting—improved generalization despite vanishing training error{, by limiting noise fitting while preserving generalization}.  The same thing happens to the conductivity of a simple superconductor \cite{london1935electromagnetic_m} below the critical temperature: $\tilde\sigma(\omega) \propto \rho_s/(i\omega)$, where $\rho_s$ is the helicity modulus, representing the wave function rigidity.  Plugging into the Ohm's Law relation, and setting the external field to zero yields a persistent dissipationless current $J(t) = J(0)$ at arbitrary time.  Thus, the quantity ${(\alpha-1)}/{\alpha}$ is directly analogous to the helicity modulus $\rho_s$  and quantifies the degree of ergodicity breaking.  Not surprisingly, the response function in linear regression has a sum rule analogous to that of the Ferrell-Glover-Tinkham in superconductivity \cite{ferrell1958conductivity_m,tinkham1959determination_m} (see Sec.~VI.C of the Supplemental Material \cite{SM}).

\textit{Discussion.}— {By going beyond previous static least-norm analyses and treating the full training dynamics with arbitrary initialization, we uncover a  dynamical phase transition underlying double descent, marked by ergodicity breaking and persistent memory of the initialization, thereby providing a mechanistic explanation of key phenomena such as benign overfitting in the overparameterized regime.} Identifying the overparameterized regime as exhibiting emergent rigidity suggests a way to analyze other reported emergent properties of neural networks such as in-context learning in large language models \cite{wei2022emergent,berti2025emergent} (but see \cite{schaeffer2023emergent} for a discussion of measurement artifacts), where the readout is associated with scaling laws \cite{kaplan2020scaling,bahri2024explaining,bordelon2024dynamical},
and grokking \cite{power2022grokking,kumar2023grokking}.  
The generalized rigidity uncovered here implies that overparameterized neural networks, and perhaps even analog physical networks, such as nonlinear contrastive local learning networks \cite{dillavou2024machine} that exhibit double descent \cite{Durian2025}, satisfy a generalized elasticity theory with scaling laws as given here and in Ref.~(\cite{goldenfeld1992dynamic}).  
%{Our work does not apply to online learning, where new data are continually introduced during training, so that the training error can never vanish, and thus where double descent does not occur.}  

Our DMFT results are a step toward a renormalization-group (RG) interpretation of phase transitions in neural networks \cite{atanasov2024scaling}, showing how collective behavior emerges in learning dynamics (as opposed to efforts \cite{beny2013deep,mehta2014exact,li2021statistical,howard2025wilsonian} to consider deep learning itself as a renormalization group process). 
It is likely that the phase transition phenomena {and the universal scaling framework} we have reported in linear regression are universal across network architectures, potentially with different universality classes for networks specified by range of interactions, nonlinearity and task complexity. {For example, random feature models have additional (multiplicative) randomness in their features, not just the labels of the data, and thus could provide a natural testbed for extending our work.  They decouple parameter count from data dimension, reflecting certain modern machine learning models, and exhibit double descent even without label noise \cite{mei2022generalization,bordelon2024dynamical}.  
Such models, even with different nonlinear activation functions, can be shown to exhibit asymptotic equivalence to linear Gaussian covariate models with additive noise~\cite{mei2022generalization}, and thus show data collapse.  However, this occurs for all values of the network capacity (and is an analogue of the so-called law of corresponding states \cite{goldenfeld2018lectures_m} for networks), not just in the critical regime around the interpolation threshold.  We are not aware of any demonstrated renormalization group universality near the interpolation threshold, and this is left for future work.}

% Under suitable conditions, their asymptotic equivalence to linear Gaussian covariate models with additive noise~\cite{mei2022generalization} provides direct evidence of universality across distinct microscopic constructions, including data collapse. 
% %However, care is required to interpret this result, since the reported data collapse reflects a law of corresponding states~\cite{goldenfeld2018lectures_m} rather than the critical point universality discussed here.}
% However, this universality applies for all values of the variable analogous to network capacity, holds because all the models studied 

%
%In this respect our work is related to these more complex machine learning contexts in the same way that infinite-range Gaussian and spherical models \cite{berlin1952spherical,binney1992theory} of magnets exhibit the qualitative features of phase transitions but whose exponents and scaling functions are quantitatively different from short-range interacting models \cite{goldenfeld2018lectures_m}. 
%
%

{\it Acknowledgments.---} We thank Yonatan Aljadeff for discussions and Doug Durian for discussions and sharing unpublished data on double descent \cite{Durian2025} in analog physical networks \cite{dillavou2024machine}. We also thank two anonymous referees for their helpful comments that improved the presentation of our work.

{\normalsize
  \begingroup
  \renewcommand{\addcontentsline}[3]{}%
  \endgroup

}

\clearpage

% End matter
\begin{center}
    \textbf{\large End Matter}
\end{center}

%%%%%%%%%%%%%%%%%%%%%%%%%%%%%%%%%%%%%%%%%%%%%%%%%%%%%%%%%%%%%%%%
%%%%%%%%%%%%%%%%%%%%%%%%%%%%%%%%%%%%%%%%%%%%%%%%%%%%%%%%%%%%%%%%%

In this End Matter, we present further details of the scaling laws near the phase transition at the interpolation threshold.  First we consider the case that there is no regularization $\lambda=0$, and compute the critical exponents.  Then we calculate how regularization stops the divergence of the generalization error, and shifts the position of its maximum.

\textit{Calculation of the critical exponents in DMFT.---}
To calculate the critical exponents $\{\sigma,\phi\}$, we first analyze the self-consistent equations for the response and correlation functions in the frequency domain (see Sec.IV of the Supplemental Material \cite{SM}). 
In the long-time limit and near criticality ($\epsilon = |1 - \alpha| \to 0$), we find that there is indeed data-collapse scaling, with the exponents $\sigma=1$ and $\phi={2}$, together with the explicit form for the universal scaling function valid on both sides of the transition (see sec.~V.B.2 of the Supplemental Material \cite{SM}),
\begin{equation}
\begin{aligned}
F(x)&=- \frac{2 \sqrt{x}}{\sqrt{\pi}} \exp \left(-2 x\right)\left(\sqrt{2}-2 \exp \left(x\right)\right)\\&+\left(\left(2+4 x\right) \operatorname{Erf}(\sqrt{x})-\left(1+4 x \right) \operatorname{Erf}(\sqrt{2x} )\right).
\label{eq:scalingf}
\end{aligned}
\end{equation}
The scaling form recovers the correct asymptotics.
For $|1-\alpha|\neq0$, the scaling variable $x = {(1-\alpha)^2 \gamma t}/{4}$ diverges as $t\to\infty$, giving $F(x\to\infty)=1$ and $\varepsilon_g \sim D/|1-\alpha|$.
At criticality ($\alpha \to 1^{\pm}$), $x\to 0$ and $\varepsilon_g \sim D\tfrac{2(2 - \sqrt{2})}{\sqrt{\pi}}\sqrt{\gamma t}$.

\begin{figure}
\centering
\includegraphics[width=0.5\textwidth]{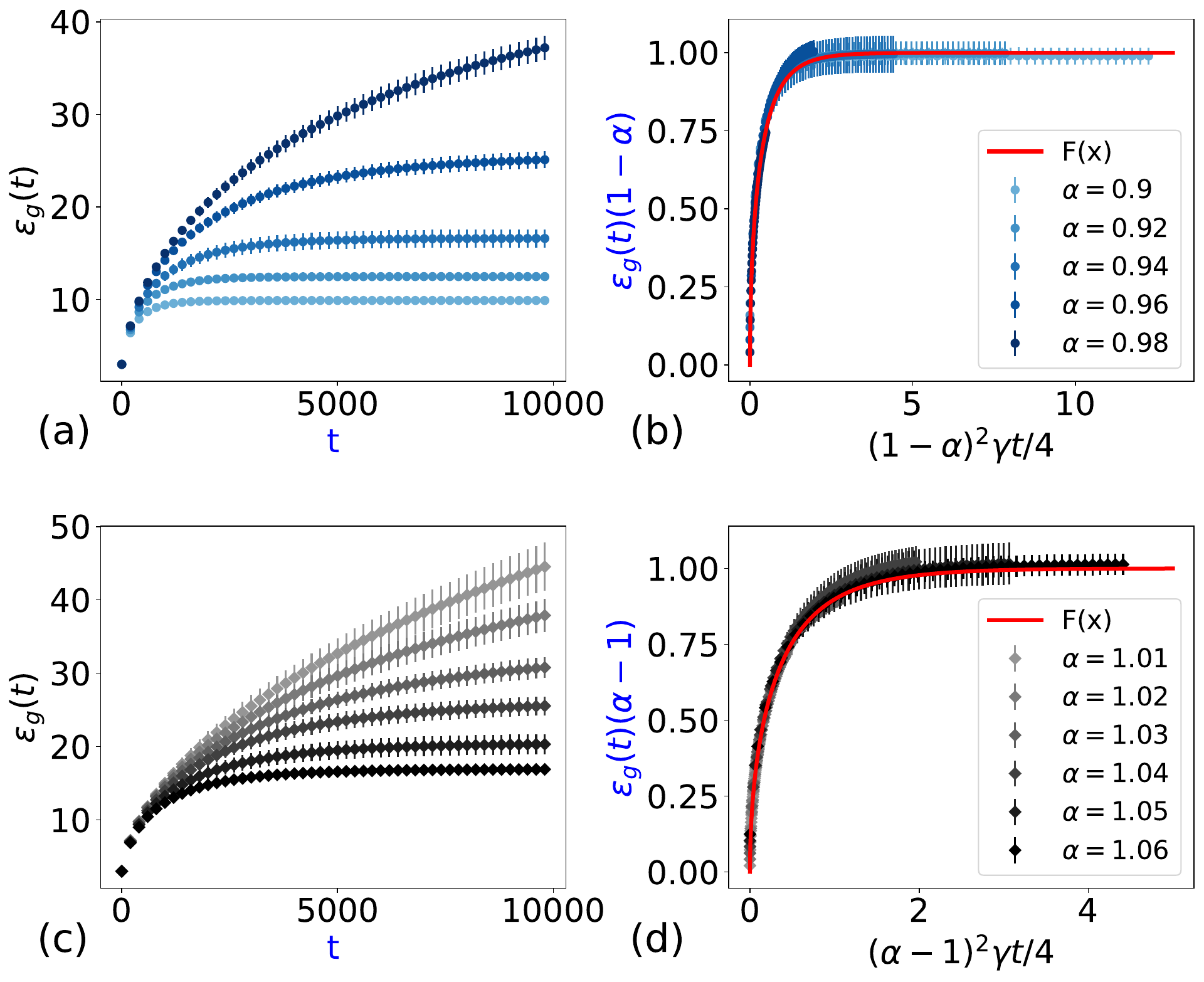}
\caption{ Comparison between theoretical data collapse and simulation results in an unregularized ($\lambda = 0$) single-layer linear neural network ($N = 10{,}000$, $d = \alpha N$, $D = 1$, $r = 1$, $\gamma = 0.5$, $\New{\zeta}=0.1$ and training time $T=10{,}000$).  The error bar characterizes the fluctuation across ten independently network realizations.
Panels (a)–(b) correspond to the under-parameterized regime ($\alpha < 1$), while panels (c)–(d) correspond to the over-parameterized regime ($\alpha > 1$).  
(a, c) Generalization error $\varepsilon_g(t)$ as a function of \new{$t$}, with markers indicating simulation results from stochastic gradient descent for various values of $\alpha$.  
(b, d) Scaled generalization errors plotted against the scaling variable, showing excellent agreement with the predicted scaling function $F(x)$ (red line).  
}
\label{datacollapse}
\end{figure}

This scaling theory characterizes double descent as a critical phenomenon and demonstrates excellent agreement with stochastic gradient descent simulations, as shown in Fig.~\ref{datacollapse}.  
Panels (a) and (c) reveal that as $\alpha \to 1^{\pm}$, the generalization error not only increases in magnitude but also converges more slowly, reflecting a hallmark of critical slowing down near the transition point.  
Despite these diverging timescales, all curves collapse onto a single universal scaling function $F(x)$ when plotted against the appropriate rescaled time variable, as shown in panels (b) and (d), validating the predicted scaling structure.

\textit{Regularization as finite-size scaling in learning dynamics.}---
An important technique in machine learning is to use $\ell_2$ regularization to bias the dynamics toward small-norm solutions, which are smoother and more stable, and often yield improved generalization\cite{hastie2009elements}. 
This comes at the cost of a tradeoff between minimizing the training error and constraining the weight norm.  Regularization acts like an external relevant field at the interpolation threshold, and rounds the divergence, with scaling laws that can be obtained by standard renormalization group considerations (see sec.~V.A.2 of the Supplemental Material \cite{SM}).  Here we present a summary of the scaling laws we obtain, and refer the interested reader to the detailed calculation in the Supplemental Material.

Introducing a weak regularization term ($\lambda \to 0$) modifies the near-critical scaling behavior of the generalization error. 
To leading order, the corresponding scaling form under weak regularization in the long-time limit near criticality can be expressed as follows (see Sec.~V.A.2 of the Supplemental Material \cite{SM}):
\begin{equation}
\varepsilon_g(\alpha \to 1^{\pm}, \lambda \to 0)
\;\approx\; 
\frac{D F\!\left({\left[\tfrac{\gamma (1 - \alpha)^2}{4} + \lambda\right]t}\right)}
{|1 - \alpha|\,\tilde{F}\!\left(\tfrac{4\lambda}{\gamma(1 - \alpha)^2}\right)},
\label{eq:regu}
\end{equation}
where $\tilde{F}(x) = \sqrt{1 + x}$ serves as a static scaling factor that regularizes the $|1 - \alpha|^{-1}$ singularity, with $x = 4\lambda / [\gamma (1 - \alpha)^2]$, while $F(x)$ denotes the same universal dynamical scaling function as in Eq.~(\ref{eq:scalingf}), capturing the time evolution of the generalization error near criticality.
Eq.~(\ref{eq:regu}) recovers the unregularized scaling form of Eq.~(\ref{eq:udc}) when taking $\lambda = 0$, reproducing the divergence at the interpolation threshold $\alpha_c=1$. To leading order, the scaling form is symmetric about $\alpha = 1$, reflecting the mirror-like behavior of the under- and over-parameterized regimes near criticality.

In the long-time limit, after the system reaches steady state, the generalization error asymptotically approaches
\begin{equation}
\varepsilon_g(t \to \infty) \sim \frac{D}{|1 - \alpha|\,\tilde{F}\!\left(\tfrac{4\lambda}{\gamma(1 - \alpha)^2}\right)},
\end{equation}
defining the static data-collapse formula near the interpolation threshold. 

\begin{figure}
\centering
\includegraphics[width=0.5\textwidth]{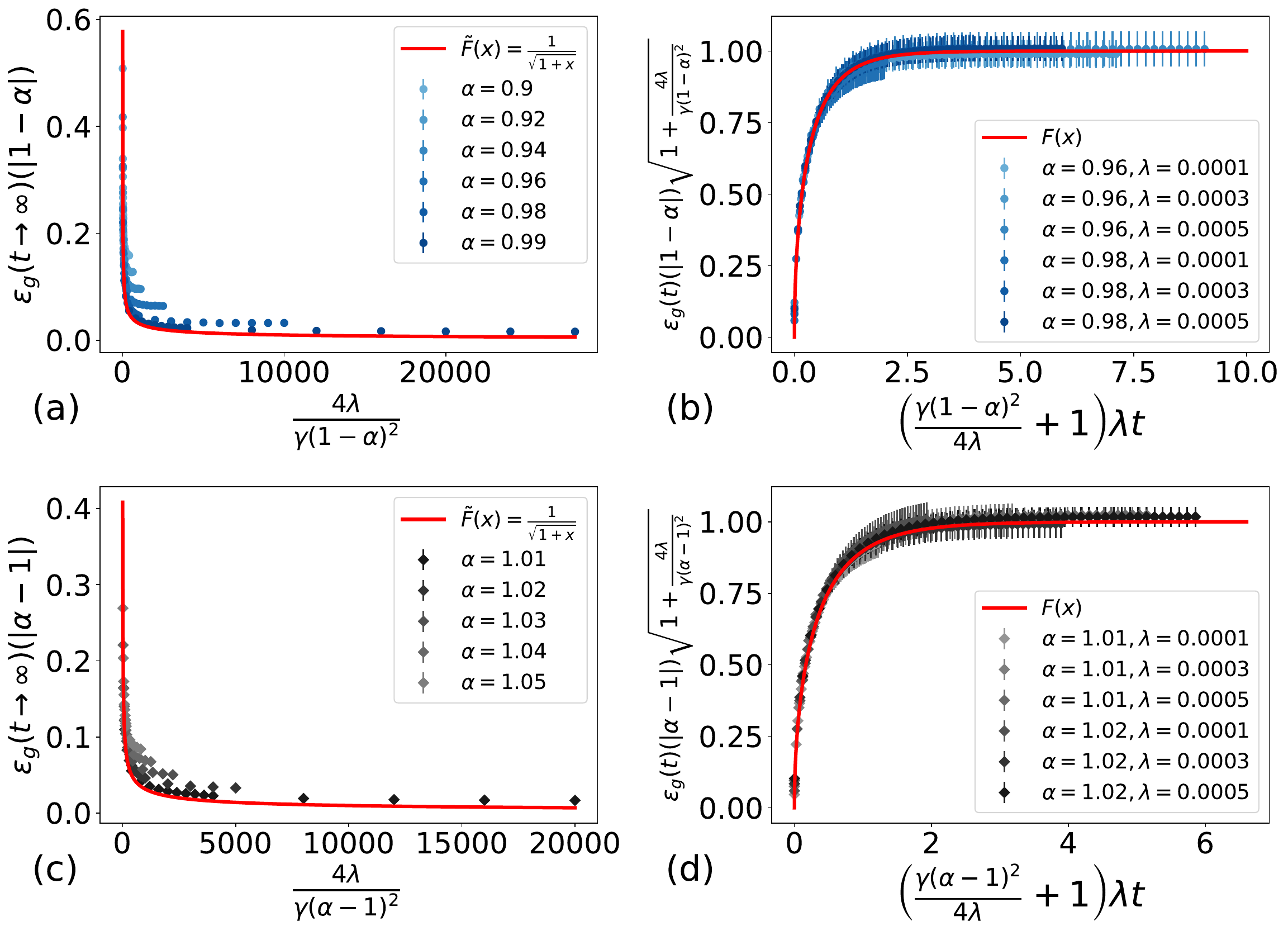}
\caption{Comparison between theoretical scaling predictions and simulation results in a regularized single-layer linear neural network ($N = 10{,}000$, $d = \alpha N$, $D = 1$, $r = 1$, $\gamma = 1$).  
Error bars represent fluctuations across ten independent realizations.  
Panels (a)–(b): under-parameterized regime ($\alpha < 1$);  
panels (c)–(d): over-parameterized regime ($\alpha > 1$).  
(a, c) Steady-state generalization error $\varepsilon_g(t \to \infty)$, rescaled by $|1 - \alpha|$, plotted against the dimensionless regularization variable $4\lambda / \gamma(1 - \alpha)^2$.  
Markers denote different values of $\alpha$, with $\lambda \in [0.01, 1.0]$.  
Only markers in the weakly regularized regime collapse onto the theoretical prediction (red).  
(b, d) Full temporal dynamics: rescaled generalization errors versus the scaling variable ${\left({\gamma(1 - \alpha)^2}/{4 \lambda} + 1\right)\lambda t}$ collapse onto the universal curve $F(x)$ (red), validating the dynamical scaling form.
}
\label{datacollapsel}
\end{figure}

The scaling predictions are quantitatively confirmed in Fig.~\ref{datacollapsel}. Panels~(a, c) show that the asymptotic generalization error collapses onto the theoretical curve $\tilde{F}(x)$ when $\varepsilon_g(t \to \infty)|1-\alpha|$ is plotted against $x = 4\lambda / \gamma(1 - \alpha)^2$, with excellent agreement in the weakly regularized regime. Deviations at larger $\lambda$ reflect subleading corrections beyond the scaling limit. Panels (b, d) demonstrate collapse of the full temporal dynamics onto the universal function $F(x)$ under the scaling variable ${[\gamma(1 - \alpha)^2 / 4\lambda + 1]\, \lambda t}$. The collapse is sharper for $\alpha < 1$, indicating an asymmetry in the subleading corrections to scaling, despite the symmetry of the leading-order theory across the interpolation threshold. Together, the static and dynamical scaling forms offer a unified description of generalization dynamics near criticality.

\begin{figure} 
\centering 
\includegraphics[width=0.5\textwidth]{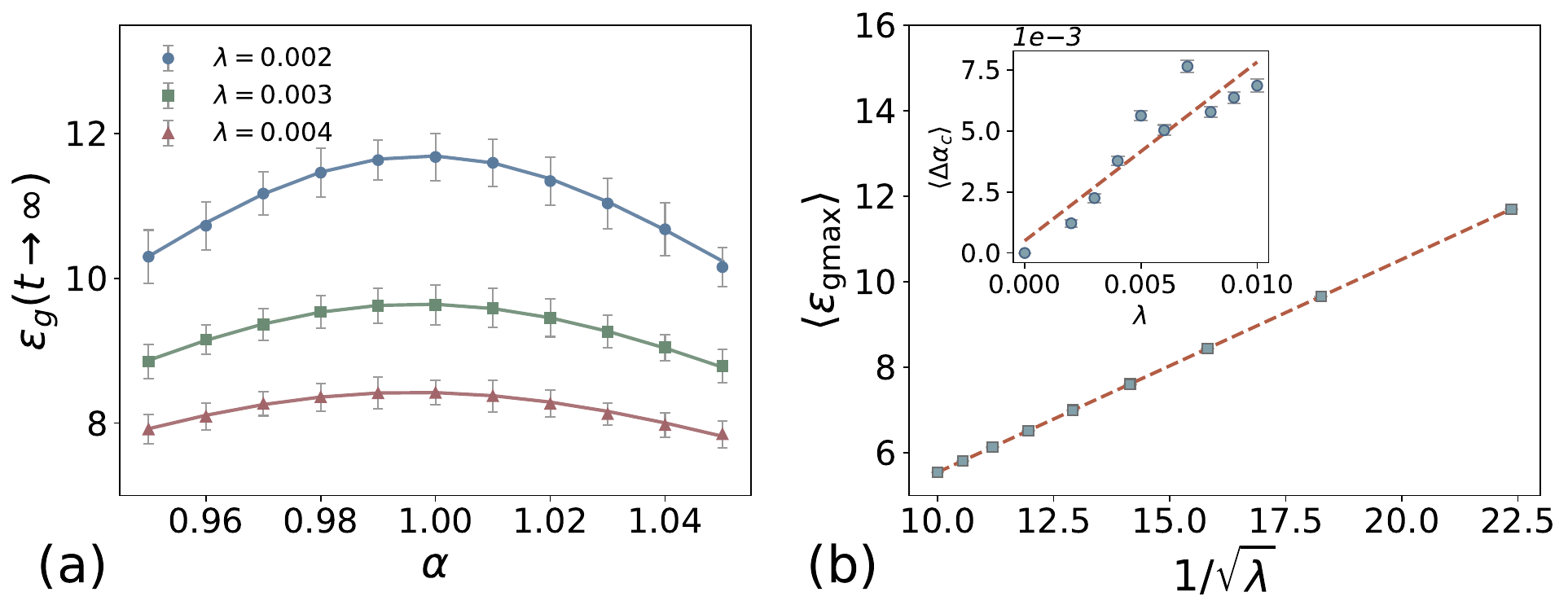}
\caption{Regularization rounds the divergence of the generalization error. 
Simulations are performed for a regularized single-layer linear network ($N = 10{,}000$, $d=\alpha N$, $D=1$, $r=0.1$, $\gamma=1$). 
(a) Steady state generalization error $\varepsilon_g(t\to\infty)$ versus $\alpha$ for different $\lambda$. 
Markers denote simulation results (50 runs) and solid lines are the exact theoretical values from Eq.~(\ref{eq:reguexact}). 
(b) Peak generalization error $\varepsilon_{\text{gmax}}$ versus $\lambda$, showing $\varepsilon_{\text{gmax}}\propto \lambda^{-1/2}$. 
Inset: The corresponding peak shift, $\Delta\alpha_c = 1 - \alpha_c$, increases linearly with $\lambda$.
Error bars in (b) and in the inset represent standard error of the mean estimated via bootstrap resampling ($B = 1,000$; see sec.~III.B of the Supplemental Material \cite{SM}). Dashed line represents a linear fit to the data points.}
\label{regularization} 
 \end{figure}

The static scaling function reveals that regularization smooths the divergence at $\alpha_c = 1$, with the peak of the generalization error scaling as $\varepsilon_{\text{gmax}} \propto \lambda^{-1/2}$ [Fig.~\ref{regularization}(b)]. The leading order behavior is symmetric across the threshold, indicating that the maximum value of generalization error lies at $\alpha_c=1$, while the subleading corrections shift the peak slightly leftward by $\Delta\alpha_c \propto \lambda$ (see Sec.~V.A.3 of the Supplemental Material \cite{SM}), thereby breaking the symmetry about $\alpha = 1$ [inset of Fig.~\ref{regularization}(b); Figs.~\ref{datacollapsel}(b,d)].

In this sense, $\lambda$ acts as a finite-size parameter in statistical mechanics: it cuts off the divergence, shifts the apparent critical point, and smooths the sharp transition into a rounded crossover \cite{goldenfeld2018lectures}. 

For finite $\lambda$, the steady-state generalization error can be computed exactly in terms of the regularized susceptibility $\chi_\lambda = \int_0^\infty R(\tau)d\tau$:
\begin{equation}
\varepsilon_{\mathrm{gen}}(t\to\infty) =
\frac{D + r \lambda^2 \chi_\lambda^2}{1 - \alpha (\lambda \chi_\lambda - 1)^2},
\label{eq:reguexact}
\end{equation}
which reduces to the scaling law in Eq.~(\ref{eq:regu}) as $\lambda \to 0$ near criticality.

Stochasticity influences learning dynamics in qualitatively distinct ways depending on whether the system is regularized. In the unregularized case ($\lambda = 0$), it acts only as a temporal rescaling, slowing convergence without affecting the asymptotic generalization performance. In contrast, when regularization is present ($\lambda \neq 0$), stochasticity fundamentally reshapes training dynamics: it allows the dynamics to escape sharp minima favored by deterministic optimization and explore flatter regions of the loss landscape that yield superior generalization \cite{hochreiter1997flat_m,baldassi2021unveiling_m}. In this regime, the noise introduced by SGD plays a constructive role, mediating a dynamical balance between data fitting and norm minimization, and steering the system toward flatter, more robust solutions that exhibit improved generalization (see Sec.~V.C of the Supplemental Material \cite{SM}).

\clearpage
\onecolumngrid

% Supplemental material title block
\begin{center}
{\bfseries Supplemental Material for}\\[0.5em]
{\bfseries\textit{Broken Ergodicity and the Violation of the Fluctuation--Dissipation Theorem Lead to Generalization Beyond Overfitting in Machine Learning}}\\[1.0em]

Chan Li$^1$ and Nigel Goldenfeld$^{1,2}$\\[0.5em]

{\small
$^1$Department of Physics, University of California San Diego, 9500 Gilman Drive, La Jolla, California 92093, USA\\
$^2$Hal\i c\i o\u{g}lu Data Science Institute, University of California San Diego, 9500 Gilman Drive, La Jolla, California 92093, USA
}\\[0.5em]
\end{center}

% Supplemental material abstract
\begin{quote}
\new{This Supplementary Material contains details of the technical calculations reported in the main paper regarding the dynamic mean field theory for double descent in linear regression. It also includes some background material for readers unfamiliar with several aspects of the mathematical connections of machine learning to models in statistical and condensed matter physics, including simple models of superconductivity and ergodicity breaking.}
\end{quote}

% Supplemental material numbering

\setcounter{section}{0}
\renewcommand{\thesection}{\Roman{section}}

\setcounter{subsection}{0}
\renewcommand{\thesubsection}{\thesection.\Alph{subsection}}

\setcounter{subsubsection}{0}
\renewcommand{\thesubsubsection}{\thesubsection.\arabic{subsubsection}}

\setcounter{figure}{0}
\renewcommand{\thefigure}{S\arabic{figure}}

\setcounter{table}{0}
\renewcommand{\thetable}{S\arabic{table}}

\setcounter{equation}{0}
\renewcommand{\theequation}{S\arabic{equation}}

\setcounter{secnumdepth}{3}

% Supplemental material table of contents
\tableofcontents

\clearpage

% Supplemental material body

% \begin{abstract}
% \new{This Supplementary Material contains details of the technical calculations reported in the main paper regarding the dynamic mean field theory for double descent in linear regression.  It also includes some background material for readers unfamiliar with several aspects of the mathematical connections of machine learning to models in statistical and condensed matter physics, including simple models of superconductivity and ergodicity breaking.}
% \end{abstract}

\section{Details of the Dynamical Mean-Field Theory}
\label{sec:DMFT}
We consider stochastic gradient descent (SGD) dynamics for a teacher-student linear regression model in high dimensions. The student weights $\hat{\boldsymbol{\beta}}(t) \in \mathbb{R}^d$ are updated to minimize the regularized loss function
$$
\mathcal{L} = \frac{1}{2} \sum_{\mu=1}^N s^\mu(t)\left(\hat{y}^\mu - y^\mu\right)^2 + \frac{\lambda}{2} \|\hat{\boldsymbol{\beta}}(t)\|_2^2,
$$
where $s^\mu(t) \in \{0,1\}$ is a binary indicator variable drawn independently at each iteration, with $\mathbb{P}(s^\mu = 1) = \gamma$, controlling the stochastic sampling of training examples. The target labels are generated by a fixed teacher through
$$
y^\mu = \frac{1}{\sqrt{d}} \sum_{i=1}^d \beta_i x_i^\mu + \epsilon^\mu,
$$
where $\beta_i$ are the ground-truth weights and $\epsilon^\mu$ is zero-mean Gaussian noise with variance $\langle \epsilon^\mu \epsilon^\nu \rangle = D \delta_{\mu \nu}$. The student prediction is $\hat{y}^\mu =\frac{1}{\sqrt{d}} \sum_{k=1}^d \hat{\beta}_k x_k^\mu$, leading to the following discrete-time update rule:
\begin{equation}
\begin{aligned}
\hat{\beta}_i(t+1) &= \hat{\beta}_i(t) - \New{\zeta} \frac{\partial \mathcal{L}}{\partial \hat{\beta}_i(t)} \\
&= \hat{\beta}_i(t) - \New{\zeta} \left[ \sum_{\mu=1}^N s^\mu(t) \frac{x_i^\mu}{\sqrt{d}} \left( \sum_{k=1}^d \frac{x_k^\mu}{\sqrt{d}} \left( \hat{\beta}_k(t) - \beta_k \right) - \epsilon^\mu \right) + \lambda \hat{\beta}_i(t) \right].
\end{aligned}
\label{eq:update}
\end{equation}

To probe the response properties, we introduce an external source term $ j_i(t)$ into the dynamical equation. In the limit $\New{\zeta} \to 0$, this leads to the continuous-time formulation:
\begin{equation}
\partial_t {\hat{\beta}}_i(t) = - \sum_{\mu=1}^N s^\mu(t) \frac{x_i^\mu}{\sqrt{d}} \left( \sum_{k=1}^d \frac{x_k^\mu}{\sqrt{d}} \left( \hat{\beta}_k(t) - \beta_k \right) - \epsilon^\mu \right) - \lambda \hat{\beta}_i(t)+j_i(t).
\label{eq:continuous}
\end{equation}

We begin by introducing the generating functional formalism and outline the derivation following standard procedures. The core idea is to reformulate the original high-dimensional dynamics as a path integral \cite{chow2015path,hertz2016path,zou2024introduction}. Specifically, we adopt the Martin–Siggia–Rose–Janssen–De Dominicis (MSRJD) formalism \cite{martin1973statistical,janssen1976lagrangean,de1976technics} to express the stochastic gradient flow as a dynamical partition function, or moment-generating functional:
\begin{equation}
Z_{\mathrm{dyn}} = \int \mathcal{D} \boldsymbol{\hat{\beta}}(t)\, \prod_{i=1}^d \delta\left[-\partial_t{\hat{\beta}}_i(t) - \sum_{\mu=1}^N s^\mu(t) \frac{x_i^\mu}{\sqrt{d}} \left( \sum_{k=1}^d \frac{x_k^\mu}{\sqrt{d}} \left( \hat{\beta}_k(t) - \beta_k \right) - \epsilon^\mu \right) - \lambda \hat{\beta}_i(t)+j_i(t) \right] = 1.
\label{eq:partition}
\end{equation}
Here, $\mathcal{D} \boldsymbol{\hat{\beta}}(t)$ denotes the path integral measure over dynamical trajectories originating from the initial condition $\boldsymbol{\hat{\beta}}(0)$. Since the dynamical partition function satisfies $Z_{\mathrm{dyn}} = 1$, we can directly perform an average over the disorder, including the input dataset $\{\mathbf{x}^\mu\}_{\mu=1}^{N}$, the initial condition $\boldsymbol{\hat{\beta}}(0)$, the stochastic sampling variables $s_{\mu}(t)$, and the label noise $\{\epsilon^\mu\}_{\mu=1}^{N}$. This average is denoted by angle brackets $\langle \cdot \rangle$. Here we can write:
\begin{equation}
Z_{\operatorname{d y n}}[\mathbf{j}, \tilde{\mathbf{j}}]=\left\langle\int \mathcal{D}  \hat{\boldsymbol{\beta}}(t) \mathcal{D} {\tilde{\boldsymbol{\beta}}}(t) \exp{\left(S_{\operatorname{d y n}}+\sum_{i=1}^d \int \tilde{j}_i(t) \hat{\beta}_i(t) d t+\sum_{i=1}^N \int j_i(t) \tilde{\beta}_i(t) d t\right)}\right\rangle, \\
\end{equation}
where we have defined the action:
\begin{equation}
S_{\operatorname{d y n}}=\sum_i^d \int_0^{\infty} d t i \tilde{\beta}_i(t)\left(-\partial_t{\hat{\beta}}_i(t) - \sum_{\mu=1}^N s^\mu(t) \frac{x_i^\mu}{\sqrt{d}} \left( \sum_{k=1}^d \frac{x_k^\mu}{\sqrt{d}} \left( \hat{\beta}_k(t) - \beta_k \right) - \epsilon^\mu \right) - \lambda \hat{\beta}_i(t) \right), \\
\label{eq:action}
\end{equation}
and we have introduced a set of response fields $\tilde{\boldsymbol{\beta}}(t)$, along with two source terms $\mathbf{j}(t)$ and $\tilde{\mathbf{j}}(t)$. The source $\mathbf{j}(t)$ couples linearly to the dynamical field $\hat{\boldsymbol{\beta}}(t)$ and serves as a weak external perturbation applied to the system. It enables the computation of linear response functions, such as $R(t, t^{\prime}) = \delta \langle \hat{\beta}(t) \rangle / \delta j(t^{\prime})$, evaluated in the limit $\mathbf{j} \to 0$. The auxiliary source $\tilde{\mathbf{j}}(t)$ couples to the response field $\hat{\boldsymbol{\beta}}(t)$ and is introduced to generate correlation functions such as $\langle \hat{\beta}_i(t)\, \hat{\beta}_j(t^{\prime}) \rangle$, via functional differentiation. As $\tilde{\mathbf{j}}(t)$ does not correspond to a physical perturbation, it is set to zero after the relevant observables are obtained.

Next, to perform the quenched averages, we define two random fields $h_{\mu}(t)$ and $H_{\mu}(t)$:
\begin{equation}
\begin{aligned}
H^\mu(t) & = \frac{1}{\sqrt{d}} \sum_{i=1}^d x_i^\mu \left( \hat{\beta}_i(t) - \beta_i \right) -  \epsilon^\mu, \\
\tilde{h}^\mu(t) &= \frac{1}{\sqrt{d}} \sum_{i=1}^d i  \tilde{\beta}_i(t) x_i^\mu.
\label{eq: random2}
\end{aligned}
\end{equation}
The statistical properties of the random fields are governed by the statistics of the training data and the correlation functions between fields. While our theoretical framework is applicable to input data with arbitrary distributions, we focus on the standard Gaussian case for simplicity. In this setting, the inputs have zero mean, $\langle x_i^\mu \rangle = 0$, and are uncorrelated, $\langle x_i^\mu x_j^\nu \rangle = \delta_{ij} \delta_{\mu\nu}$, where $\delta_{ij}$ and $\delta_{\mu\nu}$  denote the Kronecker delta.
\begin{equation}
\begin{aligned}
\langle \tilde{h}^\mu(t) \rangle &= \langle H^\mu(t) \rangle=0,\\
\langle  H^\mu(t) H^\nu(t^\prime)   \rangle& = \frac{\delta_{\mu\nu}}{d} \sum_{i=1}^d \left( \hat{\beta}_i(t) - \beta_i \right) \left( \hat{\beta}_i(t^\prime) - \beta_i \right)+\delta_{\mu\nu}D,\\
& = \delta_{\mu\nu} \left(Q_1(t,t^\prime)+D\right),\\
\langle  H^\mu(t) h^\nu(t^\prime)   \rangle& =\frac{ \delta_{\mu\nu}}{d} \sum_{i=1}^d i\tilde{\beta}_i(t^\prime)\left( \hat{\beta}_i(t) - \beta_i \right)= \delta_{\mu\nu} \left(Q_2(t,t^\prime) - Q_4(t^\prime)\right) ,\\
\langle \tilde{h}^\mu(t)  \tilde{h}^\nu(t^\prime)   \rangle& = \frac{\delta_{\mu\nu}}{d}   \sum_{i=1}^d i  \tilde{\beta}_i(t) i  \tilde{\beta}_i(t^\prime) = \delta_{\mu\nu}Q_3(t, t^\prime),\\
\label{eq: statistics}
\end{aligned}
\end{equation}
We have defined four auxiliary overlap functions:
\begin{itemize}
  \item $Q_1(t,t^{\prime}) = \frac{1}{d} \sum_{i=1}^d \left( \hat{\beta}_i(t) - \beta_i \right) \left( \hat{\beta}_i(t^{\prime}) - \beta_i \right)$ quantifies the temporal correlation of deviations between student and teacher weights;
  \item $Q_2(t,t^{\prime}) = \frac{1}{d} \sum_{i=1}^d i \tilde{\beta}_i(t^{\prime}) \hat{\beta}_i(t)$ couples the response field to the current model state;
  \item $Q_3(t,t^{\prime}) = \frac{1}{d} \sum_{i=1}^d i \tilde{\beta}_i(t) \, i \tilde{\beta}_i(t^{\prime})$ encodes the autocorrelation of the response field;
  \item $Q_4(t) = \frac{1}{d} \sum_{i=1}^d i \tilde{\beta}_i(t) \beta_i$ represents the overlap between the response field and the teacher.
\end{itemize}
The physical significance of these observables will become clear in the subsequent analysis. Next, we hope to do the average over the random fields, which can be written explicitly:
\begin{equation}
\begin{aligned}
Z_{\mathrm{dyn}}[\mathbf{j}, \tilde{\mathbf{j}}] 
&= \left\langle 
\int \mathcal{D} \hat{\boldsymbol{\beta}}(t) \, \mathcal{D} \tilde{\boldsymbol{\beta}}(t) 
\exp \left[ 
\int_0^{\infty} dt \left( 
- \sum_{i=1}^d i \tilde{\beta}_i(t) {\hat{\beta}}_i(t) 
- \sum_{\mu=1}^N s^\mu(t) \tilde{h}^\mu(t) H^\mu(t) 
- \lambda \sum_{i=1}^d i \tilde{\beta}_i(t) \hat{\beta}_i(t) 
\right) \right. \right. \\
& \left. \left. 
+ \int_0^{\infty} dt \left( 
\sum_{i=1}^d \tilde{j}_i(t) \hat{\beta}_i(t) 
+ \sum_{i=1}^d j_i(t) \tilde{\beta}_i(t) 
\right) 
\right] 
\right\rangle_{s_{\mu}(t), h^\mu(t), H^\mu(t)}.
\end{aligned}
\end{equation}
Here we consider the case that training data samples are uncorrelated, the dynamical partition function can be further simplified as:
\begin{equation}
\begin{aligned}
Z_{\operatorname{d y n}}[\mathbf{j}, \tilde{\mathbf{j}}]&=\int \mathcal{D} \hat{\boldsymbol{\beta}}(t) \mathcal{D} {\tilde{\boldsymbol{\beta}}}(t) \left[\exp{\left( \int_0^{\infty} d t \left(-\sum_i^di \tilde{\beta}_i(t)\partial_t {\hat{\beta}}_i(t) - \lambda \sum_{i=1}^d i \tilde{\beta}_i(t) \hat{\beta}_i(t) \right)+\sum_{i=1}^d \int \tilde{j}_i(t) \hat{\beta}_i(t) d t+\sum_{i=1}^N \int j_i(t) \tilde{\beta}_i(t) d t\right)}\right.\\
&\times \left.\left\langle \exp{\left(-  s(t)\tilde{h}(t)H(t)\right)}\right\rangle^N_{s_{\mu}(t), h^\mu(t), H^\mu(t)}\right].
\end{aligned}
\end{equation}
To evaluate the average over the stochastic indicator $s(t)$, we define
$I = \left\langle \exp \left( - \int_0^{\infty} dt s(t)  \tilde{h}(t)  H(t) \right) \right\rangle_{s(t)}$,
where the binary variable $s(t) \in \{0, 1\}$ denotes whether a training sample is selected at time $t$. We discretize time as $t_n$ with $n = 0, 1, 2, \dots$, and assume that the sampling variable $s(t_n)$ is temporally uncorrelated, taking the value $1$ with probability $\gamma$ and $0$ with probability $1 - \gamma$.
\begin{equation}
\begin{aligned}
I & =\left\langle\exp \left(-\int_0^{\infty} d t s(t) \tilde{h}(t)H(t)\right)\right\rangle \\
& =\left\langle\exp \left(-\sum_n \Delta t s\left(t_n\right) \tilde{h}\left(t_n\right) H\left(t_n\right)\right)\right\rangle \\
& =\left\langle\exp \left(-\Delta t s\left(t_n\right) \tilde{h}\left(t_n\right) H\left(t_n\right)\right)\right\rangle^n \\
& =\left(\gamma \exp \left(-\Delta t \tilde{h}\left(t_n\right) H\left(t_n\right)\right)+(1-\gamma)\right)^n.
\end{aligned}
\end{equation}
Next, we take the continuous-time limit $\Delta t \to 0$ and expand the exponential function to leading order,
\begin{equation}
\begin{aligned}
I & =\left(\gamma \exp \left(-\Delta t \tilde{h}\left(t_n\right) H\left(t_n\right)\right)+(1-\gamma)\right)^n \\
& \approx\left(\gamma\left(1+\left(-\Delta t \tilde{h}\left(t_n\right) H\left(t_n\right)\right)\right)+(1-\gamma)\right)^n \\
& =\left(1-\Delta t \gamma \tilde{h}\left(t_n\right) H\left(t_n\right)\right)^n \\
& \approx \exp \left(-\sum_n \Delta t \gamma \tilde{h}\left(t_n\right) H\left(t_n\right)\right) \\
& =\exp \left(-\gamma \int_0^{\infty} d t \tilde{h}(t) H(t)\right).
\end{aligned}
\label{eq:avs}
\end{equation}
In the following analysis, we rescale the response field such that $\gamma \tilde{h}(t) \rightarrow \tilde{h}(t)$ for notational simplicity. Based on the central limit theorem, in the limit of large input dimension $d$, the random fields $H^\mu(t)$ and $\tilde{h}^\mu(t)$ become jointly Gaussian with zero mean. Their second-order statistics are fully determined by the overlap functions $Q_1$, $Q_2$, $Q_3$, and $Q_4$, and are encoded in the covariance matrix:
\begin{equation}
\boldsymbol{\Sigma}\left(t, t^{\prime}\right)=\left(\begin{array}{cc}
Q_1\left(t, t^{\prime}\right)+D & \gamma Q_2\left(t, t^{\prime}\right)-\gamma Q_4\left(t^{\prime}\right) \\
\gamma Q_2\left(t^{\prime}, t\right)-\gamma Q_4\left( t\right) & \gamma^2 Q_3\left(t, t^{\prime}\right).
\end{array}\right)
\label{eq: covariance}
\end{equation}
Given the statistical properties of random fields, we are able to calculate the average $\left\langle \exp \left(- \int_0^{\infty} d t \tilde{h}(t) H(t)\right)\right\rangle_{h(t), H(t)}$, which can be written as:
\begin{equation}
\begin{aligned} 
\left\langle \exp \left(- \int_0^{\infty} d t \tilde{h}(t) H(t)\right)\right\rangle_{h(t), H(t)}& =\int \mathcal{D} \tilde{h} \mathcal{D} H \exp \left(-\int_0^{\infty} d t \tilde{h}(t) H(t)\right) \mathcal{P}[\tilde{h}, H],\\
& = \frac{1}{\left|\left(\begin{array}{cc}
\gamma Q_2^\top - \gamma Q_4+\delta &\gamma^2 Q_3 \\
Q_1+D& \gamma Q_2-\gamma Q_4^\top+\delta,
\end{array}\right)\right|},
\end{aligned}
\end{equation}
where $\mathcal{P}[\tilde{h}, H]$ denotes the joint Gaussian probability measure of the two random fields, $|\cdot|$ represents the functional determinant of operator-valued kernels, and the superscript $\top$ indicates the transpose of a kernel. After performing the quenched average, the results can be reorganized and the dynamical partition function takes the form:
\begin{equation}
\begin{aligned}
Z_{\operatorname{d y n}}[\mathbf{j}, \tilde{\mathbf{j}}]&=\int \mathcal{D} \hat{\boldsymbol{\beta}}(t) \mathcal{D} {\tilde{\boldsymbol{\beta}}}(t) \left[\exp{\left( \int_0^{\infty} d t \left(-\sum_i^di \tilde{\beta}_i(t)\partial_t {\hat{\beta}}_i(t)- \lambda \sum_{i=1}^d i \tilde{\beta}_i(t) \hat{\beta}_i(t)  \right)+\sum_{i=1}^d \int \tilde{j}_i(t) \hat{\beta}_i(t) d t+\sum_{i=1}^N \int j_i(t) \tilde{\beta}_i(t) d t\right)}\right.\\
&\left.\times  \frac{1}{\left|\left(\begin{array}{cc}
\gamma Q_2^\top - \gamma Q_4+\delta &\gamma^2 Q_3 \\
Q_1+D& \gamma Q_2-\gamma Q_4^\top+\delta.
\end{array}\right)\right|}\right].
\label{eq:avparti}
\end{aligned}
\end{equation}
To enforce the definitions of the observables, we introduce the associated observables into Eq.\ref{eq:avparti} using the Fourier integral representation of Dirac delta functions, which impose these constraints at the level of the path integral.
\begin{equation}
\begin{aligned}
& \delta\left(-dQ_1\left(t, t^{\prime}\right)+\sum_i\left( \hat{\beta}_i(t) - \beta_i \right) \left( \hat{\beta}_i(t^{\prime}) - \beta_i \right)\right) \\
& \quad=\frac{1}{2 \pi i} \int \mathcal{D} \hat{Q}_1\left(t, t^{\prime}\right) \exp \left[\iint \hat{Q}_1\left(t, t^{\prime}\right)\left(-dQ_1\left(t, t^{\prime}\right)+\sum_i\left( \hat{\beta}_i(t) - \beta_i \right) \left( \hat{\beta}_i(t^{\prime}) - \beta_i \right)\right) \mathrm{d} t \mathrm{~d} t^{\prime}\right],\\
& \delta\left(-dQ_2\left(t, t^{\prime}\right)+ \sum_k \left(\hat{\beta}_k(t)\tilde{\beta}_k(t^\prime)\right)\right) \\
& \quad=\frac{1}{2 \pi i} \int \mathcal{D} \hat{Q}_1\left(t, t^{\prime}\right) \exp \left[\iint \hat{Q}_2\left(t, t^{\prime}\right)\left(-dQ_2\left(t, t^{\prime}\right)+ \sum_k \left(\hat{\beta}_k(t)\tilde{\beta}_k(t^\prime)\right)\right) \mathrm{d} t \mathrm{~d} t^{\prime}\right]\\
& \delta\left(-dQ_3\left(t, t^{\prime}\right)+ \sum_k \tilde{\beta}_k(t)i\tilde{\beta}_k(t^\prime)\right) \\
& \quad=\frac{1}{2 \pi i} \int \mathcal{D} \hat{Q}_3\left(t, t^{\prime}\right) \exp \left[\iint \hat{Q}_3\left(t, t^{\prime}\right)\left(-dQ_3\left(t, t^{\prime}\right)+\sum_k i\tilde{\beta}_k(t)i\tilde{\beta}_k(t^\prime)\right) \mathrm{d} t \mathrm{~d} t^{\prime}\right],\\
& \delta\left(-dQ_4\left(t\right)+\sum_k i\tilde{\beta}_k(t){\beta}_k\right) \\
& \quad=\frac{1}{2 \pi i} \int \mathcal{D} \hat{Q}_4\left(t\right) \exp \left[\int \hat{Q}_4\left(t\right)\left(-dQ_4\left(t\right)+\sum_k i\tilde{\beta}_k(t){\beta}_k\right) \mathrm{d} t \right].\\
\end{aligned}
\end{equation}
The complete form of the dynamical partition function is:
\begin{equation}
\begin{aligned}
Z_{\operatorname{d y n}}& \propto \int \mathcal{D} \boldsymbol{\mathcal{B}}(t) \int \mathcal{D} \mathcal{Q}(t,t^{\prime}) \exp \left(- i \tilde{\boldsymbol{\beta}} \partial_t {\hat{\boldsymbol{\beta}}}- \lambda dQ_2(t,t) -dQ_1\cdot \hat{Q}_1 -dQ_2 \cdot \hat{Q}_2-dQ_3\cdot \hat{Q}_3\right.\\& -dQ_4\cdot \hat{Q}_4 -\frac{N}{2}\log\operatorname{det}{\left(\begin{array}{cc}
\gamma Q_2^\top - \gamma Q_4+\delta &\gamma^2 Q_3 \\
Q_1+D& \gamma Q_2-\gamma Q_4^\top+\delta
\end{array}\right)}+\tilde{\boldsymbol{j}}  \hat{\boldsymbol{\beta}}+ \boldsymbol{j} \tilde{\boldsymbol{\beta}}\\
&+\int\int dt dt^{\prime} \sum_k\left( \hat{\beta}_k(t) - \beta_k \right) \hat{Q}_1(t,t^{\prime}) \left( \hat{\beta}_k(t^\prime) - \beta_k \right)+  i\hat{\beta}_k(t) \hat{Q}_2(t,t^{\prime})  \tilde{\beta}_k(t^\prime)\\
&+\left.\int\int dt dt^{\prime} \sum_k i\tilde{\beta}_k(t) \hat{Q}_3(t,t^{\prime}) i\tilde{\beta}_k(t^\prime)+\int dt \sum_k i\tilde{\beta}_k(t)\hat{Q}_4(t)\beta_k\right),
\end{aligned}
\end{equation}
We denote the path integral measure over the dynamical fields as $\mathcal{D} \boldsymbol{\mathcal{B}}(t) = \mathcal{D} \hat{\boldsymbol{\beta}}(t)\, \mathcal{D} \tilde{\boldsymbol{\beta}}(t)$, and introduce the notation $\mathcal{D} \mathcal{Q}(t,t^{\prime})$:
\begin{equation}
\mathcal{D} \mathcal{Q}(t,t^{\prime}) = \left(\frac{d}{2\pi}\right)^4 \mathcal{D} Q_1(t, t^{\prime})\, \mathcal{D} \hat{Q}_1(t, t^{\prime})\, \mathcal{D} Q_2(t, t^{\prime})\, \mathcal{D} \hat{Q}_2(t, t^{\prime})\, \mathcal{D} Q_3(t, t^{\prime})\, \mathcal{D} \hat{Q}_3(t, t^{\prime})\, \mathcal{D} Q_4(t)\, \mathcal{D} \hat{Q}_4(t).
\end{equation}
To simplify notation, we define the shorthand inner products: $\mathbf{f} \cdot \mathbf{g} = \sum_{i=1}^d \int f_i(t)\, g_i(t)\, \mathrm{d}t$, $\hat{Q}_1 \cdot Q_1 = \iint \hat{Q}_1(t, t^{\prime})\, Q_1(t, t^{\prime})\, \mathrm{d}t\, \mathrm{d}t^{\prime}$, $\hat{Q}_2 \cdot Q_2 = \iint \hat{Q}_2(t, t^{\prime})\, Q_2(t, t^{\prime})\, \mathrm{d}t\, \mathrm{d}t^{\prime}$, $\hat{Q}_3 \cdot Q_3 = \iint \hat{Q}_3(t, t^{\prime})\, Q_3(t, t^{\prime})\, \mathrm{d}t\, \mathrm{d}t^{\prime}$, and $\hat{Q}_4 \cdot Q_4 = \int \hat{Q}_4(t)\, Q_4(t)\, \mathrm{d}t$. The averaged moment-generating functional factorizes across neurons, indicating that the collective dynamics of the original $d$-dimensional system reduce to an effective single-neuron mean-field description driven by correlated Gaussian noise. We compactly rewrite the disorder-averaged moment-generating functional as:
\begin{equation}
\begin{aligned}
Z_{\operatorname{dym}}&=\int \mathcal{D} \mathcal{Q} \exp (d f(Q_1, \hat{Q}_1,Q_2, \hat{Q}_2, Q_3, \hat{Q}_3, Q_4, \hat{Q}_4,j, \tilde{j}))\\
f(Q_1, \hat{Q}_1,Q_2, \hat{Q}_2, Q_3, \hat{Q}_3, Q_4, \hat{Q}_4, j, \tilde{j})& =  -Q_1 \cdot \hat{Q}_1 -Q_2 \cdot \hat{Q}_2-Q_3 \cdot \hat{Q}_3-Q_4 \cdot \hat{Q}_4- \lambda Q_2(t,t)\\&-\frac{1}{2\alpha}\log{\operatorname{det}}( \mathcal{L})+\log\left(Z_0(\hat{Q}_1, \hat{Q}_2,  \hat{Q}_3, \hat{Q}_4, j, \tilde{j})\right),\\
 \mathcal{L}& = \left(\begin{array}{cc}
\gamma Q_2^\top - \gamma Q_4+\delta &\gamma^2 Q_3 \\
Q_1+D& \gamma Q_2-\gamma Q_4^\top+\delta
\end{array}\right)\\
Z_0(\hat{Q}_1, \hat{Q}_2,  \hat{Q}_3, \hat{Q}_4, j, \tilde{j}) &= \int \mathcal{D} {\mathcal{B}}(t) \exp\left(\left(\hat{\beta}-\beta\right)^\top \hat{Q}_1 \left(\hat{\beta}-\beta\right)+ i\hat{\beta}^\top \hat{Q}_2  \tilde{\beta}\right.\\&\left.+ i\tilde{\beta}^\top \hat{Q}_3  i\tilde{\beta}+i\tilde{\beta}\cdot\hat{Q}_4\beta-i\tilde{\beta}\cdot \partial_t \hat{\beta}+\tilde{j}\cdot \hat{\beta} + j\cdot\tilde{\beta}\right), \\
\label{eq:compact}
\end{aligned}
\end{equation}
where we define the new notation that $f^T Q g=\iint f(t) Q\left(t, t^{\prime}\right) g(t^\prime) \mathrm{d} t \mathrm{~d} t^{\prime}$, $f\cdot g = \int dt f(t)g(t)$ and $\mathcal{D} {\mathcal{B}}(t) =  \mathcal{D} \hat{{\beta}}(t) \mathcal{D} \tilde{{\beta}}(t)$. Note that $Z_0$ denotes the effective moment-generating functional of the single-neuron mean-field system, $f(Q_1, \hat{Q}_1,Q_2, \hat{Q}_2, Q_3, \hat{Q}_3, Q_4, \hat{Q}_4,j, \tilde{j})$ indicates the dynamical action. The variables $\hat{\beta}$, $\tilde{\beta}$, and $\beta$ represent the mean-field counterparts of their original high-dimensional forms.
In the thermodynamic limit $d \to \infty$, $N \to \infty$, with $\alpha = d/N$ held fixed, we apply the saddle-point approximation to obtain:
\begin{equation}
\begin{aligned}
Z_{\operatorname{dym}}&\approx \exp (d f(Q^\star_1, \hat{Q}^\star_1,Q^\star_2, \hat{Q}^\star_2, Q^\star_3, \hat{Q}^\star_3, Q^\star_4, \hat{Q}^\star_4,j, \tilde{j})),\\ 
\end{aligned}
\end{equation}
where $\{\mathcal{Q}^\star, \hat{\mathcal{Q}}^\star\}$ represent the values of observables that maximize the dynamical action $f(Q_1, \hat{Q}_1,Q_2, \hat{Q}_2, Q_3, \hat{Q}_3, Q_4, \hat{Q}_4,j, \tilde{j})$. We thus have the first four equations:
\begin{equation}
\begin{aligned}
\frac{\delta f(Q_1, \hat{Q}_1, Q_2, \hat{Q}_2, Q_3, \hat{Q}_3, Q_4, \hat{Q}_4, \mathbf{j}, \tilde{\mathbf{j}})}{\delta \hat{Q}_1\left(t, t^{\prime}\right)}&=0 
\quad \Rightarrow \quad Q_1^{\star}\left(t, t^{\prime}\right) = \left\langle \left(\hat{\beta}(t)-\beta\right) \left(\hat{\beta}(t^{\prime})-\beta\right) \right\rangle_{Z_0} ,\\
\frac{\delta f(Q_1, \hat{Q}_1, Q_2, \hat{Q}_2, Q_3, \hat{Q}_3, Q_4, \hat{Q}_4, \mathbf{j}, \tilde{\mathbf{j}})}{\delta \hat{Q}_2\left(t, t^{\prime}\right)}&=0 
\quad \Rightarrow \quad Q_2^{\star}\left(t, t^{\prime}\right) = \left\langle \hat{\beta}(t) \cdot i\tilde{\beta}(t^{\prime}) \right\rangle_{Z_0} ,\\
\frac{\delta f(Q_1, \hat{Q}_1, Q_2, \hat{Q}_2, Q_3, \hat{Q}_3, Q_4, \hat{Q}_4, \mathbf{j}, \tilde{\mathbf{j}})}{\delta \hat{Q}_3\left(t, t^{\prime}\right)}&=0 
\quad \Rightarrow \quad Q_3^{\star}\left(t, t^{\prime}\right) = \left\langle i\tilde{\beta}(t) \cdot i\tilde{\beta}(t^\prime) \right\rangle_{Z_0} ,\\
\frac{\delta f(Q_1, \hat{Q}_1, Q_2, \hat{Q}_2, Q_3, \hat{Q}_3, Q_4, \hat{Q}_4, \mathbf{j}, \tilde{\mathbf{j}})}{\delta \hat{Q}_4\left(t\right)}&
\quad \Rightarrow \quad Q_4^{\star}\left(t\right) = \left\langle i\tilde{\beta}(t) \cdot {\beta} \right\rangle_{Z_0},\\
\end{aligned}
\end{equation}
where:
\begin{equation}
\begin{aligned}
\left\langle \mathcal{O}(\hat{\beta}, \tilde{\beta}) \right\rangle_{Z_0} 
= \frac{1}{Z_0} \int \mathcal{D} \mathcal{B}(t)\, \mathcal{O}(\hat{\beta}, \tilde{\beta})\, \exp \bigg[
& - \left(\hat{\beta} - \beta \right)^\top \hat{Q}_1 \left(\hat{\beta} - \beta \right)
+ i\, \hat{\beta}^\top \hat{Q}_2\, \tilde{\beta}
+ i\, \tilde{\beta}^\top \hat{Q}_3\, \tilde{\beta} \\
& + i\tilde{\beta}\cdot \hat{Q}_4\, \beta
- i\, \tilde{\beta} \cdot \partial_t{\hat{\beta}}
+ \tilde{j} \cdot \hat{\beta}
+ j \cdot \tilde{\beta}
\bigg].
\end{aligned}
\label{eq:Z0_observable}
\end{equation}
This average computes the expectation of an observable $\mathcal{O}(\hat{\beta}, \tilde{\beta})$ under the effective single-neuron dynamics defined by the mean-field action, capturing the typical behavior of the system in the thermodynamic limit. Next, we are able to know the physical meaning of the dynamical observables.  The dynamical observables $Q_1$ through $Q_4$ are defined as self-averaging order parameters that capture key statistical features of the learning dynamics:

\begin{itemize}
    \item \textbf{Two-time correlation of weight deviation:}
    \begin{equation}
        Q_1(t, t^{\prime}) = \frac{1}{d} \sum_{i=1}^d \left\langle \left( \hat{\beta}_i(t) - \beta_i \right)\left( \hat{\beta}_i(t^{\prime}) - \beta_i \right) \right\rangle 
        \longrightarrow \left\langle \left( \hat{\beta}(t) - \beta \right)\left( \hat{\beta}(t^{\prime}) - \beta \right) \right\rangle_{Z_0}
    \end{equation}
    This quantity measures the temporal correlation of the deviation between the student and teacher weights. It characterizes how the mismatch with the teacher evolves and persists over time, serving as a dynamical generalization of the overlap. The correlation function is always symmetric, indicating that $Q_1(t, t^{\prime}) = Q_1(t^\prime,t)$.

    \item \textbf{Linear response function:}
    \begin{equation}
        Q_2(t, t^{\prime}) = \frac{1}{d} \sum_{i=1}^d \left\langle \hat{\beta}_i(t) \cdot i\tilde{\beta}_i(t^{\prime}) \right\rangle 
        \longrightarrow \left\langle \hat{\beta}(t) \cdot i\tilde{\beta}(t^{\prime}) \right\rangle_{Z_0}
    \end{equation}
    This observable corresponds to the retarded response function $R(t, t^{\prime}) = \delta \langle \hat{\beta}(t) \rangle / \delta j(t^{\prime})$, quantifying the sensitivity of the system to infinitesimal perturbations. It reveals how information and perturbations propagate across time in the learning trajectory. Owing to causality in the system's dynamics, the response function $R(t, t^{\prime})$ is nonzero only when the perturbation at time $t^{\prime}$ precedes the response at time $t$, i.e., for $t > t^{\prime}$.

    \item \textbf{Auto-correlation of the response field:}
    \begin{equation}
        Q_3(t, t^{\prime}) = \frac{1}{d} \sum_{i=1}^d \left\langle i\tilde{\beta}_i(t) \cdot i\tilde{\beta}_i(t^{\prime}) \right\rangle 
        \longrightarrow \left\langle i\tilde{\beta}(t) \cdot i\tilde{\beta}(t^{\prime}) \right\rangle_{Z_0}
    \end{equation}
    $Q_3(t, t^{\prime}) = \left\langle i\tilde{\beta}(t) i\tilde{\beta}(t^{\prime}) \right\rangle_{Z_0}$ quantifies the intrinsic fluctuations of the response field. It characterizes temporal correlations in the system's sensitivity to perturbations, and encodes higher-order dynamical effects beyond linear response. Interestingly, this term always vanishes, as $\left.\frac{\delta^n}{\delta j\left(t_1\right) \cdots \delta j\left(t_n\right)} Z_{0}[\mathrm{j}, 0]\right|_{\mathrm{j}=0}=0$, which indicates that the response fields don't propagate.

    \item \textbf{Fluctuation-teacher alignment:}
    \begin{equation}
        Q_4(t) = \frac{1}{d} \sum_{i=1}^d \left\langle i \tilde{\beta}_i(t) \cdot \beta_i \right\rangle 
        \longrightarrow \left\langle i \tilde{\beta}(t) \cdot \beta \right\rangle_{Z_0}
    \end{equation}
    This term reflects the alignment between the response field and the teacher weights. It captures how the response fields in the learning process are correlated with the ground truth. Similarly, as the ground truth $\beta$ is independent of $\hat{\beta}$ and $\tilde{\beta}$, $Q_4(t) = \beta \left\langle i\tilde{\beta}(t) \right\rangle_{Z_0} = 0$.
\end{itemize}
Continuing the saddle-point analysis, we obtain the remaining set of self-consistency equations by taking functional derivatives with respect to the order parameters $Q_1$, $Q_2$, $Q_3$, and $Q_4$:
\begin{equation}
\begin{aligned}
\frac{\delta f(Q_1, \hat{Q}_1, Q_2, \hat{Q}_2, Q_3, \hat{Q}_3, Q_4, \hat{Q}_4, \mathbf{j}, \tilde{\mathbf{j}})}{\delta {Q}_1\left(t, t^{\prime}\right)}&=0 
\quad \Rightarrow \quad \hat{Q}_1^{\star}\left(t, t^{\prime}\right) = -\frac{1}{2\alpha}\frac{\frac{\partial \operatorname{det}(\mathcal{L})}{\partial Q_1}}{\operatorname{det}(\mathcal{L})}  ,\\
\frac{\delta f(Q_1, \hat{Q}_1, Q_2, \hat{Q}_2, Q_3, \hat{Q}_3, Q_4, \hat{Q}_4, \mathbf{j}, \tilde{\mathbf{j}})}{\delta {Q}_2\left(t, t^{\prime}\right)}&=0 
\quad \Rightarrow \quad \hat{Q}_2^{\star}\left(t, t^{\prime}\right) = -\lambda\delta(t,t^\prime)-\frac{1}{2\alpha}\frac{\frac{\partial \operatorname{det}(\mathcal{L})}{\partial Q_2}}{\operatorname{det}(\mathcal{L})}  ,\\
\frac{\delta f(Q_1, \hat{Q}_1, Q_2, \hat{Q}_2, Q_3, \hat{Q}_3, Q_4, \hat{Q}_4, \mathbf{j}, \tilde{\mathbf{j}})}{\delta {Q}_3\left(t, t^{\prime}\right)}&=0 
\quad \Rightarrow \quad \hat{Q}_3^{\star}\left(t, t^{\prime}\right)= -\frac{1}{2\alpha}\frac{\frac{\partial \operatorname{det}(\mathcal{L})}{\partial Q_3}}{\operatorname{det}(\mathcal{L})} ,\\
\frac{\delta f(Q_1, \hat{Q}_1, Q_2, \hat{Q}_2, Q_3, \hat{Q}_3, Q_4, \hat{Q}_4, \mathbf{j}, \tilde{\mathbf{j}})}{\delta {Q}_4\left(t\right)}&
\quad \Rightarrow \quad \hat{Q}_4^{\star}\left(t\right) = -\frac{1}{2\alpha}\frac{\frac{\partial \operatorname{det}(\mathcal{L})}{\partial Q_4}}{\operatorname{det}(\mathcal{L})}. \\
\end{aligned}
\end{equation}
Hence, solving the full set of saddle-point equations reduces to evaluating the functional determinant $\det(\mathcal{L})$ and its functional derivatives with respect to the overlap functions: $\partial_{Q_1} \det(\mathcal{L})$, $\partial_{Q_2} \det(\mathcal{L})$, $\partial_{Q_3} \det(\mathcal{L})$, and $\partial_{Q_4} \det(\mathcal{L})$. Based on the identity that:
\begin{equation}
\log  \operatorname{det}\left(\begin{array}{cc}
A & B \\
C & D
\end{array}\right) =  \log  \operatorname{det}\left(AD - ACA^{-1} B\right).\\
\end{equation}
The functional determinant $\det(\mathcal{L})$ can be written in the form of:
\begin{equation}
\log \det(\mathcal{L}) = \log  \operatorname{det}\left(\left(\gamma Q_2^{\top}-\gamma Q_4+\delta\right)*\left(\gamma Q_2-\gamma Q_4^{\top}+\delta\right) - \left(\gamma Q_2^{\top}-\gamma Q_4+\delta\right)*\left(Q_1+D\right)*\left(\gamma Q_2^{\top}-\gamma Q_4+\delta\right)^{-1} *\gamma^2Q_3  \right),
\end{equation}
where the operator $ * $ denotes convolution over intermediate times, defined as $A * B(t,t^\prime)=\iint A \left(t, t^{\prime\prime}\right)B\left(t^{\prime\prime}, t^\prime\right)  \mathrm{~d} t^{\prime\prime}$. Here, we can redefine $\mathcal{L}$ as $\left(\gamma Q_2^{\top}-\gamma Q_4+\delta\right)*\left(\gamma Q_2-\gamma Q_4^{\top}+\delta\right) - \left(\gamma Q_2^{\top}-\gamma Q_4+\delta\right)*\left(Q_1+D\right)*\left(\gamma Q_2^{\top}-\gamma Q_4+\delta\right)^{-1} *\gamma^2Q_3 $.  We also recall the identity $\frac{\partial \log \det \boldsymbol{B}}{\partial x} = \operatorname{Tr}\left(\boldsymbol{B}^{-1} \frac{\partial \boldsymbol{B}}{\partial x}\right)$, which allows us to compute the required derivatives of the functional determinant. Consequently, we obtain the following expressions for $\frac{\partial \log \operatorname{det}(\mathcal{L})}{\partial Q_1\left(t, t^{\prime}\right)}$ and $\frac{\partial \log \operatorname{det}(\mathcal{L})}{\partial Q_3\left(t, t^{\prime}\right)}$ first:
\begin{equation}
\begin{aligned}
&\frac{\partial \log \operatorname{det}(\mathcal{L})}{\partial Q_1\left(t, t^{\prime}\right)}  =\operatorname{Tr}\left(\mathcal{L}^{-1} \frac{\partial \mathcal{L}}{\partial Q_1\left(t, t^{\prime}\right)}\right), \\
& =-\int d t_1 \int d t_2 \int d t_3 \mathcal{L}^{-1}\left(t_1, t_2\right)\left(\gamma Q^\top_2-\gamma Q_4+\delta \right)\left(t_2, t\right)\left(\gamma Q^\top_2-\gamma Q_4+\delta \right)^{-1}\left(t^{\prime}, t_3\right) \gamma^2 Q_3\left(t_3, t_1\right).
\end{aligned}
\end{equation}
\begin{equation}
\begin{aligned}
&\frac{\partial \log \operatorname{det}(\mathcal{L})}{\partial Q_3\left(t, t^{\prime}\right)}  =\operatorname{Tr}\left(\mathcal{L}^{-1} \frac{\partial \mathcal{L}}{\partial Q_3\left(t, t^{\prime}\right)}\right), \\&= -\int d t_1 \int d t_2 \int d t_3 \int d t_4 \mathcal{L}^{-1}\left(t_1, t_2\right)\left(\gamma Q_2^\top-\gamma Q_4+\delta \right)\left(t_2, t_3\right) \left(Q_1\left(t_3, t_4\right)+D\right) \left(\gamma Q_2^\top-\gamma Q_4+\delta \right)^{-1}\left(t_4, t\right) \gamma^2 \delta\left({t_1, t^{\prime}}\right) \\
& = -\int d t_2 \int d t_3 \int d t_4 \mathcal{L}^{-1}\left(t^{\prime}, t_2\right)\left(\gamma Q_2^\top-\gamma Q_4+\delta \right)\left(t_2, t_3\right) \left(r+Q_1\left(t_3, t_4\right)\right) \left(\gamma Q_2^\top-\gamma Q_4+\delta \right)^{-1}\left(t_4, t\right) \gamma^2,\\
& = -\gamma^2 \mathcal{L}^{-1}*\left(\gamma Q_2^\top-\gamma Q_4+\delta \right)*\left(Q_1+D\right)*\left(\gamma Q_2^\top-\gamma Q_4+\delta \right)^{-1}(t^\prime,t),\\
\end{aligned}
\end{equation}
The calculations of $\frac{\partial \log \operatorname{det}(\mathcal{L})}{\partial Q_2\left(t, t^{\prime}\right)}$ and $\frac{\partial \log \operatorname{det}(\mathcal{L})}{\partial Q_4\left(t, t^{\prime}\right)}$ are a little tricky.  We first decompose the expression for $\log \det(\mathcal{L})$ into three analyticly tractable components:
\begin{equation}
\begin{aligned}
& \operatorname{Tr}\left(\mathcal{L}^{-1} \frac{\partial \mathcal{L}}{\partial Q_2\left(t, t^{\prime}\right)}\right)=\int d t_3 \int d t_0 \int d t_1 \delta\left({t_0, t_3}\right) \mathcal{L}^{-1}\left(t_0, t_1\right) \frac{\partial \mathcal{L}\left(t_1, t_3\right)}{\partial Q_2\left(t, t^{\prime}\right)}=P_1+P_2+P_3, \\
\quad \\
& P_1(t,t^\prime)=\gamma^2 \int d t_3 \int d t_0 \int d t_1 \delta\left(t_0, t_3\right) \mathcal{L}^{-1}\left(t_0, t_1\right) \frac{\partial\left[\left(Q^\top_2- Q_4\right) *\left( Q_2-  Q_4^\top\right) \right]\left(t_1, t_3\right)}{\partial Q_2\left(t, t^{\prime}\right)}, \\
& =\gamma^2 \int d t_3 \int d t_0 \int d t_1 \delta\left(t_0, t_3\right) \mathcal{L}^{-1}\left(t_0, t_1\right)\left(\delta\left({t_1, t^\prime}\right)\delta\left(t_2,t\right) (Q_2\left(t_2, t_3\right)-Q_4(t_3))+(Q_2\left(t_2, t_1\right) - Q_4(t_1)) \delta\left(t_2, t\right)\delta\left(t_3,t^\prime\right)\right), \\
& =\gamma^2 \int d t_0 \mathcal{L}^{-1}\left(t_0, t^\prime\right) (Q_2\left(t, t_0\right)-Q_4(t_0))+\gamma^2 \int d t_1 \mathcal{L}^{-1}\left(t^\prime, t_1\right) (Q_2\left(t, t_1\right)-Q_4(t_1)), \\
& = \gamma^2 (\mathcal{L}^{-1})^\top *(Q^\top_2-Q_4)(t^\prime, t)+\gamma^2 \mathcal{L}^{-1} *(Q^\top_2-Q_4)(t^\prime,t), \\
& = \gamma^2\left(Q_2 - Q_4^\top\right)*\mathcal{L}^{-1}(t,t^\prime)+\gamma^2\left(Q_2 - Q_4^\top\right)*(\mathcal{L}^{-1})^\top(t,t^\prime),\\
\end{aligned}
\end{equation}
\begin{equation}
\begin{aligned}
& P_2=\gamma \int d t_3 \int d t_0 \int d t_1 \delta\left(t_0, t_3\right) \mathcal{L}^{-1}\left(t_0, t_1\right) \frac{\partial\left(Q_2+Q_2^{\top}\right)\left(t_1, t_3\right)}{\partial Q_2\left(t, t^{\prime}\right)} \\
& =\gamma \int d t_3 \int d t_0 \int d t_1 \delta\left(t_0, t_3\right) \mathcal{L}^{-1}\left(t_0, t_1\right) \frac{\partial\left(Q_2\left(t_1, t_3\right)+Q_2\left(t_3, t_1\right)\right)}{\partial Q_2\left(t, t^{\prime}\right)}, \\
& =\gamma \int d t_3 \int d t_0 \int d t_1 \delta\left(t_0, t_3\right) \delta\left({t_1, t}\right) \delta\left({t_3, t^\prime}\right) \mathcal{L}^{-1}\left(t_0, t_1\right) \\
& +\gamma \int d t_3 \int d t_0 \int d t_1 \delta\left(t_0, t_3\right) \delta\left({t_3, t}\right) \delta\left({t_1, t^\prime} \right)\mathcal{L}^{-1}\left(t_0, t_1\right), \\
& =\gamma(\mathcal{L}^{-1}\left(t^{\prime}, t\right)+\mathcal{L}^{-1}\left(t, t^{\prime}\right)),\\
\end{aligned}
\end{equation}
\begin{equation}
\begin{aligned}
& P_3=-\gamma^2 \int d t_3 \int d t_0 \int d t_1 \delta\left(t_0, t_3\right) \mathcal{L}^{-1}\left(t_0, t_1\right)\\& \frac{\partial\left[\left(\gamma Q_2^\top-\gamma Q_4+\delta \right) * \left(D+Q_1\right) *\left(\gamma Q_2^\top-\gamma Q_4+\delta \right)^{-1} Q_3\right]\left(t_1, t_3\right)}{\partial Q_2\left(t, t^{\prime}\right)} \\
& =-\gamma^2 \int d t_3 \int d t_0 \int d t_1 \int d t_2 \int d t_4 \int d t_5 \delta\left(t_0, t_3\right) \mathcal{L}^{-1}\left(t_0, t_1\right) Q_3\left(t_5, t_3\right)\\
& \frac{\partial\left[\left(\gamma Q_2^\top-\gamma Q_4+\delta \right)\left(t_1, t_2\right) \left(D+Q_1\right)\left(t_2, t_4\right)\left(\gamma Q_2^\top-\gamma Q_4+\delta \right)^{-1}\left(t_4, t_5\right) \right]}{\partial Q_2\left(t, t^{\prime}\right)} \\
& =-\gamma^3 \int d t_0 \int d t_4 \int d t_5 \mathcal{L}^{-1}\left(t_0, t^\prime\right) \times\left(D+Q_1\right)\left(t, t_4\right)\left(\gamma Q_2^\top-\gamma Q_4+\delta \right)^{-1}\left(t_4, t_5\right) Q_3\left(t_5, t_0\right) \\
& +\gamma^3 \int d t_0 \int d t_1 \int d t_2 \int d t_4 \int d t_5 \mathcal{L}^{-1}\left(t_0, t_1\right) \\
& \times[\left(\gamma Q_2^\top-\gamma Q_4+\delta \right)\left(t_1, t_2\right) \left(D+Q_1\right)\left(t_2, t_4\right)\left(\gamma Q_2^\top-\gamma Q_4+\delta \right)^{-1}\left(t_4, t^\prime \right)\left(\gamma Q_2^\top-\gamma Q_4+\delta \right)^{-1}\left(t, t_5\right) Q_3\left(t_5, t_0\right)], \\
\end{aligned}
\end{equation}
where $\delta(\cdot, \cdot)$ denotes the Dirac delta function, and we use the identity $\frac{\partial\left(\mathbf{X}^{-1}\right)_{k l}}{\partial X_{i j}}=-\left(\mathbf{X}^{-1}\right)_{k i}\left(\mathbf{X}^{-1}\right)_{j l}$ in the last equation. Similarly, we can calculate the $\frac{\partial \log \operatorname{det}(\mathcal{L})}{\partial Q_4\left(t, t^{\prime}\right)}$:
\begin{equation}
\begin{aligned}
& \operatorname{Tr}\left(\mathcal{L}^{-1} \frac{\partial \mathcal{L}}{\partial Q_4\left(t\right)}\right)=\int d t_3 \int d t_0 \int d t_1 \delta\left({t 0, t_3}\right) \mathcal{L}^{-1}\left(t_0, t_1\right) \frac{\partial \mathcal{L}\left(t_1, t_3\right)}{\partial Q_4\left(t\right)}=P^\prime_1+P^\prime_2+P^\prime_3, \\
\quad \\
& P_1^\prime=\gamma^2 \int d t_3 \int d t_0 \int d t_1 \delta\left({t_0, t_3}\right) \mathcal{L}^{-1}\left(t_0, t_1\right) \frac{\partial\left[\left(Q_2^\top- Q_4\right) *\left( Q_2 -  Q_4^\top\right) \right]\left(t_1, t_3\right)}{\partial Q_4\left(t\right)}, \\
& = \gamma^2 \int dt_2 \int d t_3 \int d t_0 \int d t_1 \delta\left({t_0, t_3}\right) \mathcal{L}^{-1}\left(t_0, t_1\right) \frac{\partial\left[\left(Q_2(t_2, t_1)- Q_4(t_1)\right)\left( Q_2(t_2, t_3) -  Q_4(t_3)\right) \right]}{\partial Q_4\left(t\right)},\\
& = -\gamma^2 \int dt_2  \int d t_0 \mathcal{L}^{-1}\left(t_0, t\right) \left( Q_2(t_2, t_0) -  Q_4(t_0)\right)-\gamma^2  \int dt_2  \int d t_1 \mathcal{L}^{-1}\left(t, t_1\right) \left( Q_2(t_2, t_1) -  Q_4(t_1)\right),\\
& = -\gamma^2 (\mathcal{L}^{-1})^\top *(Q^\top_2-Q_4)(t, t^\prime)-\gamma^2 \mathcal{L}^{-1} *(Q^\top_2-Q_4)(t, t^\prime),\\
& =- \int dt^\prime P^\top_1(t,t^\prime),\\
\quad \\
& P_2^\prime=-\gamma \int d t_3 \int d t_0 \int d t_1 \delta\left({t_0, t_3}\right) \mathcal{L}^{-1}\left(t_0, t_1\right) \frac{\partial\left[Q_4(t_1)+Q_4(t_3)\right]}{\partial Q_4\left(t\right)},\\
& = -\gamma \int d t_0   \mathcal{L}^{-1}\left(t_0, t\right)-\gamma \int d t_1   \mathcal{L}^{-1}\left(t, t_1\right),\\
& =- \int dt^\prime P^\top_2(t,t^\prime),\\ 
\quad \\
&P_3^\prime(t) = -\int dt^\prime P^\top_3(t,t^\prime).\\
\end{aligned}
\end{equation}
Therefore, we can write down the rest of the saddle-point equations:
\begin{equation}
\begin{aligned}
\hat{Q}_1^{\star}\left(t, t^{\prime}\right) &= \frac{1}{2\alpha} \int d t_1 \int d t_2 \int d t_3 \left(\mathcal{L}^\star\right)^{-1}\left(t_1, t_2\right)\left(\gamma \left(Q_2^\star\right)^\top-\gamma Q^\star_4+\delta \right)\left(t_2, t\right)\left(\gamma \left(Q_2^\star\right)^\top-\gamma Q^\star_4+\delta \right)^{-1}\left(t^{\prime}, t_3\right) \gamma^2 Q^\star_3\left(t_3, t_1\right),\\
\hat{Q}_2^{\star}\left(t, t^{\prime}\right) &= -\lambda\delta(t,t^\prime)-\frac{1}{2 \alpha}(P^\star_1+P^\star_2+P^\star_3),\\
\hat{Q}_3^{\star}\left(t, t^{\prime}\right) &= \frac{\gamma^2}{2 \alpha}\int d t_2 \int d t_3 \int d t_4 \mathcal{L}^{-1}\left(t^{\prime}, t_2\right)\left(\gamma (Q^\star_2)^\top-\gamma Q^\star_5+\delta \right)\left(t_2, t_3\right)\left(D+Q^\star_1\right)\left(t_3, t_4\right)\left(\gamma (Q^\star_2)^\top-\gamma Q^\star_5+\delta \right)^{-1}\left(t_4, t\right),\\
\hat{Q}_4^{\star}\left(t, t^{\prime}\right) &= -\frac{1}{2 \alpha}(\left(P_1^\prime\right)^\star+\left(P_2^\prime\right)^\star+\left(P_3^\prime\right)^\star),
\end{aligned}
\end{equation}
Here, the superscript $\star$ denotes the saddle-point solution of the corresponding observable, and $\mathcal{L}^\star$ is the functional kernel evaluated at these saddle-point values. The terms $\{P_1^{\star}, P_2^{\star}, P_3^{\star}\}$,and $\{\left(P_1^\prime\right)^\star, \left(P_2^\prime\right)^\star, \left(P_3^\prime\right)^\star\}$ arise from the decomposition of the derivative with respect to $Q_2$ and $Q_4$, as introduced in the previous analysis. 

Next, we simplify the saddle-point equations by using the following identities derived from the statistical properties of the dynamical observables:
\begin{equation}
\begin{aligned}
Q_1^\star(t,t^{\prime}) &= C(t,t^\prime) = \left\langle \left(\hat{\beta}(t)-\beta\right) \left(\hat{\beta}(t^{\prime})-\beta\right) \right\rangle_{Z_0}, \\
Q_2^\star(t,t^{\prime}) &= R(t,t^\prime) = \left\langle \hat{\beta}(t) \cdot i\tilde{\beta}(t^{\prime}) \right\rangle_{Z_0}, \\
R(t,t^{\prime}) &= 0 \quad \text{for } t < t^{\prime}, \qquad R(t,t) = 0 \quad \text{(Ito convention)}, \\
Q_3^\star(t,t^{\prime}) &= \left\langle i \tilde{\beta}(t) i \tilde{\beta}(t^{\prime}) \right\rangle_{Z_0} = 0, \\
Q_4^\star(t) &= \left\langle i \tilde{\beta}(t) \cdot \beta \right\rangle_{Z_0}=0.
\end{aligned}
\end{equation}
Applying these constraints, the other saddle-point solutions reduce to:
\begin{equation}
\begin{aligned}
\left(\mathcal{L}^\star\right)^{-1}& = \left(\gamma R+\delta\right)^{-1}*\left(\gamma R^{\top}+\delta\right)^{-1},\\
\hat{Q}_1^\star(t,t^{\prime}) &= 0, \\
\hat{Q}_2^\star(t,t^{\prime}) &= -\lambda\delta(t,t^{\prime}) - \frac{1}{\alpha} \left(\gamma^2 Q^\star_2 * \left(\mathcal{L}^\star\right)^{-1} + \gamma \left(\mathcal{L}^\star\right)^{-1}\right)(t,t^{\prime}) \\
&= -\lambda \delta(t,t^{\prime}) - \frac{\gamma}{\alpha} \left(\gamma R^\top + \delta\right)^{-1}(t,t^{\prime}), \\
\hat{Q}_3^\star(t,t^{\prime}) &= \frac{\gamma^2}{2\alpha} \int dt_2 dt_3 dt_4 \left(\mathcal{L}^\star\right)^{-1}(t^{\prime},t_2) (\gamma R^\top + \delta)(t_2, t_3)\left(D+C\right)(t_3, t_4) (\gamma R^\top + \delta)^{-1}(t_4, t) \\
&= \frac{1}{2\alpha} \gamma^2 (\gamma R + \delta)^{-1} * \left(D+C\right) * (\gamma R^\top + \delta)^{-1}(t,t^{\prime}), \\
\hat{Q}_4^\star(t) &= \int dt^{\prime}\frac{\gamma}{\alpha} \left(\gamma R + \delta\right)^{-1}(t,t^{\prime}).
\end{aligned}
\end{equation}
After doing the saddle point analysis, we can write down the dynamical partition function as:
\begin{equation}
\begin{aligned}
Z_{\operatorname{dym}}&\approx \exp (d f(Q^\star_1, \hat{Q}^\star_1,Q^\star_2, \hat{Q}^\star_2, Q^\star_3, \hat{Q}^\star_3, Q^\star_4, \hat{Q}^\star_4,j, \tilde{j})) = \overline{Z}^d,\\ 
f(Q^\star_1, \hat{Q}^\star_1,Q^\star_2, \hat{Q}^\star_2, Q^\star_3, \hat{Q}^\star_3, Q^\star_4, \hat{Q}^\star_4,j, \tilde{j})& = -R\cdot \left( -\lambda \delta(t,t^{\prime}) - \frac{\gamma}{\alpha} \left(\gamma R^\top + \delta\right)^{-1}\right) \\&- \frac{1}{2\alpha}\log\operatorname{det} \left(\left(\gamma R+\delta\right)^{-1}*\left(\gamma R^{\top}+\delta\right)^{-1}\right)\\&+ \int \mathcal{D} {\mathcal{B}}(t) \exp\left( i\hat{\beta}^\top \left( -\lambda \delta - \frac{\gamma}{\alpha} \left(\gamma R^\top + \delta\right)^{-1}\right)  \tilde{\beta}\right.\\&\left.+ i\tilde{\beta}^\top \left( \frac{1}{2\alpha} \gamma^2 (\gamma R + \delta)^{-1} * \left(D+C\right) * (\gamma R^\top + \delta)^{-1}\right)  i\tilde{\beta}\right.\\&+\left.i\tilde{\beta}\cdot\left( \int dt^{\prime}\frac{\gamma}{\alpha} \left(\gamma R + \delta\right)^{-1}\right)\beta-i\tilde{\beta}\cdot \partial_t \hat{\beta}+\tilde{j}\cdot \hat{\beta} + j\cdot\tilde{\beta}\right).
\end{aligned}
\end{equation}
More specifically, we know that the first two terms in $f(Q^\star_1, \hat{Q}^\star_1,Q^\star_2, \hat{Q}^\star_2, Q^\star_3, \hat{Q}^\star_3, Q^\star_4, \hat{Q}^\star_4,j, \tilde{j})$ has to vanish because of the causality in the system’s dynamics:
\begin{equation}
\begin{aligned}
-R\cdot \left( -\lambda \delta(t,t^{\prime}) - \frac{\gamma}{\alpha} \left(\gamma R^\top + \delta\right)^{-1}\right)&= \iint dt dt^\prime R(t,t^\prime) \frac{\gamma}{ \alpha}\left(\gamma R^\top +\delta \right)^{-1}(t,t^\prime)=0,\\
- \frac{1}{2\alpha}\log\operatorname{det} \left(\left(\gamma R+\delta\right)^{-1}*\left(\gamma R^{\top}+\delta\right)^{-1}\right)& = 0.
\end{aligned}
\end{equation}
In conclusion, the single-neuron partition function can be further simplified as:
\begin{equation}
\begin{aligned}
\overline{Z}& = \int D \mathcal{B}(t) \exp(-\frac{i \gamma \tilde{\beta}^{\top}}{\alpha} \left(\gamma R +\delta \right)^{-1}\hat{\beta}- \lambda i\hat{\beta}(t)\cdot\tilde{\beta}(t)  \\&+\frac{\gamma^2}{2\alpha}i \tilde{\beta}^{\top} (\gamma R+\delta )^{-1}*(D+C)*(\gamma R^\top+\delta )^{-1}i \tilde{\beta}-i \tilde{\beta}\cdot \partial_t \hat{\beta}+\tilde{j}\cdot \hat{\beta} + j\cdot\tilde{\beta}\\&+i \tilde{\beta}\cdot\int dt^\prime \frac{\gamma}{ \alpha}\left(\gamma R +\delta \right)^{-1}(t,t^\prime)\beta),\\
& = \int D \mathcal{B}(t) \exp(-S\left[\hat{\beta}, \tilde{\beta}\right]),\\
S\left[\hat{\beta}, \tilde{\beta}\right]& =\frac{i \gamma \tilde{\beta}^{\top}}{\alpha} \left(\gamma R +\delta \right)^{-1}\hat{\beta}+ \lambda i\hat{\beta}(t)\cdot\tilde{\beta}(t)  \\&-\frac{\gamma^2}{2\alpha}i \tilde{\beta}^{\top} (\gamma R+\delta )^{-1}*(D+C)*(\gamma R^\top+\delta )^{-1}i \tilde{\beta}+i \tilde{\beta}\cdot \partial_t \hat{\beta}-\tilde{j}\cdot \hat{\beta} - j\cdot\tilde{\beta}\\&-i \tilde{\beta}\cdot\int dt^\prime \frac{\gamma}{ \alpha}\left(\gamma R +\delta \right)^{-1}(t,t^\prime)\beta.
\end{aligned}
\end{equation}
From the effective partition function, we can derive out the effective dynamics of a single weight $\hat{\beta}(t)$. We assume that the effective dynamics follow the ordinary equation for a stochastic process:
\begin{equation}
\partial_t \hat{\beta}(t)=-\frac{\partial H}{\partial \hat{\beta}}+\eta(t),
\end{equation}
where $\eta(t)$ denotes an effective Gaussian noise with zero mean and covariance $\langle \eta(t)\eta(t') \rangle = \hat{D}(t,t')$. The partition function for this stochastic process can be written as:
\begin{equation}
\begin{aligned}
1 \equiv Z &= \int \mathcal{D} \hat{\beta} \, P(\hat{\beta}) = \int \mathcal{D} \hat{\beta} \, \mathcal{D} \eta \, P(\eta) \, \delta\left( -\partial_t \hat{\beta}(t) - \partial_{\hat{\beta}} H + \eta(t) \right) \\
&= \left( \det(2\pi D) \right)^{-1/2} \int \mathcal{D} \tilde{\beta} \, \mathcal{D} \hat{\beta} \, \mathcal{D} \eta \\
&\quad \times \exp \left[ -\frac{1}{2} \int dt \, dt^{\prime} \, \eta(t) \hat{D}^{-1}(t, t^{\prime}) \eta(t^{\prime}) 
- i \int dt \, \tilde{\beta}(t) \partial_t \hat{\beta}(t)
- i \int dt \, \tilde{\beta}(t) \partial_{\hat{\beta}} H 
+ i \int dt \, \tilde{\beta}(t) \eta(t) \right] \\
&= \left( \det(2\pi D) \right)^{-1/2} \int \mathcal{D} \tilde{\beta} \, \mathcal{D} \hat{\beta} \, \mathcal{D} \eta \\
&\quad \times \exp \left[ -\frac{1}{2} \int dt \, dt^{\prime} \left( \eta(t) - \int ds \, \hat{D}(t, s) i\tilde{\beta}(s) \right) \hat{D}^{-1}(t, t^{\prime}) 
\left( \eta(t^{\prime}) - \int ds \, \hat{D}(t^{\prime}, s) i\tilde{\beta}(s) \right) \right. \\
&\quad \left. + \frac{1}{2} \int dt \, dt^{\prime} \, i\tilde{\beta}(t) \hat{D}(t, t^{\prime}) i\tilde{\beta}(t^{\prime}) 
- i \int dt \, \tilde{\beta}(t) \partial_t \hat{\beta}(t)
- i \int dt \, \tilde{\beta}(t) \partial_{\hat{\beta}} H \right] \\
&= \int \mathcal{D} \tilde{\beta} \, \mathcal{D} \hat{\beta} \,
\exp \left[ \frac{1}{2} \int dt \, dt^{\prime} \, i\tilde{\beta}(t) \hat{D}(t, t^{\prime}) i\tilde{\beta}(t^{\prime}) 
- i \int dt \, \tilde{\beta}(t) \partial_t \hat{\beta}(t)
- i \int dt \, \tilde{\beta}(t) \partial_{\hat{\beta}} H \right] \\
&= \int \mathcal{D} \tilde{\beta} \, \mathcal{D} \hat{\beta} \,
\exp \left( -S_0[\hat{\beta}, \tilde{\beta}] \right), \\
S_0[\hat{\beta}, \tilde{\beta}] &= -\frac{1}{2} \int dt \, dt^{\prime} \, i\tilde{\beta}(t) \hat{D}(t, t^{\prime}) i\tilde{\beta}(t^{\prime}) 
+ i \int dt \, \tilde{\beta}(t) \partial_t \hat{\beta}(t)
+ i \int dt \, \tilde{\beta}(t) \partial_{\hat{\beta}} H.
\end{aligned}
\end{equation}
Comparing the resulting action $S_0\left[\hat{\beta}, \tilde{\beta}\right]$ with the corresponding terms in the effective single-neuron action $S\left[\hat{\beta}, \tilde{\beta}\right]$, we can directly identify the underlying stochastic dynamics of the representative weight variable $\hat{\beta}(t)$. The corresponding terms are:
\begin{equation}
\begin{aligned}
\hat{D}(t,t^\prime) &= \frac{\gamma^2}{\alpha}(\gamma R+\delta )^{-1}*(D+C)*(\gamma R^\top+\delta )^{-1}(t,t^\prime),\\
\partial_{\hat{\beta}} H& =  \frac{ \gamma }{\alpha} \int dt^\prime \left(\gamma R +\delta \right)^{-1}(t,t^\prime)\hat{\beta}(t^\prime) -\int dt^\prime \frac{\gamma}{ \alpha}\left(\gamma R +\delta \right)^{-1}(t,t^\prime)\beta-\lambda \hat{\beta}(t),\\
& =  \frac{ \gamma }{\alpha} \int dt^\prime \left(\gamma R +\delta \right)^{-1}(t, t^\prime)\left(\hat{\beta}(t^\prime)-\beta\right)+\lambda \hat{\beta}(t).
\end{aligned}
\end{equation}
In conclusion, the effective dynamics for each weight $\hat{\beta}$ is:
\begin{equation}
\begin{aligned}
\partial_t{\hat{\beta}}(t)&=-\frac{\partial H}{\partial \hat{\beta}}+\eta(t),\\
& =-\frac{ \gamma }{\alpha} \int dt^\prime \left(\gamma R +\delta \right)^{-1}(t, t^\prime)\left(\hat{\beta}(t^\prime)-\beta\right)-\lambda \hat{\beta}(t)+\eta(t),\\
\label{eq:dynamics}
\end{aligned}
\end{equation}
where $\eta(t)$ is an effective Gaussian noise with zero mean and covariance $\langle \eta(t)\eta(t') \rangle = \hat{D}(t,t')$. The dynamics of the effective single-variable system is fully determined by the dynamical order parameters: the response function $R\left(t, t^{\prime}\right)$ and the correlation function $C\left(t, t^{\prime}\right)$. These observables are themselves obtained from a set of self-consistent equations, which can be derived by the following relations:
\begin{equation}
\begin{aligned}
& \left\langle\frac{\partial \hat{\beta}(t)}{\partial \eta\left(t^{\prime}\right)}\right\rangle\propto  \int \mathcal{D} \eta \exp \left[-\frac{1}{2} \eta^\top \hat{D}^{-1} \eta\right] \frac{\partial}{\partial \eta\left(t^{\prime}\right)} \int \mathcal{D} \hat{\beta} \mathcal{D} \tilde{\beta} \hat{\beta}(t) \exp \left[i\tilde{\beta}\cdot\left(-\partial_t \hat{\beta}-\partial_{\hat{\beta}} H\right)+i \tilde{\beta} \cdot\eta\right], \\
& =\left\langle i\hat{\beta}(t) \tilde{\beta}\left(t^{\prime}\right)\right\rangle=R\left(t, t^{\prime}\right).
\end{aligned}
\end{equation}
And
\begin{equation}
\begin{aligned}
&\left\langle \hat{\beta}(t) \, \eta(t^{\prime}) \right\rangle \\
&= \left. \int \mathcal{D} \eta \, \mathcal{D} \tilde{\beta} \, \mathcal{D} \hat{\beta} \; 
\hat{\beta}(t) \, \eta(t^{\prime}) 
\exp \left[ -\frac{1}{2} \int dt \, dt^{\prime\prime} \, \eta(t) \hat{D}^{-1}(t, t^{\prime\prime}) \eta(t^{\prime\prime}) 
+ i \int dt \, \tilde{\beta}(t) \left( -\partial_t \hat{\beta}(t) - \partial_{\hat{\beta}} H + \eta(t) \right)
+ \int dt \, j(t) \eta(t)
\right] \right|_{j = 0} \\
&= \left. \int \mathcal{D} \eta \, \mathcal{D} \tilde{\beta} \, \mathcal{D} \hat{\beta} \; 
\hat{\beta}(t) \, \frac{\delta}{\delta j(t^{\prime})} 
\exp \left[ -\frac{1}{2} \int dt \, dt^{\prime\prime} \, \eta(t) \hat{D}^{-1}(t, t^{\prime\prime}) \eta(t^{\prime\prime}) 
+ i \int dt \, \tilde{\beta}(t) \left( -\partial_t \hat{\beta}(t) - \partial_{\hat{\beta}} H + \eta(t) \right)
+ \int dt \, j(t) \eta(t)
\right] \right|_{j = 0} \\
&= \left. \frac{\delta}{\delta j(t^{\prime})} 
\int \mathcal{D} \tilde{\beta} \, \mathcal{D} \hat{\beta} \; 
\hat{\beta}(t) \, \exp \left[ \frac{1}{2} \int dt \, dt^{\prime\prime} \, i\tilde{\beta}(t) \hat{D}(t, t^{\prime\prime}) i\tilde{\beta}(t^{\prime\prime}) 
+ i \int dt \, \tilde{\beta}(t) \left( -\partial_t \hat{\beta}(t) - \partial_{\hat{\beta}} H \right) \right. \right. \\
&\quad \left. \left. + \frac{1}{2} \int dt \, dt^{\prime\prime} \left( j(t) \hat{D}(t, t^{\prime\prime}) j(t^{\prime\prime}) 
+ i j(t) \hat{D}(t, t^{\prime\prime}) \tilde{\beta}(t^{\prime\prime}) 
+ i \tilde{\beta}(t) \hat{D}(t, t^{\prime\prime}) j(t^{\prime\prime}) \right) \right] \right|_{j = 0} \\
&= \int \mathcal{D} \tilde{\beta} \, \mathcal{D} \hat{\beta} \;
\hat{\beta}(t) \int dt^{\prime\prime} \, \hat{D}(t^{\prime}, t^{\prime\prime}) \, i \tilde{\beta}(t^{\prime\prime}) 
\exp \left[ -S(\hat{\beta}, \tilde{\beta}) \right] \\
&= \int dt^{\prime\prime} \, \hat{D}(t^{\prime}, t^{\prime\prime}) \, R(t, t^{\prime\prime}).
\end{aligned}
\end{equation}
Based on the relations above, we can get the self-consistent equations for the response function:
\begin{equation}
\begin{aligned}
& \frac{\partial R\left(t_1, t_2\right)}{\partial t_1}=\frac{\partial}{\partial t_1}\left\langle\frac{\delta \hat{\beta}\left(t_1\right)}{\delta \eta \left(t_2\right)}\right\rangle=\left\langle\frac{\delta \partial_t{\hat{\beta}}\left(t_1\right)}{\delta \eta\left(t_2\right)}\right\rangle \\
& = -\frac{ \gamma}{\alpha} \int dt^\prime\left(\gamma R +\delta \right)^{-1}(t_1,t^\prime) R(t^\prime, t_2)-\lambda R(t_1, t_2)+\delta (t_1, t_2).
\end{aligned}
\end{equation}
To simplify the analysis of the correlation dynamics, we express the two-time correlation function as
\begin{equation}
C(t_1, t_2) = \langle (\hat{\beta}(t_1) - \beta)(\hat{\beta}(t_2) - \beta) \rangle 
= \hat{C}(t_1, t_2) - Q_5(t_1) - Q_5(t_2) + r,
\end{equation}
where we define the following observables:
\begin{equation}
\begin{aligned}
Q_5(t) &= \langle \hat{\beta}(t)\, \beta \rangle, \\
\hat{C}(t_1, t_2) &= \langle \hat{\beta}(t_1)\, \hat{\beta}(t_2) \rangle, \\
r &= \langle \beta^2 \rangle.
\end{aligned}
\end{equation}
Note that the correlation function is symmetric: $C(t_1, t_2) = C(t_2, t_1)$.

The observable $Q_5(t) = \langle \hat{\beta}(t)\, \beta \rangle$ describes the overlap between the estimator $\hat{\beta}(t)$ and the ground truth $\beta$,  which satisfies the following dynamical equation:
\begin{equation}
\frac{dQ_5(t)}{dt} 
= \langle \partial_t{\hat{\beta}}(t)\, \beta \rangle 
= -\frac{\gamma}{\alpha} \int dt' \left( \gamma R + \delta \right)^{-1}(t, t') \left[ Q_5(t') - r \right] - \lambda Q_5(t).
\label{eq:Q5dynamics}
\end{equation}

This equation mirrors the structure of the response function, with the kernel
\begin{equation}
b(t) = \beta \frac{\gamma}{\alpha} \int_0^t dt'\, (\gamma R + \delta)^{-1}(t, t')
\end{equation}
serving as a bridge between the two. Specifically,
\begin{equation}
\begin{aligned}
\frac{d}{dt_1} \left[ \beta \int_0^{t_1} dt_2\, b(t_2)\, R(t_1, t_2) \right] 
&= -\frac{\gamma}{\alpha} \int dt'\, (\gamma R + \delta)^{-1}(t_1, t') \left[ \int_0^{t_1} dt_2\, \beta b(t_2)\, R(t', t_2) - r \right] \\
&\quad - \lambda \left[ \int_0^{t_1} dt_2\, \beta b(t_2)\, R(t_1, t_2) \right].
\end{aligned}
\end{equation}

This confirms that the function $\beta \int_0^{t_1} dt_2\, b(t_2)\, R(t_1, t_2)$ satisfies the same differential equation as $Q_5(t_1)$ in Eq.~\eqref{eq:Q5dynamics}. Without loss of generality, we adopt the standard convention $\langle \hat{\beta}(0) \rangle = 0$, commonly used in the machine learning literature. Under this assumption, we obtain:
\begin{equation}
Q_5(0) = \langle \hat{\beta}(0)\, \beta \rangle = \beta \int_0^{t_1} dt_2\, b(t_2)\, R(0, t_2) = 0.
\end{equation}

By uniqueness of solutions to differential equations, the two expressions must coincide for all $t_1$, and therefore:
\begin{equation}
Q_5(t_1) = \beta \int_0^{t_1} dt_2\, b(t_2)\, R(t_1, t_2).
\end{equation}

Next, the dynamics of the full two-time covariance $\hat{C}(t_1, t_2) = \langle \hat{\beta}(t_1)\, \hat{\beta}(t_2) \rangle=\hat{C}(t_2, t_1)$ is governed by:
\begin{equation}
\begin{aligned}
\frac{\partial \hat{C}(t_1, t_2)}{\partial t_1} 
&= -\frac{\gamma}{\alpha} \int dt'\, (\gamma R + \delta)^{-1}(t_1, t') \left[ \hat{C}(t', t_2) - Q_5(t_2) \right] 
+ \langle \eta(t_1)\, \hat{\beta}(t_2) \rangle - \lambda\, \hat{C}(t_1, t_2) \\
&= -\frac{\gamma}{\alpha} \int dt'\, (\gamma R + \delta)^{-1}(t_1, t') \left[ \hat{C}(t', t_2) - Q_5(t_2) \right] \\
&\quad + \int dt''\, \hat{D}(t_1, t'') R(t_2, t'') - \lambda\, \hat{C}(t_1, t_2).
\end{aligned}
\end{equation}
Finally, we express the correlation function in terms of the covariance $\hat{C}(t_1, t_2)$ and the response function as:
\begin{equation}
\begin{aligned}
C(t_1, t_2) 
&= \hat{C}(t_1, t_2) - Q_5(t_1) - Q_5(t_2) + r \\
&= \hat{C}(t_1, t_2) - \beta \int_0^{t_1} dt'\, b(t')\, R(t_1, t') - \beta \int_0^{t_2} dt'\, b(t')\, R(t_2, t') + r,
\end{aligned}
\end{equation}
where $\hat{C}(t_1, t_2) = \langle \hat{\beta}(t_1) \hat{\beta}(t_2) \rangle$, $Q_5(t) = \langle \hat{\beta}(t)\, \beta \rangle$, and $r = \langle \beta^2 \rangle$.

This expression shows that the correlation function $C(t_1, t_2)$ can be constructed from the dynamical evolution of the full two-time covariance $\hat{C}(t_1, t_2)$, the response function $R(t, t')$, and the kernel $b(t)$ that captures the influence of the target $\beta$. 

\section{Generalization and Training Errors in the Thermodynamic Limit}
\subsection{Generalization Error}
\label{sec:generalization}
The generalization error is defined as the mean squared deviation between the predicted output and the ground truth over unseen data:
\begin{equation}
\varepsilon_{\mathrm{g}}(t) = \left\langle \left( H^0(t) \right)^2 \right\rangle,
\end{equation}
where the average is taken over the random input vector $\vec{x}^0 \in \mathbb{R}^d$ and label noise $\epsilon^0$. The random fields for the test data are defined as:
\begin{equation}
\begin{aligned}
H^0(t) &= \frac{1}{\sqrt{d}} \sum_{i=1}^d x^0_i \left( \hat{\beta}_i(t) - \beta_i \right) - \epsilon^0.
\end{aligned}
\end{equation}
Under the assumptions that the test input $\vec{x}^0$ is independently drawn from the same distribution as the training data, specifically, that its components are i.i.d. with zero mean and unit variance. And the label noise $\epsilon^0$ is also drawn from the same distribution as the training residual error, which is independent Gaussian noise with variance $D$, we obtain:
\begin{equation}
\begin{aligned}
\left\langle H^0(t) \right\rangle &= 0, \
\left\langle H^0(t) H^0(t^{\prime}) \right\rangle &= Q_1^\star(t, t^{\prime}) + D,
\end{aligned}
\end{equation}
where $Q_1^\star(t, t^{\prime}) = C(t, t^{\prime})$ denotes the two-time correlation function of the student weights deviation evaluated at the saddle point.

Consequently, the generalization error takes the compact form:
\begin{equation}
\varepsilon_{\mathrm{g}}(t) = \left( Q_1^\star(t, t) + D \right).
\end{equation}
While this is the standard definition of generalization error, its absolute value does not necessarily reflect the quality of generalization performance. This is because the mean-squared error (MSE) also depends on the norm of the teacher weights per dimension. A more meaningful indicator is the alignment between the student and teacher weights, which we will examine in greater detail in the asymptotic regime in Section~\ref{sec:longtimel0}.
\subsection{Two Formulations for Computing the Training Error}
\label{sec:training}
To proceed, we next calculate the training error, where the data used in validation is the same as the data used in training, therefore the training error should be defined as:
\begin{equation}
\begin{aligned}
&S_{\operatorname{d y n}}=\sum_i^d \int_0^{\infty} d t i \tilde{\beta}_i(t)\left(-\partial_t {\hat{\beta}}_i(t) - \sum_{\mu=1}^N s^\mu(t) \frac{x_i^\mu}{\sqrt{d}} \left( \sum_{k=1}^d \frac{x_k^\mu}{\sqrt{d}} \left( \hat{\beta}_k(t) - \beta_k \right) - \epsilon^\mu \right) - \lambda \hat{\beta}_i(t) \right), \\
& \varepsilon_{t}(t)=\left\langle\int \mathcal{D} \hat{\boldsymbol{\beta}}(t) \mathcal{D} \tilde{\boldsymbol{\beta}}(t) (\sum_j (\hat{\beta}_j {x}_{j}^1 - {\beta}_j(t){x}_{ j}^1) - \epsilon^1)^2 e^{S_{\operatorname{d y n}}}\right\rangle,\\
\end{aligned}
\end{equation}
where one training data point $\vec{x}^1$ is randomly selected from the entire training dataset for the purpose of evaluating the training error. Similarly, we define the random fields:
\begin{equation}
\begin{aligned}
H^\mu(t) & = \frac{1}{\sqrt{d}} \sum_{i=1}^d x_i^\mu \left( \hat{\beta}_i(t) - \beta_i \right) -  \epsilon^\mu, \\
\tilde{h}^\mu(t) &= \frac{1}{\sqrt{d}} \sum_{i=1}^d i  \tilde{\beta}_i(t) x_i^\mu.
\label{eq: random1}
\end{aligned}
\end{equation}
Next, we can decompose the dynamical action $S_{\operatorname{d y n}}$ into two parts, 
\begin{equation}
\begin{aligned}
S_{\operatorname{d y n}} &= S^{\mu>1}_{\operatorname{d y n}}+ S^{1}_{\operatorname{d y n}},\\
S^{\mu>1}_{\operatorname{d y n}}& = \sum_i^d \int_0^{\infty} d t i \tilde{\beta}_i(t)\left(-\partial_t{\hat{\beta}}_i(t) - \sum_{\mu=1}^{N-1} s^\mu(t) \frac{x_i^\mu}{\sqrt{d}} \left( \sum_{k=1}^d \frac{x_k^\mu}{\sqrt{d}} \left( \hat{\beta}_k(t) - \beta_k \right) - \epsilon^\mu \right) - \lambda \hat{\beta}_i(t) \right),\\
& =  \sum_i^d \int_0^{\infty} d t \left(-i \tilde{\beta}_i(t)\partial_t {\hat{\beta}}_i(t) -i \tilde{\beta}_i(t)\lambda{\hat{\beta}}_i(t) - \sum_{\mu=1}^{N-1}s^\mu(t) \tilde{h}^\mu(t) H^\mu(t)  \right),\\
S^{1}_{\operatorname{d y n}}& =  \sum_i^d \int_0^{\infty} d t \left( - s^1(t) \tilde{h}^1(t) H^1(t)  \right),\\
\end{aligned}
\end{equation}
Therefore, the training error can be written in the form of the random fields:
\begin{equation}
\begin{aligned}
\varepsilon_{t}(t)&=\left\langle\int \mathcal{D} \hat{\boldsymbol{\beta}}(t) \mathcal{D} \tilde{\boldsymbol{\beta}}(t) (\sum_j (\hat{\beta}_j {x}_{j}^1 - {\beta}_j(t){x}_{ j}^1) - \epsilon^1)^2 e^{S_{\operatorname{d y n}}}\right\rangle,\\
&=\int \mathcal{D} \hat{\boldsymbol{\beta}}(t) \mathcal{D} \tilde{\boldsymbol{\beta}}(t) \left\langle H^1(t)^2 e^{-\int_0^{\infty} d t s^1(t)\tilde{h}^1(t)H^1(t)}\right\rangle\left\langle e^{S^{\mu>1}_{\operatorname{d y n}}}\right\rangle,\\
\end{aligned}
\end{equation}
where the average $\langle \cdot \rangle$ now denotes the expectation over the random fields $s(t)$, $h(t)$, and $H(t)$. As in Equation~\ref{eq:avs}, we first carry out the average over $s(t)$, and for notational simplicity, we rescale the response field as $\gamma \tilde{h}(t) \to \tilde{h}(t)$.  To directly calculate the first average $ \left\langle H^1(t)^2 e^{-\int_0^{\infty} d t \tilde{h}^1(t)H^1(t)}\right\rangle$, we need to view $\tilde{h}(t), H(t)$ both as functions of time. Next, we define the covariance structure between the random fields as $\Sigma$ and $\mathbf{x}(t)=\binom{\tilde{h}(t)}{H(t)}$ the original formula can be written as:
\begin{equation}
\begin{aligned}
{\Sigma}(t,t^\prime)&=\left(\begin{array}{cc}
\boldsymbol{\Sigma}_{ \tilde{h} \tilde{h}}(t,t^\prime)  & \boldsymbol{\Sigma}_{ \tilde{h}{H}}(t,t^\prime) ,  \\
\boldsymbol{\Sigma}_{{H} \tilde{h}} (t,t^\prime)& \boldsymbol{\Sigma}_{{H}{H}} (t,t^\prime)
\end{array}\right)\\
\Sigma\Sigma^{-1}\left(t,t^\prime\right)=\int dt_1 \Sigma(t,t_1) \Sigma^{-1}(t_1, t^\prime)& = \left(\begin{array}{cc}
\delta (t-t^\prime)  & 0 ,  \\
0& \delta (t-t^\prime)\\
\end{array}\right),\\
 \langle H(t)^2 \exp{\left(-\int_{0}^{\infty}dt \tilde{h}(t)  H(t)   \right)} \rangle & =    \frac{1}{\mathcal{N}}  \int d\tilde{h}(t) \int dH(t) H^2(t)\exp{\left(-\frac{1}{2}\int \int \mathbf{x}^\top(t)\left(\Sigma^{-1}(t,t^\prime)+\delta (t-t^\prime)\mathbf{L}\right)\mathbf{x}(t^\prime)dt dt^\prime \right)},\\
\mathbf{L}&=\left(\begin{array}{cc}
0 & 1 \\
1 & 0
\end{array}\right),
 \end{aligned}
\end{equation}
where $\mathcal{N} = (2 \pi)^{-T} \operatorname{det}(\boldsymbol{\Sigma})^{-1 / 2}$ is the renormalization factor of the joint Gaussian distribution of $\mathbf{x}(t)$. To further simplify the results, we introduce the new operator $\mathbf{N}(t, t^{\prime}) = \Sigma^{-1}(t, t^{\prime}) + \delta(t - t^{\prime}) \mathbf{L}$. With this, the path integral can be evaluated, yielding:
\begin{equation}
\begin{aligned}
 \langle H^2(t)\exp{\left(-\int_{0}^{\infty}dt \tilde{h}(t)  H(t)   \right)} \rangle& = \sqrt{\frac{1}{\operatorname{det}\left(\boldsymbol{\Sigma}\right)\operatorname{det}\left(\mathbf{N}\right)}} (\mathbf{N})^{-1}_{(1,1)}(t,t),\\
 & = \sqrt{\frac{1}{\operatorname{det}\left(\mathbf{M}\right)}}\mathbf{N}^{-1}_{(1,1)}(t,t),\\
\operatorname{det}\left(\mathbf{M}\right)&=\operatorname{det}\left(\boldsymbol{\Sigma}\mathbf{N}\right)={\operatorname{det} \left(\begin{array}{cc}
 \boldsymbol{\Sigma}_{ \tilde{h}{H}}+\delta  &    \boldsymbol{\Sigma}_{ \tilde{h} \tilde{h}}  \\
 \boldsymbol{\Sigma}_{{H}{H}}  &   \boldsymbol{\Sigma}_{{H} \tilde{h}}+\delta
\end{array}\right),}\\\end{aligned}
\end{equation}
where $\mathbf{N}^{-1}_{(1,1)}$ denotes the lower-right block (corresponding to the observable $H$) of the inverse operator $\mathbf{N}^{-1}$. Recall that $\mathbf{N}(t, t^{\prime}) = \Sigma^{-1}(t, t^{\prime}) + \delta(t - t^{\prime}) \mathbf{L}$. The inverse can be computed analyticly as follows:
\begin{equation}
\begin{aligned}
\mathbf{N}&=\Sigma^{-1}+\left(\begin{array}{cc}
\mathbf{0} &\delta \\
\delta & \mathbf{0} 
\end{array}\right),\\
\Sigma \mathbf{N}& =  \delta +\left(\begin{array}{cc}
\boldsymbol{\Sigma}_{\tilde{h} \tilde{h}} & \boldsymbol{\Sigma}_{\tilde{h} H} \\
\boldsymbol{\Sigma}_{H \tilde{h}} & \boldsymbol{\Sigma}_{H H }
\end{array}\right)\left(\begin{array}{cc}
\mathbf{0} &\delta \\
\delta & \mathbf{0} 
\end{array}\right),\\
(\Sigma \mathbf{N})^{-1} & = \mathbf{N}^{-1} \Sigma^{-1} =  \left(\begin{array}{cc}
\delta +\boldsymbol{\Sigma}_{\tilde{h} H} & \boldsymbol{\Sigma}_{\tilde{h} \tilde{h}} \\
\boldsymbol{\Sigma}_{H H} & \delta +\boldsymbol{\Sigma}_{H \tilde{h} }
\end{array}\right)^{-1},\\
\mathbf{N}^{-1}& = \left(\begin{array}{cc}
 \boldsymbol{\Sigma}_{ \tilde{h}{H}}+\delta  &    \boldsymbol{\Sigma}_{ \tilde{h} \tilde{h}}  \\
 \boldsymbol{\Sigma}_{{H}{H}}  &   \boldsymbol{\Sigma}_{{H} \tilde{h}}+\delta
\end{array}\right)^{-1} \left(\begin{array}{cc}
\boldsymbol{\Sigma}_{\tilde{h} \tilde{h}} & \boldsymbol{\Sigma}_{\tilde{h} H} \\
\boldsymbol{\Sigma}_{H \tilde{h}} & \boldsymbol{\Sigma}_{H H }
\end{array}\right).
\end{aligned}
\end{equation}
Here, $\delta$ represents the Dirac delta kernel, acting as the identity under convolution in the functional space.  Note that $\Sigma_{\tilde{h}\tilde{h}}=0$ as shown in Section~\ref{sec:DMFT}, we can get the formula for $\mathbf{N}^{-1}$:
\begin{equation}
\begin{aligned}
\mathbf{N}^{-1}& = \left(\begin{array}{cc}
 \boldsymbol{\Sigma}_{ \tilde{h}{H}}+\delta  &   0 \\
 \boldsymbol{\Sigma}_{{H}{H}}  &   \boldsymbol{\Sigma}_{{H} \tilde{h}}+\delta
\end{array}\right)^{-1} \left(\begin{array}{cc}
0 & \boldsymbol{\Sigma}_{\tilde{h} H}, \\
\boldsymbol{\Sigma}_{H \tilde{h}} & \boldsymbol{\Sigma}_{H H },\\
 \end{array}\right),\\
& =\left(\begin{array}{cc}
 (\boldsymbol{\Sigma}_{ \tilde{h}{H}}+\delta)^{-1}   &   0 \\
-( \boldsymbol{\Sigma}_{{H} \tilde{h}}+\delta)^{-1}*  \boldsymbol{\Sigma}_{{H}{H}} * (\boldsymbol{\Sigma}_{ \tilde{h}{H}}+\delta)^{-1}   &   (\boldsymbol{\Sigma}_{{H} \tilde{h}}+\delta)^{-1}
\end{array}\right)\left(\begin{array}{cc}
0 & \boldsymbol{\Sigma}_{\tilde{h} H}, \\
\boldsymbol{\Sigma}_{H \tilde{h}} & \boldsymbol{\Sigma}_{H H },\\
 \end{array}\right),\\
\end{aligned}
\end{equation}
where we can read $\mathbf{N}^{-1}_{(11)}(t, t) = ( \boldsymbol{\Sigma}_{{H} \tilde{h}}+\delta)^{-1}*  \boldsymbol{\Sigma}_{{H}{H}} * (\boldsymbol{\Sigma}_{ \tilde{h}{H}}+\delta)^{-1} (t, t)$ from the final expression. The operator $ * $ denotes convolution over intermediate times, defined as $A * B(t,t^\prime)=\iint A \left(t, t^{\prime\prime}\right)B\left(t^{\prime\prime}, t^\prime\right)  \mathrm{~d} t^{\prime\prime}$.

The expression for the second average, $\left\langle e^{S^{\mu>1}_{\operatorname{d y n}}}\right\rangle$, derived in Section~\ref{sec:DMFT}, takes the following form:
\begin{equation}
\begin{aligned}
\left\langle e^{S^{\mu>1}_{\operatorname{d y n}}}\right\rangle = \exp \left(\int_0^{\infty} d t\left(-\sum_j^d i \tilde{\beta}_j(t) \partial_t{\hat{\beta}}_j(t)-\sum_j^d i \tilde{\beta}_j(t) \lambda{\hat{\beta}}_j(t)\right)-\frac{N-1}{2} \log \operatorname{det}\left(\mathbf{M}\right)\right).
\end{aligned}
\end{equation}
Taken together, the training error in the thermodynamic limit can be written as:
\begin{equation}
\begin{aligned}
\varepsilon_{t}(t)&=\left\langle\int \mathcal{D} \hat{\boldsymbol{\beta}}(t) \mathcal{D} \tilde{\boldsymbol{\beta}}(t) (\sum_j (\hat{\beta}_j {x}_{j}^1 - {\beta}_j(t){x}_{ j}^1) - \epsilon^1)^2 e^{S_{\operatorname{d y n}}}\right\rangle,\\
&=\int \mathcal{D} \hat{\boldsymbol{\beta}}(t) \mathcal{D} \tilde{\boldsymbol{\beta}}(t) \left\langle H^1(t)^2 e^{-\int_0^{\infty} d t s^1(t)\tilde{h}^1(t)H^1(t)}\right\rangle\left\langle e^{S^{\mu>1}_{\operatorname{d y n}}}\right\rangle,\\
& =\int \mathcal{D} \hat{\boldsymbol{\beta}}(t) \mathcal{D} \tilde{\boldsymbol{\beta}}(t)  \sqrt{\frac{1}{\operatorname{det}\left(\mathbf{M}\right)}}\mathbf{N}^{-1}_{(1,1)}(t,t) \exp \left(\int_0^{\infty} d t\left(-\sum_j^d i \tilde{\beta}_j(t) \partial_t {\hat{\beta}}_j(t)-\sum_j^d i \tilde{\beta}_j(t) \lambda{\hat{\beta}}_j(t)\right)-\frac{N-1}{2} \log \operatorname{det}\left(\mathbf{M}\right)\right),\\
& =\int \mathcal{D} \hat{\boldsymbol{\beta}}(t) \mathcal{D} \tilde{\boldsymbol{\beta}}(t)  \sqrt{\frac{1}{\operatorname{det}\left(\mathbf{M}\right)}}\mathbf{N}^{-1}_{(1,1)}(t,t) \sqrt{\operatorname{det}\left(\mathbf{M}\right)}\exp \left(\int_0^{\infty} d t\left(-\sum_j^d i \tilde{\beta}_j(t) \partial_t {\hat{\beta}}_j(t)-\sum_j^d i \tilde{\beta}_j(t) \lambda{\hat{\beta}}_j(t)\right)-\frac{N}{2} \log \operatorname{det}\left(\mathbf{M}\right)\right),\\
& = \left\langle  \int \mathcal{D} \hat{\boldsymbol{\beta}}(t) \mathcal{D} \tilde{\boldsymbol{\beta}}(t) \left( ( \boldsymbol{\Sigma}_{{H} \tilde{h}}+\delta)^{-1}*  \boldsymbol{\Sigma}_{{H}{H}} * (\boldsymbol{\Sigma}_{ \tilde{h}{H}}+\delta)^{-1} (t, t)\right)e^{S_{\operatorname{d y n}}}\right\rangle,\\
& = \left(\gamma R + \delta \right)^{-1} *\left(D + C \right)* \left(\gamma R^\top + \delta \right)^{-1}  (t,t).
\end{aligned}
\end{equation}
To compute the training error, one must first simulate the dynamics governed by the effective action $S[\hat{\beta}, \tilde{\beta}]$. Once the system reaches a fixed point characterized by the observables $R$ and $C$, the training error can then be evaluated accordingly.

We can also calculate the training error using another formulation, by adding an external leg $j(t)$ coupled to $ (\sum_j (\hat{\beta}_j (t) {x}_{j}^1 - {\beta}_j{x}_{ j}^1 -\epsilon^1))$:
\begin{equation}
\begin{aligned}
&S_{\operatorname{d y n}}=\sum_i^d \int_0^{\infty} d t i \tilde{\beta}_i(t)\left(-\partial_t{\hat{\beta}}_i(t) - \sum_{\mu=1}^N s^\mu(t) \frac{x_i^\mu}{\sqrt{d}} \left( \sum_{k=1}^d \frac{x_k^\mu}{\sqrt{d}} \left( \hat{\beta}_k(t) - \beta_k \right) - \epsilon^\mu \right) - \lambda \hat{\beta}_i(t)+\int_0^{\infty} d t j(t)  (\sum_j (\hat{\beta}_j (t) {x}_{j}^1 - {\beta}_j{x}_{ j}^1 -\epsilon^1)) \right), \\
& \varepsilon_{t}(t)=\left\langle\int D \hat{\boldsymbol{\beta}}(t) D \tilde{\boldsymbol{\beta}}(t) \left(\frac{\partial^2 e^{S_{\operatorname{d y n}}}}{\partial (j(t))^2}\right)|_{j (t)=0}\right\rangle.\\
& = \left\langle\int D \hat{\boldsymbol{\beta}}(t) D \tilde{\boldsymbol{\beta}}(t) \left(\frac{\delta^2 e^{S_{\operatorname{d y n}}}}{\delta (j(t))^2}\right)\right\rangle|_{j (t)=0}.\\
\end{aligned}
\end{equation}
Similarly, we can decompose the dynamical action into two parts, and the action can be rearranged into:
\begin{equation}
\begin{aligned}
& S_{d y n}=\sum_j^d \int_0^{\infty} d t \left(-i \tilde{\beta}_j(t)\partial_t{\hat{\beta}}_j(t)-i \tilde{\beta}_j(t)\lambda{\hat{\beta}}_j(t)-\sum_\mu^N \tilde{h}_\mu(t) H_{\mu}(t)  \right) +\int_0^{\infty} d t  j (t) H_{1}(t),\\
& = \sum_j^d \int_0^{\infty} d t \left(-i \tilde{\beta}_j(t)\partial_t{\hat{\beta}}_j(t)-i \tilde{\beta}_j(t)\lambda{\hat{\beta}}_j(t)-\sum_\mu^{N-1} \tilde{h}_\mu(t) H_{\mu}(t)  \right) +\int_0^{\infty} d t  (j (t)-\tilde{h}_1(t)) H_{1}(t).
\end{aligned}
\end{equation}
This formula implies that the external leg $j(t)$ serves as a translation of $\tilde{h}_1(t)$. Next, we can calculate the average term $\langle \exp{\left(\int_0^{\infty} d t  (j (t)-\tilde{h}_1(t)) H_{1}(t)\right)} \rangle$:
\begin{equation}
\begin{aligned}
\langle \exp{\left(\int_0^{\infty} d t  (j (t)-\tilde{h}_1(t)) H_{1}(t)\right)} \rangle =  \sqrt{\frac{1}{\operatorname{det}\left(\mathbf{M}\right)}} \exp{(\frac{1}{2} \int_0^\infty dt \int_0^\infty dt^\prime j(t) \mathbf{N}_{(1,1)}^{-1}(t, t^\prime)j(t^\prime))},
\end{aligned}
\end{equation}
where we already know that $\mathbf{N}_{(1,1)}^{-1} =  ( \boldsymbol{\Sigma}_{{H} \tilde{h}}+\delta)^{-1}*  \boldsymbol{\Sigma}_{{H}{H}} * (\boldsymbol{\Sigma}_{ \tilde{h}{H}}+\delta)^{-1} $. Taken together, we can get the dynamical partition function with external leg $j(t)$:
\begin{equation}
\begin{aligned}
Z_{d y n}&=  \int D \hat{\boldsymbol{\beta}}(t) D \tilde{\boldsymbol{\beta}}(t) \exp \left(\int_0^{\infty} d t\left(-\sum_j^d i \tilde{\beta}_j(t) \partial_t{\hat{\beta}}_j(t)-\sum_j^d i \tilde{\beta}_j(t) \lambda{\hat{\beta}}_j(t)\right)-\frac{N-1}{2} \log \operatorname{det}\left(\mathbf{M}\right)\right)\\&\times \sqrt{\frac{1}{\operatorname{det}\left(\mathbf{M}\right)}} \exp{(\frac{1}{2} \int_0^\infty dt \int_0^\infty dt^\prime j(t) \mathbf{N}_{(1,1)}^{-1}(t, t^\prime)j(t^\prime))},\\
& =  \int D \hat{\boldsymbol{\beta}}(t) D \tilde{\boldsymbol{\beta}}(t) \exp \left(\int_0^{\infty} d t\left(-\sum_j^d i \tilde{\beta}_j(t) \partial_t{\hat{\beta}}_j(t)-\sum_j^d i \tilde{\beta}_j(t) \lambda{\hat{\beta}}_j(t)\right)-\frac{N}{2} \log \operatorname{det}\left(\mathbf{M}\right)\right)\\&\times \exp{(\frac{1}{2} \int_0^\infty dt \int_0^\infty dt^\prime j(t) \mathbf{N}_{(1,1)}^{-1}(t, t^\prime)j(t^\prime))},\\
\varepsilon_{t}(t)& =  \int D \hat{\boldsymbol{\beta}}(t) D \tilde{\boldsymbol{\beta}}(t) \left(\frac{\delta^2 \langle e^{S_{d y n}}\rangle}{\delta (j(t))^2}\right)|_{j (t)=0},\\
& =  \int D \hat{\boldsymbol{\beta}}(t) D \tilde{\boldsymbol{\beta}}(t)) \exp \left(\int_0^{\infty} d t\left(-\sum_j^d i \tilde{\beta}_j(t) \partial_t {\hat{\beta}}_j(t)-\sum_j^d i \tilde{\beta}_j(t) \lambda{\hat{\beta}}_j(t)\right)-\frac{N}{2} \log \operatorname{det}\left(\mathbf{M}\right)\right) \mathbf{N}_{(1,1)}^{-1}(t, t),\\
& = \left\langle  \int \mathcal{D} \hat{\boldsymbol{\beta}}(t) \mathcal{D} \tilde{\boldsymbol{\beta}}(t) \left( ( \boldsymbol{\Sigma}_{{H} \tilde{h}}+\delta)^{-1}*  \boldsymbol{\Sigma}_{{H}{H}} * (\boldsymbol{\Sigma}_{ \tilde{h}{H}}+\delta)^{-1} (t, t)\right)e^{S_{\operatorname{d y n}}}\right\rangle,\\
& = \left(\gamma R + \delta \right)^{-1} *\left(D + C \right)* \left(\gamma R^\top + \delta \right)^{-1}  (t,t).
\end{aligned}
\end{equation}
Both approaches yield consistent results.

In the thermodynamic limit, we have derived compact expressions for the generalization and training errors. The generalization error is given by  
\begin{equation}
\varepsilon_{\mathrm{g}}(t) =C(t,t) + D,
\label{eq:generalization}
\end{equation}
where $ C(t,t) $ denotes the two-time correlation function of the deviations of the student weights from the teacher weights, i.e., $ C(t, t^{\prime}) = \left\langle \left( \hat{\beta}(t) - \beta \right) \left( \hat{\beta}(t^{\prime}) - \beta \right) \right\rangle $, and $ D $ is the variance of the label noise. The training error, on the other hand, takes the form  
\begin{equation}
\varepsilon_{\mathrm{train}}(t) =  \left[ \left(\gamma R + \delta \right)^{-1} *\left(D + C \right)* \left(\gamma R^\top + \delta \right)^{-1} \right](t,t),
\label{eq:training}
\end{equation}
where $ R(t,t^{\prime}) $ represents the linear response function of the student weights to external perturbations.

This comparison reveals that while the generalization error directly reflects the intrinsic fluctuations of the student weights and the noise level, the training error is further modulated by the temporal structure of the learning dynamics via the response function $ R(t,t^{\prime}) $. Specifically, the training error includes the effect of error suppression through the inverse operator $ \left(\gamma R + \delta \right)^{-1} $, which characterizes how sensitively the system responds to input perturbations during training. As a result, the training error is always less than or equal to the generalization error, since the response operator acts to filter out part of the noise and fluctuation contributions that still affect generalization. This separation quantifies the generalization gap and highlights the dynamical origin of the discrepancy between performance on training and unseen data.

\section{Details of Numerical Simulations}
\subsection{Dynamical Mean-Field Theory (DMFT) Implementation}
\label{sec:implementation}

To compare the solution of DMFT equations with the generalization performance of the actual network, we simulate the DMFT equations. The iteration scheme operates in discrete time steps, and we must specify a total duration $ L \, (\mathrm{ms}) $ and a time resolution $ \Delta t \, (\mathrm{ms}) $ for discretization. Time is measured in milliseconds, a convention commonly adopted in modeling neural dynamics in brain circuits (e.g., for rate-based models).

At the beginning of the iteration, we initialize the self-consistent matrices $ C[t, t^{\prime}] $ and $ R[t, t^{\prime}] $, both of dimension $ (L / \Delta t) \times (L / \Delta t) $. All the differential and integral operators are discretized according to the rules below, where $ \delta(t_1, t_2) $ denotes the continuous Dirac delta function, $ \delta_{ij} $ is the Kronecker delta\cite{zou2024introduction,roy2019numerical}:
\begin{equation}
\begin{aligned}
\delta(t_1, t_2) &\longrightarrow \frac{1}{\Delta t} \delta_{ij}, \\
\int_0^L dt &\longrightarrow \Delta t \sum_{i=0}^{L/\Delta t - 1}, \\
\frac{df(t)}{dt} \bigg|_{t = i \Delta t} &\longrightarrow \frac{f_{i+1} - f_i}{\Delta t}, \\
\int_0^L dt_1 \int_0^L dt_2 \, A(t_1, t_2) B(t_2, t_1) &\longrightarrow (\Delta t)^2 \sum_{i,j} A_{ij} B_{ji},\\
\left(\gamma R+\delta\right)^{-1}\left(t_1, t_2\right) &\longrightarrow \left(\gamma R+\frac{I}{\Delta t}\right)^{-1}\left[t_1, t_2\right] \frac{1}{(\Delta t)^2}
\end{aligned}
\end{equation}
Based on the rules above, we can discretize the self-consistent equations for the response function $R(t,t^\prime)$ and the correlation function $C(t,t^\prime)$ accordingly:
\begin{equation}
\begin{aligned}
& \frac{\partial R\left(t_1, t_2\right)}{\partial t_1}= -\frac{ \gamma}{\alpha} \int dt^\prime\left(\gamma R +\delta \right)^{-1}(t_1,t^\prime) R(t^\prime, t_2)+\delta (t_1, t_2)-\lambda R(t_1, t_2),\\
&R\left[t_1+1, t_2\right] = R\left[t_1, t_2\right] -\frac{ \gamma}{\alpha}  \sum_{t^\prime}^{t_1}\left(\gamma R +\frac{I}{\Delta t} \right)^{-1}\left[t_1,t^\prime\right] R\left[t^\prime, t_2\right]+\delta_{t_1, t_2}-\lambda R \left[t_1, t_2\right],\\
\frac{\partial Q_5(t)}{\partial t} & =-\frac{\gamma}{\alpha} \int d t^{\prime}(\gamma R^\star+\delta)^{-1}\left(t, t^{\prime}\right)\left(Q_5\left(t^{\prime}\right)-r\right)-\lambda Q_5(t) \\
Q_5\left[t+1\right] & =Q_5\left[t\right]-\frac{\gamma}{\alpha} \sum_{t^{\prime}=0}^t(\gamma R^\star+\delta)^{-1}\left[t, t^{\prime}\right]\left(Q_5\left[t^{\prime}\right]-r\right)-\lambda Q_5\left[t\right] \\
\frac{\partial \hat{C}\left(t_1, t_2\right)}{\partial t_1} & =-\frac{\gamma}{\alpha} \int d t^{\prime}(\gamma R^\star+\delta)^{-1}\left(t_1, t^{\prime}\right)\left(\hat{C}\left(t^{\prime}, t_2\right)-Q^\star_5\left(t_2\right)\right)\\&+\int d t^{\prime \prime} \frac{\gamma^2}{\alpha}(\gamma R^\star+\delta)^{-1}  *\left((r+D)+\hat{C}-Q_5^{\star} -\left(Q_5^{\star}\right)^{\top}\right) *\left(\gamma (R^\star)^{\top}+\delta\right)^{-1}\left(t_1, t^{\prime \prime}\right) R^\star\left(t_2, t^{\prime \prime}\right) \\
\hat{C}\left[t_1+1, t_2\right] & =C\left[t_1, t_2\right]-\frac{\gamma}{\alpha} \sum_{t^{\prime}}\left(\gamma R^\star+\frac{I}{\Delta t}\right)^{-1}\left[t_1, t^{\prime}\right]\left(\hat{C}\left[t^{\prime}, t_2\right]-Q^\star_5\left[t_2\right]\right) \\
& +\frac{\gamma^2}{\alpha} \sum_{t^{\prime \prime}}\left(\gamma R^\star+\frac{I}{\Delta t}\right)^{-1} *\left((r+D)+C-Q_5^{\star} - \left(Q_5^{\star}\right)^{\top}\right)  *\left(\gamma (R^\star)^{\top}+\frac{I}{\Delta t}\right)^{-1}\left[t_1, t^{\prime \prime}\right] R^\star\left[t_2, t^{\prime \prime}\right] \\
C\left[t_1, t_2\right] &= \hat{C}^\star\left[t_1, t_2\right] - Q_5^\star\left[t_1\right]- Q_5^\star\left[t_2\right]+r.
\end{aligned}
\end{equation}
where $ I $ denotes the identity matrix, corresponding to the Kronecker delta in matrix form; $ \mathcal{O}[t_1, t_2] $ represents the matrix elements used in the simulation; $ (\cdot)^{-1} $ indicates the matrix inverse; $R^\star, Q_5^\star, \hat{C}^\star$ represents the converged value for the observables; and $ * $ denotes matrix multiplication here. Symmetry of the correlation function is enforced by setting $C\left[t_1+1,t_2\right] =C\left[t_2, t_1+1\right]$. 

\begin{figure}[htbp] 
	\centering
	\includegraphics[scale = 0.35]{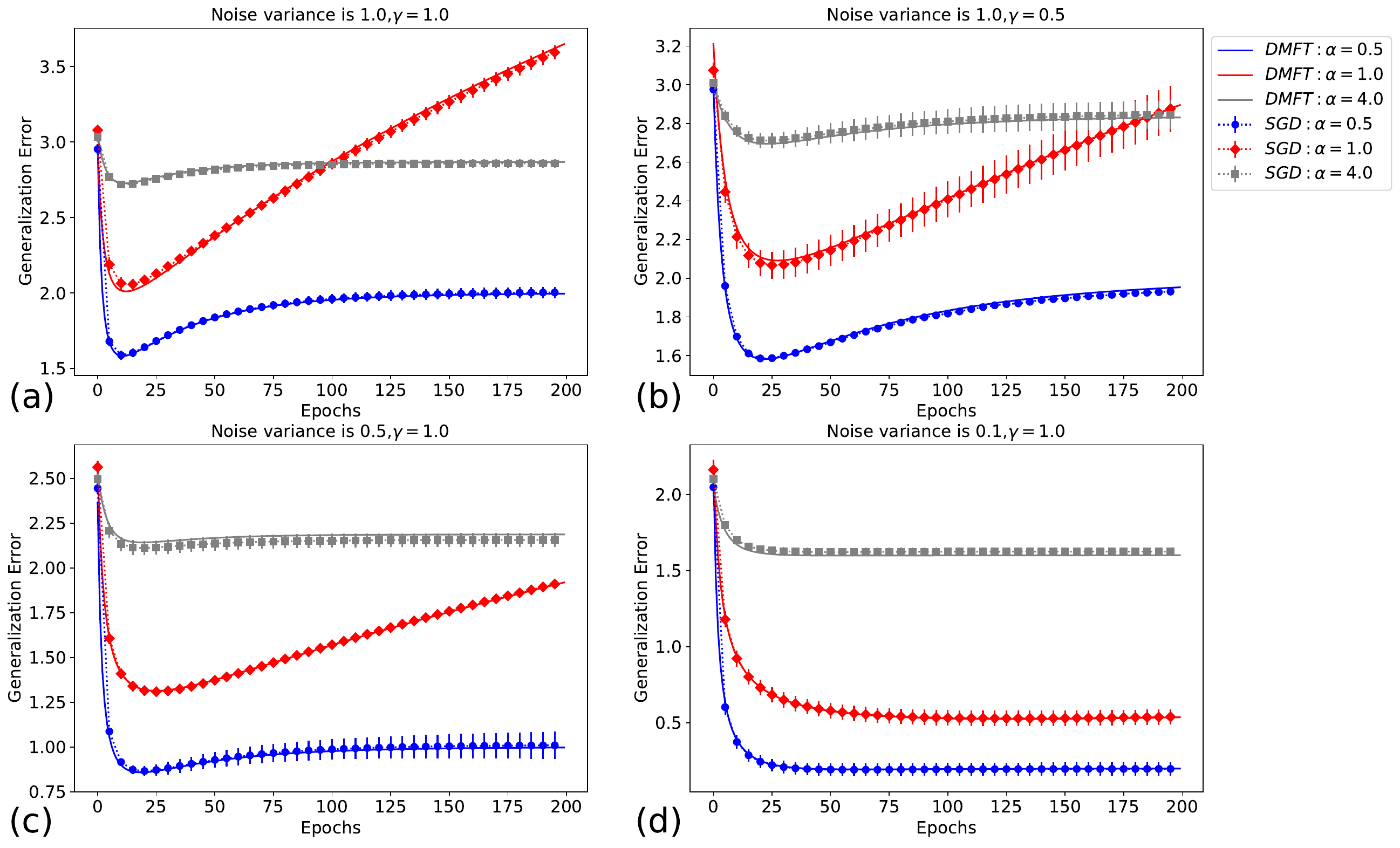}
	\caption{Comparison of generalization error between DMFT predictions and simulation results across different hyper-parameter settings for $\lambda=0$. For DMFT, we set $dt = 0.1$ and total time $T = 20$, giving $T/dt = 200$ steps. The response function $R(t, t')$ is initialized to zero, and the covariance $\hat{C}(0,0) = \frac{1}{d} \sum_i \hat{\beta}_i^2(0)$, with all other elements zero. In simulations, we fix $N = 1,000$ and $P = \alpha N$, use learning rate $\New{\zeta} = 0.1$, and initialize student weights $\hat{\boldsymbol{\beta}}(0)$ as i.i.d. standard Gaussian. Teacher weights $\boldsymbol{\beta} \sim \mathcal{N}(0, r = 1)$ are fixed across runs. Error bars are averaged over 10 trials.  
(a) $D = 1.0, \gamma = 1.0$.  
(b) $D = 1.0, \gamma = 0.5$.  
(c) $D = 0.5, \gamma = 1.0$.  
(d) $D = 0.1, \gamma = 1.0$.
}   
	\label{figs1}
\end{figure}

After the self-consistent equations converge, we obtain the saddle-point solutions $C^{\star}\left(t, t^{\prime}\right)$ and $R^{\star}\left(t, t^{\prime}\right)$, which can then be used to compute the generalization error (Equation~\ref{eq:generalization}) and training error (Equation~\ref{eq:training}) in their mean-field forms. The numerical results obtained from DMFT show excellent agreement with those from high-dimensional linear regression, solved using a teacher–student setup, as illustrated in Figure~\ref{figs1}.

\begin{figure}[htbp] 
	\centering
	\includegraphics[scale = 0.35]{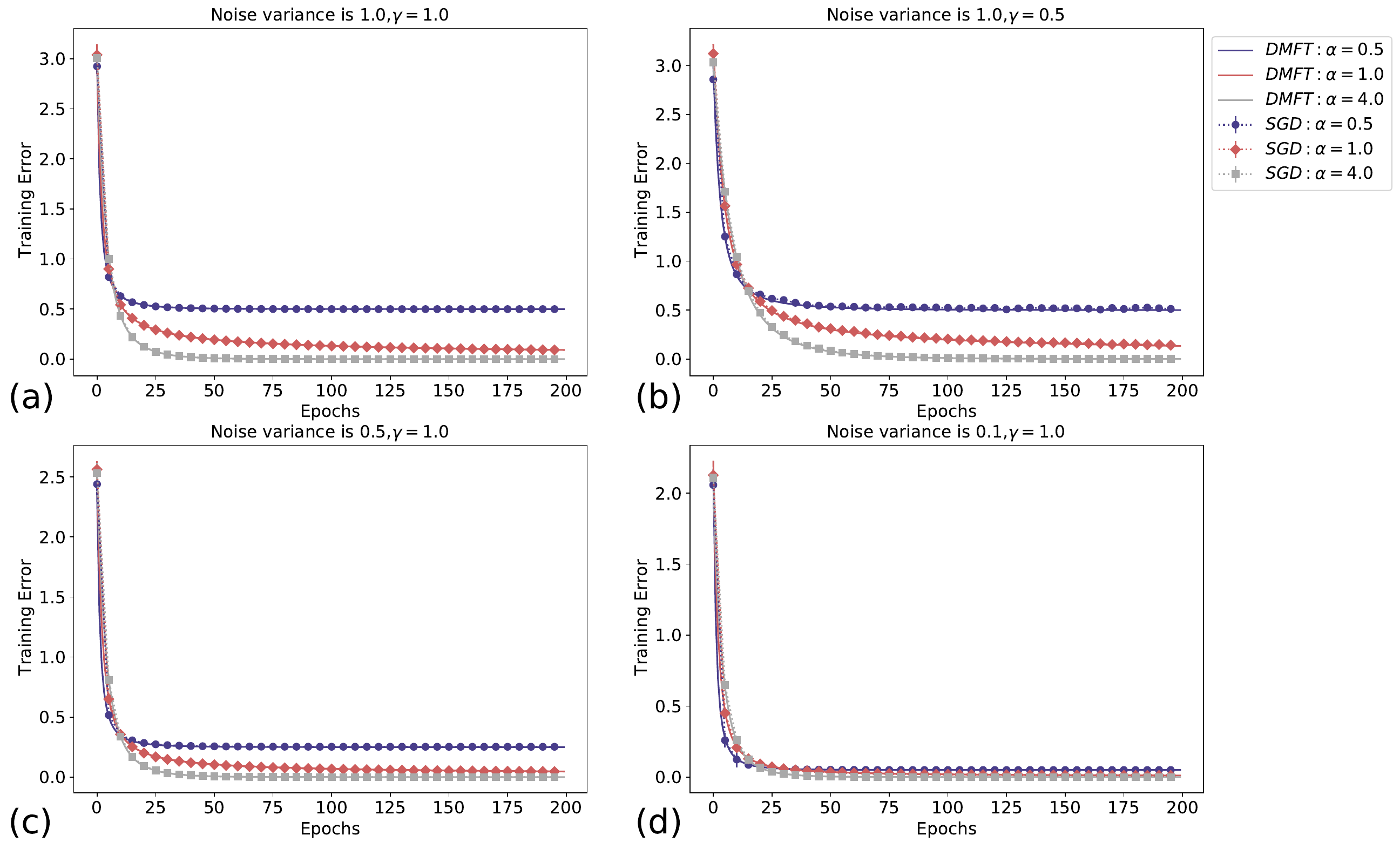}
	\caption{Comparison of training error between DMFT predictions and simulation results across different hyper-parameter settings for $\lambda=0$. Unless stated otherwise, other training conditions are the same as in Figure~\ref{figs1}.  
(a) $D = 1.0, \gamma = 1.0$.  
(b) $D = 1.0, \gamma = 0.5$.  
(c) $D = 0.5, \gamma = 1.0$.  
(d) $D = 0.1, \gamma = 1.0$.
}   
	\label{figs2}
\end{figure}

\subsection{Numerical Implementation for Results in the Main Text}

This section summarizes the numerical procedures used to produce the results shown in the main text.  
All simulations were performed on single-layer linear neural networks trained by stochastic gradient descent (SGD).  
Unless otherwise specified, the following parameters are shared across all figures:

\begin{itemize}
    \item $N$: number of training samples.  
    \item $d = \alpha N$: input dimension, where $\alpha$ controls model capacity (ratio of parameters to samples).  
    \item $D = 1$: variance of the additive Gaussian label noise.  
    \item $r$: variance of the teacher weight vector $\boldsymbol{\beta}$, representing signal strength.  
    \item $\gamma$: stochasticity parameter; each sample is included independently in each update with probability $\gamma$ ($\gamma = 1$ corresponds to full-batch training).  
    \item $\New{\zeta} = 0.1$: learning rate.  
    \item $\lambda$: L2 regularization strength (set to $0$ in unregularized cases).  
    \item Convergence criterion: $|\varepsilon_{\text{g}}(t) - \varepsilon_{\text{g}}(t-1)| < 10^{-20}$. 
    \item $T/\New{\zeta}$: total number of training iterations.
\end{itemize}

Teacher weights $\boldsymbol{\beta}$ are drawn from a Gaussian distribution with zero mean and variance $r$, and remain fixed across all independent realizations.  
Student weights $\hat{\boldsymbol{\beta}}(0)$ are initialized from a standard normal distribution.  
Unless otherwise noted, reported results are averaged over ten independent runs, and error bars indicate standard deviations across runs.

\paragraph{Figure 1.}
Comparison between numerical simulations and theoretical predictions for the generalization error $\varepsilon_{\text{g}}(t)$ and training error $\varepsilon_{\text{t}}(t)$, both in the steady state [panel~(a)] and during the learning dynamics [panel~(b)].  
Simulations are performed for an unregularized ($\lambda = 0$) network with $N = 1,000$, $r = 1$, and $\gamma = 0.5$.  

\textbf{(a)}~Asymptotic values ($T = 10^4$) of $\varepsilon_{\text{g}}$ and $\varepsilon_{\text{t}}$ plotted versus $\alpha$.  
The total number of iterations is $T/\New{\zeta} = 10^5$, ensuring convergence.  
The sampled $\alpha$ values are  
$$
[0.1, 0.2, 0.3, 0.4, 0.5, 0.6, 0.7, 0.8, 0.9, 0.92, 0.94, 0.96, 1.04, 1.06, 1.1, 1.3, 1.5, 1.8, 2.0, 3.0, 4.0, 5.0].$$  
Markers denote simulations, solid lines analytical theory. The Heaviside function $H(x)$ is defined as $H(x) = 1$ for $x>0$ and $0$ otherwise.

\textbf{(b)}~Time evolution of $\varepsilon_{\text{g}}(t)$ and $\varepsilon_{\text{t}}(t)$ for different $\alpha$.  
The DMFT integration time spacing equals the SGD learning rate, $dt = \New{\zeta} = 0.1$ ($T/dt = 200$ epochs).  
Markers show averaged simulations; solid lines indicate DMFT results.  
Initial response elements $R[i][j]$ are sampled from a standard Gaussian with $R[i][j] = 0$ for $i \le j$, and correlation elements initialized to zero except $C[0][0] = d^{-1}\sum_i (\hat{\beta}_i(0) - \beta_i)^2$.  
Further implementation details are given in Sec.~\ref{sec:implementation}.

\paragraph{Figure 2.}
Comparison between theoretical data-collapse predictions and simulations for the unregularized ($\lambda = 0$) network with $N = 10{,}000$, $r = 1$, $\gamma = 0.5$, and $T = 10{,}000$.  
Error bars show fluctuations across ten runs.

Panels~(a,b) correspond to the underparameterized regime ($\alpha < 1$); panels~(c,d) to the overparameterized regime ($\alpha > 1$).  
\textbf{(a, c)}~Temporal evolution of $\varepsilon_{\text{g}}(t)$ versus training time $t$ for different $\alpha$.  
\textbf{(b, d)}~Scaled generalization errors plotted against the theoretical scaling variable, showing excellent data collapse onto the universal scaling function $F(x)$ (red line) predicted analytically.

\paragraph{Figure 3.}
Comparison between theoretical scaling predictions and simulations for a regularized single-layer linear network.  
Parameters: $N = 10{,}000$, $r = 1$, $\gamma = 1$, and $\lambda \in [0.01, 1.0]$.  
Each point averages ten runs; error bars denote standard deviations.  

\textbf{(a, c)}~Steady-state generalization error $\varepsilon_{\text{g}}(t \to \infty)$, rescaled by $|1-\alpha|$, as a function of the dimensionless regularization variable $4\lambda / [\gamma(1-\alpha)^2]$.  
Each marker corresponds to a different $\alpha$. In the weakly regularized regime, curves collapse onto the theoretical prediction (red), confirming the static scaling law.  The total trainig time $T$ is chosen adaptively for each $(\lambda,\,\alpha)$ to ensure convergence of the generalization error.

\textbf{(b, d)}~Temporal evolution of the rescaled $\varepsilon_{\text{g}}(t)\,|1-\alpha|\sqrt{1 + 4\lambda / [\gamma(1-\alpha)^2]}$ versus the composite scaling variable ${\left[\gamma(1-\alpha)^2 / (4\lambda) + 1\right]\lambda t}$.  
Training time is fixed at $T = 10{,}000$.  
Data from different $(\alpha, \lambda)$ collapse onto the universal function $F(x)$ (red), validating the dynamical scaling near criticality.

\paragraph{Figure 4.}
Regularization rounds the divergence of the generalization error.  
Simulations are performed for a regularized network with $N = 10{,}000$, $r = 0.1$, $\gamma = 1$, and additive noise variance $D = 1$.  
Each data point averages 50 runs.  

\textbf{(a)}~Steady-state generalization error $\varepsilon_{\text{g}}(t \to \infty)$ as a function of $\alpha$ for several regularization strengths $\lambda = 0.002, 0.003, 0.004$.  
Markers denote simulations; solid lines are exact theoretical values from  
\begin{equation}
\varepsilon_{\mathrm{gen}}(t \to \infty) =
\frac{D + r \lambda^2 \chi_\lambda^2}{1 - \alpha(\lambda \chi_\lambda - 1)^2},
\label{eq:smreguexact}
\end{equation}
where $\chi_\lambda = \int_0^\infty R(\tau)\,d\tau$ is the regularized susceptibility.  
Finite $\lambda$ smooths the divergence at $\alpha_c = 1$, consistent with the theoretical scaling form.

\textbf{(b)}~Peak generalization error $\varepsilon_{\text{gmax}}$ versus $\lambda$, showing the scaling $\varepsilon_{\text{gmax}} \propto \lambda^{-1/2}$.  
Inset: critical-point shift $\Delta\alpha_c = 1 - \alpha_c$, increasing linearly with $\lambda$.  
Error bars in (b) and the inset represent standard deviations of the mean estimated via bootstrap resampling ($B = 1,000$).  
The bootstrap procedure resamples $n=50$ observations with replacement for each $(\alpha, \lambda)$ pair, fits cubic splines to the resampled mean curves, and extracts the maxima $\bar{\varepsilon}_{\max}^{*(b)}$ and their locations $\alpha_{\max}^{*(b)}$.  
The standard deviations  
\begin{equation}
\sigma_{\alpha_{\max}} = \mathrm{std}(\alpha_{\max}^{*(b)}), \quad
\sigma_{\varepsilon_{\max}} = \mathrm{std}(\bar{\varepsilon}_{\max}^{*(b)})
\end{equation}
are reported as error bars.  
The parameters scanned are  
$\alpha \in [0.95, 1.05]$ (step $0.01$) and $\lambda \in [0.002, 0.01]$ (step $0.001$).

\textit{Bootstrap procedure.}  
To quantify the statistical uncertainty in the peak height $\varepsilon_{\text{gmax}}$ and the critical shift $\Delta \alpha_c$, we use nonparametric bootstrap resampling:  \\
(1)~For each $\alpha$ and $\lambda$, begin with the original dataset containing $n = 50$ independent samples of the generalization error.  \\
(2)~Generate $B = 1,000$ bootstrap replicates by randomly sampling $n$ data points \emph{with replacement} from the original set. \\ 
(3)~For each bootstrap replicate $b$, compute the mean curve $\bar{\varepsilon}^{*(b)}(\alpha)$, fit it with a cubic spline, and extract the peak position $\alpha_{\max}^{*(b)}$ and the corresponding maximum value $\bar{\varepsilon}_{\max}^{*(b)}$.  \\
(4)~Repeat this procedure $B$ times to obtain the bootstrap ensembles  
$\{\alpha_{\max}^{*(1)}, \ldots, \alpha_{\max}^{*(B)}\}$ and  
$\{\bar{\varepsilon}_{\max}^{*(1)}, \ldots, \bar{\varepsilon}_{\max}^{*(B)}\}$.  \\
(5)~While the standard deviations of these ensembles quantify the variability within each ensemble, our goal is to estimate the uncertainty of the mean itself, which determines the accuracy of the inferred linear relation:
\begin{equation}
\sigma_{\alpha_{\max}} = \frac{\mathrm{std}(\alpha_{\max}^{*(b)})}{\sqrt{B}}, 
\qquad
\sigma_{\varepsilon_{\max}} = \frac{\mathrm{std}(\bar{\varepsilon}_{\max}^{*(b)})}{\sqrt{B}},
\end{equation}
which correspond to the error bars plotted in panel~(b) and its inset.  
This nonparametric approach is robust and particularly useful when the underlying distribution of errors is unknown or when analytic expressions for the standard error of the mean are not available.

\section{Dynamical Observables in the Frequency Domain}
\label{sec:RF}
Given the derivations in the previous sections, we arrive at the effective mean-field dynamics for a representative weight $\hat{\beta}(t)$:
\begin{equation}
\begin{aligned}
\partial_t{\hat{\beta}}(t) &= -\frac{\partial H}{\partial \hat{\beta}(t)} + \eta(t) \\
&= -\frac{ \gamma }{\alpha} \int dt^\prime \left(\gamma R + \delta \right)^{-1}(t, t^\prime)\left(\hat{\beta}(t^\prime) - \beta \right) - \lambda \hat{\beta}(t) + \eta(t),
\end{aligned}
\end{equation}
where $\eta(t)$ is an effective Gaussian noise with zero mean and covariance
\begin{equation}
\langle \eta(t)\eta(t') \rangle = \hat{D}(t,t') = \frac{\gamma^2}{\alpha} \left(\gamma R + \delta \right)^{-1} * (D + C) * \left(\gamma R^\top + \delta \right)^{-1}(t, t').
\end{equation}

The dynamical observables: the correlation function $C(t, t')$ and the linear response function $R(t, t')$ satisfy the following self-consistent equations:

\paragraph{Response function dynamics:}
\begin{equation}
\begin{aligned}
\frac{\partial R(t_1, t_2)}{\partial t_1}
&= -\frac{ \gamma }{\alpha} \int dt^\prime \left(\gamma R + \delta \right)^{-1}(t_1, t^\prime) R(t^\prime, t_2) - \lambda R(t_1, t_2) + \delta(t_1 - t_2),
\end{aligned}
\end{equation}

\paragraph{Correlation function dynamics:}
\begin{equation}
\begin{aligned}
\frac{\partial \hat{C}(t_1, t_2)}{\partial t_1} 
&= -\frac{\gamma}{\alpha} \int dt'\, (\gamma R + \delta)^{-1}(t_1, t') \left[ \hat{C}(t', t_2) - Q_5(t_2) \right] 
+ \langle \eta(t_1)\, \hat{\beta}(t_2) \rangle - \lambda\, \hat{C}(t_1, t_2) \\
&= -\frac{\gamma}{\alpha} \int dt'\, (\gamma R + \delta)^{-1}(t_1, t') \left[ \hat{C}(t', t_2) - Q_5(t_2) \right] \\
&\quad + \int dt''\, \hat{D}(t_1, t'') R(t_2, t'') - \lambda\, \hat{C}(t_1, t_2).
\label{eq:covariance}
\end{aligned}
\end{equation}
\begin{equation}
\begin{aligned}
C(t_1, t_2) 
&= \hat{C}(t_1, t_2) - Q_5(t_1) - Q_5(t_2) + r \\
&= \hat{C}(t_1, t_2) - \beta \int_0^{t_1} dt'\, b(t')\, R(t_1, t') - \beta \int_0^{t_2} dt'\, b(t')\, R(t_2, t') + r.
\label{eq:correlation}
\end{aligned}
\end{equation}
To analyze the long-time behavior of dynamical observables, it is often useful to work in the frequency domain. Two standard tools for this purpose are the Fourier and Laplace transforms. The Fourier transform is particularly well-suited for systems with time-translational invariance and observables defined over the entire real line. It is defined as:
\begin{equation}
\begin{aligned}
\tilde{f}(\omega) &= \int_{-\infty}^{\infty} f(t)\, e^{-i \omega t} \, dt, \\
f(t) &= \frac{1}{2 \pi} \int_{-\infty}^{\infty} \tilde{f}(\omega)\, e^{i \omega t} \, d\omega.
\label{eq:Fourier}
\end{aligned}
\end{equation}

In contrast, the Laplace transform is more appropriate for causal systems defined only for $t \geq 0$, as it naturally incorporates initial conditions and ensures convergence for a broader class of functions. It is defined as:
\begin{equation}
\begin{aligned}
\tilde{f}(s) &= \int_0^{\infty} f(t)\, e^{-s t} \, dt, \\
f(t) &= \frac{1}{2\pi i} \int_{\hat{\gamma} - i\infty}^{\hat{\gamma} + i\infty} \tilde{f}(s)\, e^{s t} \, ds,
\label{eq:Laplace}
\end{aligned}
\end{equation}
where $\hat{\gamma}$ is a real constant chosen so that the integration contour lies to the right of all singularities of $\tilde{f}(s)$ in the complex $s$-plane.

In the analysis that follows, we employ the Fourier transform to investigate the stationary behavior of the response function $R(t_1, t_2)$, particularly in regimes where time-translational invariance holds.  In contrast, the Laplace transform is particularly suited for initial-value problems and for situations where causality and transient dynamics are essential. For instance, in analyzing the correlation function $C(t_1,t_2)$, the Laplace framework naturally incorporates initial conditions while maintaining analytic control over the causal structure.

\subsection{Response Function and Susceptibility}
 In the steady state, the time-translational invariance holds for the observables, and we consider $\tau = t_1-t_2$ as the argument for the response function first. The self-consistent equation for the response function can be rewritten as:
\begin{equation}
\begin{aligned}
\frac{\partial R(\tau)}{\partial \tau}
&= -\frac{ \gamma }{\alpha} \int_{-\infty}^\infty du \left(\gamma R + \delta \right)^{-1}(\tau-u) R(u) - \lambda R(\tau) + \delta(\tau).
\label{eq:response}
\end{aligned}
\end{equation}
Next, we perform a Fourier transform on the ordinary differential equation introduced above:
\begin{equation}
\begin{aligned}
&i \omega \tilde{R}(\omega) = -\frac{\gamma}{\alpha} \frac{\tilde{R}(\omega)}{\gamma \tilde{R}(\omega) + 1} -\lambda \tilde{R}(\omega)+ 1, \\
&\left(i \omega+\lambda\right) \alpha \gamma \, \tilde{R}^2(\omega) + \left((i \omega+\lambda) \alpha + \gamma - \gamma \alpha\right)\tilde{R}(\omega) - \alpha = 0, \\
&\tilde{R}(\omega) = \frac{ - (\left(i \omega+\lambda\right) \alpha + \gamma(1 - \alpha)) \pm \sqrt{(\left(i \omega+\lambda\right) \alpha + \gamma(1 - \alpha))^2 + 4\left(i \omega+\lambda\right) \alpha^2 \gamma} }{2 \left(i \omega+\lambda\right) \alpha \gamma}.
\end{aligned}
\label{eq:RF}
\end{equation}
In the unregularized case ($\lambda = 0$), the expression for the response function $\tilde{R}(\omega)$ is incomplete, as the contribution from $\omega = 0$ is omitted. This missing term will be accounted for later using the Kramers–Kronig relation in Section~\ref{sec:KK}. The appropriate sign in the solution is selected to ensure that $\tilde{R}(\omega)$ satisfies the causality condition. To determine the appropriate sign, we examine the low-frequency and weak-regularization limit $\omega \to 0$, $\lambda \to 0^+$, in which the square root term can be expanded perturbatively around the small parameter $4 \left(i \omega + \lambda\right) \alpha^2 \gamma$, yielding:
\begin{equation}
\tilde{R}(\omega \to 0) \approx \frac{ - (\left(i \omega + \lambda\right) \alpha + \gamma(1 - \alpha)) \pm \left[ (\left(i \omega + \lambda\right) \alpha + \gamma(1 - \alpha)) + \frac{2 \left(i \omega + \lambda\right)\alpha^2 \gamma}{\left(i \omega + \lambda\right) \alpha + \gamma(1 - \alpha)} \right] }{2 \left(i \omega + \lambda\right) \alpha \gamma},
\end{equation}
We use the identity $\sqrt{z^2} = |z| e^{i \arg(z)}$ to evaluate the square root, where $z = \left(i \omega + \lambda\right) \alpha + \gamma(1 - \alpha)$. For $\alpha < 1$, we have the real part $ \gamma(1 - \alpha) + \lambda\alpha > 0$, implying that $\arg(z) \in (-\pi/2, \pi/2)$ and hence $\sqrt{z^2} = z$. This implies
\begin{equation}
\sqrt{\left( \left(i \omega + \lambda\right) \alpha + \gamma(1 - \alpha)\right)^2} = \left(i \omega + \lambda\right) \alpha + \gamma(1 - \alpha),
\end{equation}
and choosing the \textit{positive sign} in the solution ensures that $\tilde{R}(\omega)$ remains analytic in the lower half-plane, thus satisfying the causality condition. This leads to the following low-frequency behavior for $\alpha < 1$:
\begin{equation}
\tilde{R}(\omega \to 0) \approx \frac{\alpha}{\left(i \omega + \lambda\right)  \alpha + \gamma(1 - \alpha)}.
\label{eq:smallw}
\end{equation}
In the large-$ \tau $ limit, corresponding to the small-$ \omega $ behavior in Eq.~\eqref{eq:smallw}, the response function decays \textit{exponentially} as
$$
R(\tau) = \theta(\tau) \cdot \alpha \cdot e^{-\left(\frac{\gamma(1 - \alpha)}{\alpha}+\lambda\right) \tau},
$$
where the decay rate is governed by both the distance from the critical point $\alpha_c = 1$ and the regularization strength $\lambda$. As $\alpha \to 1$ and $\lambda \to 0^+$, the decay becomes increasingly slow, reflecting the emergence of long-time correlations and the phenomenon of \textit{critical slowing down}.

Moreover, in the regime $\alpha > 1$, selecting the \textit{positive sign} in Equation~\ref{eq:RF} is essential to ensure that the response function remains causal in the time domain. To justify this choice, we examine the long time (low-frequency) behavior of the response function using the positive-sign solution. Taking the limit $\omega\to 0, \lambda\to 0$:
\begin{equation}
\begin{aligned}
\tilde{R}(\omega \to 0) &\approx - \frac{\alpha}{(i \omega + \lambda) \alpha + \gamma(1 - \alpha)} - \frac{1}{\gamma} + \frac{\alpha - 1}{(i \omega + \lambda) \alpha}, \\
&\approx \frac{\alpha - 1}{(i \omega + \lambda) \alpha},
\label{eq:exla}
\end{aligned}
\end{equation}
where only the leading-order divergent term has been retained in the simultaneous limits $\omega \to 0$ and $\lambda \to 0^+$. In the time domain, this behavior corresponds to a slowly decaying exponential response:
$$
R(\tau \to \infty) \sim \frac{\alpha - 1}{\alpha} e^{-\lambda \tau}.
$$
As $\lambda \to 0^+$, the decay becomes progressively slower, asymptotically approaching a nonzero plateau $R(\tau \to \infty) = \frac{\alpha - 1}{\alpha}$, which characterizes the extent of ergodicity breaking in the system. The regularization parameter $\lambda$ thus plays a dual role: it ensures analyticity and therefore causality of the response function in the complex $\omega$-plane, while also controlling the degree of long-term memory in the time domain. This analysis validates the choice of the positive sign as the physically consistent and causal solution in the overparameterized regime $\alpha > 1$. Notably, in the limit $\lambda = 0$, the response function develops a simple pole at $\omega = 0$, signaling a breakdown of causality. This singular behavior and its resolution are further examined in Section~\ref{sec:transition}.

In the marginal case $\alpha = 1$, the response function remains well-defined only when $\lambda \neq 0$. At $\lambda = 0$, the system reaches the critical point $\alpha_c = 1$, where the perturbative analysis breaks down and standard notions such as causality and ergodicity become ill-defined. The corresponding response function in Fourier space takes the form
\begin{equation}
\tilde{R}(\omega) = \frac{-\left(i \omega + \lambda\right) + \sqrt{\left(i \omega + \lambda\right)^2 + 4 \left(i \omega + \lambda\right) \gamma}}{2 \left(i \omega + \lambda\right) \gamma}.
\end{equation}

Based on the preceding analysis, we adopt the positive sign in Equation~\ref{eq:RF} to ensure that the response function remains consistent with causality and yields physically meaningful behavior across all regimes. The appropriate analytic branch of the square root is determined using the identity $\sqrt{z^2} = |z| e^{i \arg(z)}$, which guarantees continuity and analyticity across the complex plane for each $\alpha$ regime. A more detailed analysis of the associated \textit{pole structure} is presented in Section~\ref{sec:pole}, where we clarify how the pole structure impacts the system's dynamical properties.

Another key macroscopic observable is the \textit{susceptibility}, defined as
\begin{equation}
\chi = \frac{\delta \langle \hat{\beta}(\tau) \rangle}{\delta j} = \int_0^{\infty} R(\tau)\, d\tau,
\end{equation}
which quantifies the total integrated response of the system to a constant (step-like) perturbation $j$. When regularization is present ($\lambda \neq 0$), this quantity can be computed directly from the zero-frequency limit of the response function:
\begin{equation}
\chi = \tilde{R}(\omega \to 0) = \frac{ - (\lambda\alpha + \gamma(1 - \alpha)) + \sqrt{(\lambda \alpha + \gamma(1 - \alpha))^2 + 4\lambda\alpha^2 \gamma} }{2 \lambda \alpha \gamma}.
\end{equation}
In this regularized setting, $\chi$ remains analytic for all values of $\alpha$, and no phase transition is observed, analogous to the rounding of singularities in finite-size systems near criticality.

However, in the unregularized case ($\lambda = 0$), the susceptibility is well-defined only when $\alpha < 1$, and takes the form
\begin{equation}
\chi_{\alpha<1} = \tilde{R}_{\alpha<1}(\omega \to 0) = \frac{\alpha}{\gamma(1 - \alpha)}.
\end{equation}
This expression diverges as $\alpha \to 1$, signaling the presence of a \textit{critical point} at the interpolation threshold. The divergence reflects the emergence of long-range temporal correlations and heightened sensitivity to perturbations near criticality.

It is important to emphasize that the susceptibility becomes \textit{ill-defined} for $\alpha \geq 1$ in the absence of regularization, as the system enters a non-equilibrium regime where stationary linear response theory no longer applies. A more detailed treatment of this regime will be provided in Section~\ref{sec:transition}.

\subsection{Correlation Function}
\label{sec:solveC}
Based on the dynamics of the covariance function $\hat{C}(t_1, t_2)$ in Equation~\ref{eq:covariance}, we can rewrite the dynamics by defining a new operator $M$:
\begin{equation}
\begin{aligned}
\left(\frac{\partial }{\partial t_1}+M \right)\hat{C}(t_1, t_2)
& = \frac{\gamma}{\alpha} \int d t^{\prime}(\gamma R+\delta)^{-1}\left(t_1, t^{\prime}\right)Q_5\left(t_2\right)\\&+ \int d t^{\prime \prime} \frac{\gamma^2}{\alpha}(\gamma R+\delta )^{-1}*\left(D+r-Q_5-Q_5^\top + \hat{C} \right)*(\gamma R^\top+\delta )^{-1}\left(t_1, t^{\prime \prime}\right) R\left(t_2, t^{\prime \prime}\right),\\
& = b(t_1)\int_0^{t_2} b(t^\prime) R(t_2, t^\prime)\\&+ \int d t^{\prime \prime} \frac{\gamma^2}{\alpha}(\gamma R+\delta )^{-1}*\left(D+r-Q_5-Q_5^\top + \hat{C} \right)*(\gamma R^\top+\delta )^{-1}\left(t_1, t^{\prime \prime}\right) R\left(t_2, t^{\prime \prime}\right),\\
M \hat{C}\left(t_1, t_2\right)&=\frac{\gamma}{\alpha} \int_0^t d t_0(\gamma R+\delta)^{-1}\left(t_1, t_0\right)\hat{C}\left(t_0, t_2\right)+\lambda \hat{C}\left(t_1, t_2\right).
\end{aligned}
\end{equation}
To get the general solution $\hat{C}_g(t_1, t_2)$ for the correlation function, we need to calculate the Green function, which is the inverse of the operator $\left(\partial t_1+M\right)$, satisfying:
\begin{equation}
\begin{aligned}
\left(\partial_t+M\right) G(t, s) & =\partial_t G(t, s)+\frac{\gamma}{\alpha} \int_0^t d t_0(\gamma R+\delta)^{-1}\left(t, t_0\right) G\left(t_0, s\right)+\lambda G(t, s)=\delta(t-s), \\
G(t, s) & =R(t, s),
\end{aligned}
\end{equation}
where the Green function is exactly the response function. Therefore, the general solution for the correlation function can be written as:
\begin{equation}
\begin{aligned}
\hat{C}_g(t_1, t_2)& =  \int d t^{\prime \prime}\int ds R(t_1, s)\left(\hat{D}(s,t^{\prime\prime})+b(s)b(t^{\prime \prime})\right) R\left(t_2, t^{\prime \prime}\right),\\
\end{aligned}
\end{equation}
where $\hat{D}=\frac{\gamma^2}{\alpha}(\gamma R+\delta )^{-1}*\left(D+r-Q_5-Q_5^\top + \hat{C} \right)*(\gamma R^\top+\delta )^{-1}\left(t_1, t^{\prime \prime}\right) R\left(t_2, t^{\prime \prime}\right)$ denotes the noise correlation kernel. Next, we need to add the special solutions $\hat{C}^\star(t_1, t_2)$ to the general solution, which yields the complete solution for the correlation function:
\begin{equation}
\left(\frac{\partial }{\partial t_1}+M \right)\hat{C}^\star(t_1, t_2) =\frac{\partial \hat{C}^\star(t_1, t_2) }{\partial t_1}+ \frac{\gamma}{\alpha} \int_0^t d t_0(\gamma R+\delta)^{-1}\left(t_1, t_0\right) \hat{C}^\star\left(t_0, t_2 \right)+\lambda \hat{C}^\star\left(t_1, t_2\right)=0,
\label{eq:special}
\end{equation}
where the initial condition of the response function is given by $\hat{C}(0,0) = \frac{1}{d}\sum_i \left(\hat{\beta}_i(0)\right)^2$. To find the solution of the this equation in the presence of causal initial conditions ($t \geq 0$), it is advantageous to work in the frequency domain. In particular, the Laplace transform is more appropriate than the Fourier transform, as it naturally accommodates initial values and ensures convergence for causal functions. We adopt the Laplace transform conventions in Equation~\ref{eq:Laplace}.

Applying the Laplace transform for $t_1$ to both sides of the Equation~\ref{eq:special}, we obtain the Laplace transform $\hat{C}_{\mathcal{L}}^\star$:
\begin{equation}
\begin{aligned}
s\, \hat{C}_{\mathcal{L}}^\star(s, t_2) - \hat{C}^\star(0, t_2) &= -\frac{\gamma}{\alpha} \int_0^{\infty} dt\, e^{-s t} \int_0^t dt_0\, \left[ \gamma R + \delta \right]^{-1}(t_1 - t_0)\, \hat{C}^\star(t_0, t_2)-\lambda \hat{C}_{\mathcal{L}}^\star(s,t_2) \\
&= -\frac{\gamma}{\alpha} \frac{\tilde{C}^\star(s, t')}{\gamma \tilde{R}(s) + 1}-\lambda \hat{C}_{\mathcal{L}}^\star(s,t_2),\\
\hat{C}_{\mathcal{L}}^\star(s, t_2)& = \frac{\hat{C}^\star(0, t_2) }{s+\frac{\gamma}{\alpha} \frac{1}{\gamma \tilde{R}(s) + 1}+\lambda}
\end{aligned}
\end{equation}
where in the second line we have used the convolution theorem and defined $\tilde{R}(s)$ as the Laplace transform of $R(t)$.  To further simplify this expression, we can perform the Laplace transform to the self-consistent equation for the response function in Equation~\ref{eq:response}, yielding:
\begin{equation}
\tilde{R}(s) = \frac{1}{s + \frac{\gamma}{\alpha} \cdot \frac{1}{\gamma \tilde{R}(s) + 1} + \lambda},
\label{eq:Rlaplace1}
\end{equation}
Using this relation, the Laplace transform of the correlation function satisfies
\begin{equation}
\hat{C}_{\mathcal{L}}^\star(s, t_2) = \hat{C}^\star(0, t_2)\, \tilde{R}(s),
\end{equation}
where $\hat{C}^\star(0, t_2)$ acts as the initial condition with respect to $t_1$. In particular, by setting $t_2 = 0$, we obtain
\begin{equation}
\hat{C}^\star(t_1, 0) = \hat{C}^\star(0, 0)\, R(t_1), \quad \text{for } t_1 > 0,
\end{equation}
where $\hat{C}^\star(0, 0)$ denotes the initial variance. 

Therefore, the special solution takes the factorized form
\begin{equation}
\hat{C}^\star(t_1, t_2) = \hat{C}(0, 0)\, R(t_1)\, R(t_2), \quad \text{for } t_1, t_2 > 0,
\label{eq:Cfactorized}
\end{equation}
 This expression reflects the fact that all temporal correlations are fully mediated by the response function, with the initial variance $\hat{C}(0, 0)$ setting the amplitude of fluctuations. The evolution of correlations at later times is entirely governed by how these initial fluctuations propagate forward via $R(t)$.

Taken together, using Equation~\ref{eq:correlation}, the solution for the correlation function now can be written as:
\begin{equation}
\begin{aligned}
C(t_1, t_2) &=\hat{C}(t_1, t_2)-Q_5(t_1)-Q_5(t_2)+r,\\&= \int d t^{\prime \prime}\int ds R(t_1, s)\left(\hat{D}(s,t^{\prime\prime})+b(s)b(t^{\prime \prime})\right) R\left(t_2, t^{\prime \prime}\right)\\&-Q_5(t_1)-Q_5(t_2) + \hat{C}(0, 0)\, R(t_1)\, R(t_2),\quad \text{for } t_1, t_2 > 0,
\end{aligned}
\end{equation}
where $Q_5\left(t_1\right)=\beta \int_0^{t_1} d t_2 b\left(t_2\right) R\left(t_1, t_2\right)$ and $b(t)=\beta \frac{\gamma}{\alpha} \int_0^t d t^{\prime}(\gamma R+\delta)^{-1}\left(t, t^{\prime}\right)$.
Next, we introduce two independent Laplace variables, $s_1$ and $s_2$, corresponding to the time variables $t_1$ and $t_2$, respectively. Applying the Laplace transform to both sides of the two-time correlation equation yields
\begin{equation}
\begin{aligned}
\tilde{C}(s_1, s_2) &= \int_0^\infty \int_0^\infty dt_1\, dt_2\, e^{-s_1 t_1 - s_2 t_2} C(t_1, t_2) \\
&= \frac{\gamma^2}{\alpha} \frac{\tilde{R}(s_1)}{\gamma \tilde{R}(s_1) + 1} \left( \frac{D}{s_1 s_2} + \tilde{C}(s_1, s_2) \right) \frac{\tilde{R}(s_2)}{\gamma \tilde{R}(s_2) + 1} +\frac{r}{s_1s_2}+ r \frac{\gamma^2}{\alpha^2} \frac{1}{s s^{\prime}} \frac{1}{\gamma \tilde{R}(s)+1} \frac{1}{\gamma \tilde{R}\left(s^{\prime}\right)+1} \tilde{R}(s) \tilde{R}\left(s^{\prime}\right)\\&+ \hat{C}(0, 0)\, \tilde{R}(s_1)\, \tilde{R}(s_2)-r\frac{\gamma}{\alpha}\frac{1}{s_1} \frac{\tilde{R}(s_1)}{\gamma\tilde{R}(s_1)+1}-r\frac{\gamma}{\alpha}\frac{1}{s_2} \frac{\tilde{R}(s_2)}{\gamma\tilde{R}(s_2)+1},\\
&= \frac{\gamma^2}{\alpha} \frac{\tilde{R}(s_1)}{\gamma \tilde{R}(s_1) + 1} \left( \frac{D}{s_1 s_2} +\frac{\frac{r}{\alpha}}{s_1 s_2}+ \tilde{C}(s_1, s_2) \right) \frac{\tilde{R}(s_2)}{\gamma \tilde{R}(s_2) + 1} +\frac{r}{s_1s_2}\\&+ \hat{C}(0, 0)\, \tilde{R}(s_1)\, \tilde{R}(s_2)-r\frac{\gamma}{\alpha}\frac{1}{s_1s_2} \frac{\tilde{R}(s_1)}{\gamma\tilde{R}(s_1)+1}-r\frac{\gamma}{\alpha}\frac{1}{s_1s_2} \frac{\tilde{R}(s_2)}{\gamma\tilde{R}(s_2)+1},\\
\end{aligned}
\end{equation}
which can be rearranged into a closed-form expression:
\begin{equation}
\begin{aligned}
&\tilde{C}(s_1, s_2)=\\ & \frac{ \displaystyle  \frac{\gamma^2}{\alpha} \frac{\tilde{R}(s_1)}{\gamma \tilde{R}(s_1) + 1} \left( \frac{D}{s_1 s_2} +\frac{\frac{r}{\alpha}}{s_1 s_2}\right) \frac{\tilde{R}(s_2)}{\gamma \tilde{R}(s_2) + 1} +\frac{r}{s_1s_2}+ \hat{C}(0, 0)\, \tilde{R}(s_1)\, \tilde{R}(s_2)-r\frac{\gamma}{\alpha}\frac{1}{s_1s_2} \frac{\tilde{R}(s_1 )}{\gamma\tilde{R}(s_1 )+1}-r\frac{\gamma}{\alpha}\frac{1}{s_1 s_2} \frac{\tilde{R}(s_2)}{\gamma\tilde{R}(s_2)+1}}{\displaystyle 1 - \frac{\gamma^2}{\alpha} \frac{\tilde{R}(s_1)}{\gamma \tilde{R}(s_1) + 1} \cdot \frac{\tilde{R}(s_1s_2)}{\gamma \tilde{R}(s_2) + 1}}.
\end{aligned}
\end{equation}

To simplify this expression further, we use the self-consistent relation for the Laplace-transformed response function from Eq.~\ref{eq:Rlaplace1}:
\begin{equation}
\frac{1}{s}\frac{\gamma}{\alpha} \cdot \frac{\tilde{R}(s)}{\gamma \tilde{R}(s) + 1} =\frac{1}{s}\left( 1 - (s + \lambda)\, \tilde{R}(s)\right).
\end{equation}
Substituting this into the denominator of the expression for $\tilde{C}(s_1, s_2)$, which yields a compact form:
\begin{equation}
\begin{aligned}
\tilde{C}(s_1, s_2) &= \frac{ \displaystyle  \frac{\gamma^2}{\alpha} \frac{\tilde{R}(s_1)}{\gamma \tilde{R}(s_1) + 1} \left( \frac{D}{s_1 s_2} +\frac{\frac{r}{\alpha}}{s_1 s_2}\right) \frac{\tilde{R}(s_2)}{\gamma \tilde{R}(s_2) + 1} + \hat{C}(0, 0)\, \tilde{R}(s_1)\, \tilde{R}(s_2)-r\frac{\left( 1 - (s_2 + \lambda)\, \tilde{R}(s_2)\right)}{s_1s_2} +r\frac{\left(  (s_1 + \lambda)\, \tilde{R}(s_1)\right)}{s_1 s_2} }{\displaystyle 1 - \frac{\gamma^2}{\alpha} \frac{\tilde{R}(s_1)}{\gamma \tilde{R}(s_1) + 1} \cdot \frac{\tilde{R}(s_1s_2)}{\gamma \tilde{R}(s_2) + 1}},\\
& = -\left( \frac{D}{s_1 s_2} +\frac{\frac{r}{\alpha}}{s_1 s_2}\right) + \frac{ \displaystyle  \frac{D}{s_1 s_2} +\frac{\frac{r}{\alpha}-r}{s_1 s_2} + \hat{C}(0, 0)\, \tilde{R}(s_1)\, \tilde{R}(s_2)+r\frac{\left( (s_2 + \lambda)\, \tilde{R}(s_2)\right)}{s_1s_2} +r\frac{\left( (s_1 + \lambda)\, \tilde{R}(s_1)\right)}{s_1 s_2} }{\displaystyle 1-\alpha+\alpha\left((s_1+\lambda) \tilde{R}(s_1)+\left(s_2+\lambda\right) \tilde{R}\left(s_2\right)-(s_1+\lambda)\left(s_2+\lambda\right) \tilde{R}(s_1) \tilde{R}\left(s_2\right)\right)},\\
& = - \frac{D}{s_1 s_2} + \frac{ \displaystyle  \frac{D}{s_1 s_2} +\hat{C}(0, 0)\, \tilde{R}(s_1)\, \tilde{R}(s_2)+\frac{r(s_1+\lambda)\left(s_2+\lambda\right)}{s_1s_2} \tilde{R}(s_1) \tilde{R}\left(s_2\right) }{\displaystyle 1-\alpha+\alpha\left((s_1+\lambda) \tilde{R}(s_1)+\left(s_2+\lambda\right) \tilde{R}\left(s_2\right)-(s_1+\lambda)\left(s_2+\lambda\right) \tilde{R}(s_1) \tilde{R}\left(s_2\right)\right)}.\\
\end{aligned}
\end{equation}

\subsection{Generalization error and Training error}
\paragraph{Generalization Error.}
As defined in Section~\ref{sec:generalization}, the generalization error in the time domain is given by
\begin{equation}
\varepsilon_{g}(t) = C(t, t) + D,
\end{equation}
where $C(t, t)$ denotes the two-time correlation function evaluated at equal times and $D$ is the intrinsic label noise variance. For analytical convenience, we consider a more general form:
\begin{equation}
\varepsilon_{g}(t) = \lim_{t_1 \to t,\, t_2 \to t} \varepsilon_{g}(t_1, t_2), \quad \text{with} \quad \varepsilon_{g}(t_1, t_2) = C(t_1, t_2) + D.
\end{equation}
Taking the Laplace transform with respect to both time variables $t_1$ and $t_2$, we obtain
\begin{equation}
\begin{aligned}
\mathcal{L}\{\varepsilon_{g}(t_1, t_2)\}(s_1, s_2) &= \tilde{C}(s_1, s_2) + \frac{D}{s_1 s_2} \\
&=  \frac{ \displaystyle  \frac{D}{s_1 s_2} +\hat{C}(0, 0)\, \tilde{R}(s_1)\, \tilde{R}(s_2)+\frac{r(s_1+\lambda)\left(s_2+\lambda\right)}{s_1s_2} \tilde{R}(s_1) \tilde{R}\left(s_2\right) }{\displaystyle 1-\alpha+\alpha\left((s_1+\lambda) \tilde{R}(s_1)+\left(s_2+\lambda\right) \tilde{R}\left(s_2\right)-(s_1+\lambda)\left(s_2+\lambda\right) \tilde{R}(s_1) \tilde{R}\left(s_2\right)\right)}.
\end{aligned}
\end{equation}

\paragraph{Training Error.}
As defined in Section~\ref{sec:training}, the training error can be expressed in terms of the two-time kernel convolution structure:
\begin{equation}
\begin{aligned}
\varepsilon_{\text{train}}(t) &= \lim_{t_1 \to t,\, t_2 \to t} \varepsilon_{\text{train}}(t_1, t_2), \\
\varepsilon_{\text{train}}(t_1, t_2) &= \left[(\gamma R + \delta)^{-1} * (D + C) * (\gamma R^\top + \delta)^{-1} \right](t_1, t_2),
\end{aligned}
\end{equation}
where the asterisks denote temporal convolution and $R^\top$ represents time-reversed response.

To facilitate analysis, we take the Laplace transform of both time variables $t_1$ and $t_2$. Using the convolution theorem, we obtain the transformed expression:
\begin{equation}
\begin{aligned}
\varepsilon_{\text{train}}(s_1, s_2) &= \frac{1}{\gamma \tilde{R}(s_1) + 1} \left( \frac{D}{s_1 s_2} + \tilde{C}(s_1, s_2) \right) \frac{1}{\gamma \tilde{R}(s_2) + 1},\\
&= \frac{1}{\gamma \tilde{R}(s_1) + 1} \varepsilon_{g}(s_1, s_2)  \frac{1}{\gamma \tilde{R}(s_2) + 1}.\\
\label{eq:trainL}
\end{aligned}
\end{equation}

The time-dependent training and generalization errors can be recovered by taking the inverse Laplace transforms of $\varepsilon_{\text{train}}(s_1, s_2)$ and $\varepsilon_{g}(s_1, s_2)$, respectively, followed by the coincident-time limit $t_1, t_2 \to t$.

\section{Scaling and Data Collapse of the Generalization Error}

In this section, we analyze the scaling behavior and data collapse properties of key macroscopic observables—specifically, the response function and the generalization error. Our focus lies in the asymptotic regime of long times ($t \to \infty$) and near the interpolation threshold ($\alpha \to 1$), where critical behavior emerges. The analysis distinguishes between the unregularized case ($\lambda = 0$) and the regularized case ($\lambda \neq 0$), where the regularization strength $\lambda$ acts as a relevant variable that smooths the singularity at the critical point $\alpha = 1$. We derive explicit scaling forms in both regimes, showing how $\lambda$ controls the rounding of the transition and demonstrating that the divergence of the generalization error at criticality is governed by the underlying susceptibility.
\subsection{Regularized Dynamics: $\lambda\neq 0$}
\label{sec:regularized}
For analytical convenience and consistency, we adopt the Laplace transform throughout this section to investigate the scaling behavior of the response function and the generalization error, as well as the asymptotic long-time behavior of the generalization error. In the presence of regularization ($\lambda \neq 0$), the dynamics are regularized, and all key observables admit closed-form expressions in the Laplace frequency domain.

The Laplace-transformed response function $\tilde{R}(s)$ is given by:
\begin{equation}
\tilde{R}(s) = \frac{ -\left[\alpha(s + \lambda) + \gamma(1 - \alpha)\right] + \sqrt{\left[\alpha(s + \lambda) + \gamma(1 - \alpha)\right]^2 + 4\alpha^2 \gamma(s + \lambda)} }{ 2\alpha \gamma(s + \lambda) }.
\label{eq:Rlaplace2}
\end{equation}
Since $\lambda \neq 0$, this expression is well-defined at $s = 0$, and no additional contribution is ignored.

The Laplace-transformed two-time correlation function $\tilde{C}(s_1, s_2)$ takes the form:
\begin{equation}
\tilde{C}(s_1, s_2) = -\frac{D}{s_1 s_2} +  \frac{ \displaystyle  \frac{D}{s_1 s_2} +\hat{C}(0, 0)\, \tilde{R}(s_1)\, \tilde{R}(s_2)+\frac{r(s_1+\lambda)\left(s_2+\lambda\right)}{s_1s_2} \tilde{R}(s_1) \tilde{R}\left(s_2\right) }{\displaystyle 1-\alpha+\alpha\left((s_1+\lambda) \tilde{R}(s_1)+\left(s_2+\lambda\right) \tilde{R}\left(s_2\right)-(s_1+\lambda)\left(s_2+\lambda\right) \tilde{R}(s_1) \tilde{R}\left(s_2\right)\right)}.
\end{equation}

From this, the Laplace-domain generalization error $\varepsilon_{g}(s_1, s_2)$ is:
\begin{equation}
\varepsilon_{g}(s_1, s_2) =  \frac{ \displaystyle  \frac{D}{s_1 s_2} +\hat{C}(0, 0)\, \tilde{R}(s_1)\, \tilde{R}(s_2)+\frac{r(s_1+\lambda)\left(s_2+\lambda\right)}{s_1s_2} \tilde{R}(s_1) \tilde{R}\left(s_2\right) }{\displaystyle 1-\alpha+\alpha\left((s_1+\lambda) \tilde{R}(s_1)+\left(s_2+\lambda\right) \tilde{R}\left(s_2\right)-(s_1+\lambda)\left(s_2+\lambda\right) \tilde{R}(s_1) \tilde{R}\left(s_2\right)\right)}.
\label{eq:gerrorlaplace}
\end{equation}

Finally, the Laplace-domain training error $\varepsilon_{\text{train}}(s_1, s_2)$ is given by:
\begin{equation}
\begin{aligned}
\varepsilon_{\text{train}}(s_1, s_2) &= \frac{1}{\gamma \tilde{R}(s_1) + 1} \left( \frac{D}{s_1 s_2} + \tilde{C}(s_1, s_2) \right) \frac{1}{\gamma \tilde{R}(s_2) + 1} \\
&= \frac{1}{\gamma \tilde{R}(s_1) + 1}\, \varepsilon_{g}(s_1, s_2)\, \frac{1}{\gamma \tilde{R}(s_2) + 1}.
\end{aligned}
\label{eq:terrorlaplace}
\end{equation}
\subsubsection{Data Collapse for the Response Function}
\label{sec:unResponse}
To analyze the scaling behavior of the response function near criticality, we consider the long-time limit ($t \to \infty$), which corresponds to the small-$s$ regime in Laplace space. Simultaneously, we approach the interpolation threshold by letting $\alpha \to 1$, and define $\epsilon = |1 - \alpha|$ to quantify the distance to criticality. Finally, we take the small regularization parameter $\lambda$, which reflects common practice in machine learning settings where regularization is weak or absent.

In this regime, we retain only the leading contributions in the small parameters $s$, while neglecting all higher-order terms such as $\mathcal{O}(s^2)$. This approximation isolates the dominant behavior of the Laplace-transformed response function $\tilde{R}(s)$ near criticality and in the long-time limit.

\begin{equation}
\begin{aligned}
\tilde{R}(s) &\approx -\frac{1}{2\gamma} - \frac{1-\alpha}{2\alpha (s + \lambda)} + \frac{1}{2 \alpha \gamma(s + \lambda)} \sqrt{\alpha^2(\lambda^2 + 2s\lambda) + \gamma^2 (1-\alpha)^2 + 2\gamma\alpha(s + \lambda) + 2\alpha^2 \gamma (s + \lambda)}.
\end{aligned}
\end{equation}

We now group the terms in the square root into a linearized form:
\begin{equation}
\begin{aligned}
\tilde{R}(s) &\approx -\frac{1}{2\gamma} - \frac{1-\alpha}{2\alpha (s + \lambda)} + \frac{1}{2 \alpha \gamma(s + D)} \sqrt{A s + B},
\end{aligned}
\end{equation}
where we define:
\begin{equation}
\begin{aligned}
A &= 2\lambda\alpha^2 + 2\gamma\alpha + 2\alpha^2 \gamma, \\
B &= \gamma^2 (1-\alpha)^2 + \lambda\left[\alpha^2 \lambda + 2\gamma\alpha + 2\alpha^2 \gamma\right], \\
D &= \lambda.
\end{aligned}
\end{equation}

To simplify further, we observe that the expression inside the square root can be written as:
\begin{equation}
A s + B = A(s + D) + (B - A D),
\end{equation}
with
\begin{equation}
B - A D = \gamma^2 (1-\alpha)^2 - \lambda^2\alpha^2.
\end{equation}

This structure reveals that the square-root term depends on the combination $s + \lambda$, indicating that the regularization parameter $\lambda$ acts as a shift in the Laplace frequency domain. As a result, $\lambda$ effectively delays or slows down the dynamics by setting a characteristic time scale $\tau \sim 1/\lambda$, below which temporal correlations are suppressed. This highlights the role of regularization as a relevant parameter that modifies the temporal structure of fluctuations near criticality. We also note:
\begin{equation}
\frac{B}{A} = \frac{B - A D}{A} + \lambda.
\end{equation}
Based on the known inverse Laplace transformations:
\begin{equation}
\begin{aligned}
\mathcal{L}^{-1}\left[\frac{1}{s+D}\right]& = \exp{\left(-Dt\right)},\\
\mathcal{L}^{-1}\left[\frac{\sqrt{As+B}}{s+D}\right]& =\frac{e^{-D t}\left(\sqrt{A} e^{\left(-\frac{B}{A}+D\right) t} \sqrt{(B-A D) t}+(B-A D) \sqrt{\pi} \operatorname{tErf}\left[\frac{\sqrt{(B-A D) t}}{\sqrt{A}}\right]\right)}{\sqrt{B-A D} \sqrt{\pi} t}.\\
\end{aligned}
\end{equation}
The scaling form for the response function in time domain can be written as:
\begin{equation}
\begin{aligned}
R(\tau>0)& = \mathcal{L}^{-1}\left[\tilde{R}(s) \right]=  -\frac{(1-\alpha) e^{-\lambda \tau}}{2\alpha}+ e^{-\lambda \tau} \frac{\left(\sqrt{A} e^{\left(-\frac{B}{A} + D \right) \tau} \sqrt{(B-AD) \tau}+(B-AD) \sqrt{\pi} \tau \operatorname{Erf}\left[\frac{\sqrt{(B-AD) \tau}}{\sqrt{A}}\right]\right)}{2\alpha\gamma\sqrt{B-AD} \sqrt{\pi} \tau},\\
& = e^{-\lambda \tau} \left(-\frac{(1-\alpha) }{2\alpha}+  \frac{\left(\sqrt{A} e^{\left(-\frac{B}{A} + D \right) \tau} \sqrt{(B-AD) \tau}+(B-AD) \sqrt{\pi} \tau \operatorname{Erf}\left[\frac{\sqrt{(B-AD) \tau}}{\sqrt{A}}\right]\right)}{2\alpha\gamma\sqrt{B-AD} \sqrt{\pi} \tau}\right)
\end{aligned}
\end{equation}
where 
$B - A D = \gamma^2 (1-\alpha)^2 - \lambda^2\alpha^2$ 
and 
$A = 2\lambda\alpha^2 + 2\gamma\alpha + 2\alpha^2 \gamma$. 
To simplify the expression near criticality, we expand the square-root term
$$
\sqrt{B-AD} 
= \gamma\epsilon \sqrt{1 - \frac{\lambda^2(1-\epsilon)^2}{\gamma^2\epsilon^2}},
$$
with $\epsilon = |1-\alpha|$. In the limit $\lambda \to 0$, the ratio $\lambda^2/(\gamma^2\epsilon^2)$ vanishes, and we may neglect subleading contributions. This gives
\begin{equation}
\sqrt{B-AD} \;\approx\; \gamma \epsilon.
\end{equation}
Furthermore, in the same regime we may approximate
$A \approx 2\lambda + 4\gamma$. 
Substituting these expressions into the response function yields the scaling form
\begin{equation}
R(\tau>0) \;\approx\; e^{-\lambda \tau}\,\epsilon\, 
F_R\!\left(\frac{\gamma \epsilon \sqrt{\tau}}{\sqrt{\,2\lambda+4\gamma\,}}\right),
\end{equation}
where the scaling function is given explicitly by
\begin{equation}
F_R(x) = \frac{1}{2}\left( \frac{e^{-x^2}}{\sqrt{\pi}\,x} 
+ \operatorname{Erf}(x) 
- \operatorname{Sign}(1-\alpha) \right).
\label{eq:Rscaling}
\end{equation}

\begin{figure}[htbp] 
	\centering
	\includegraphics[scale = 0.5]{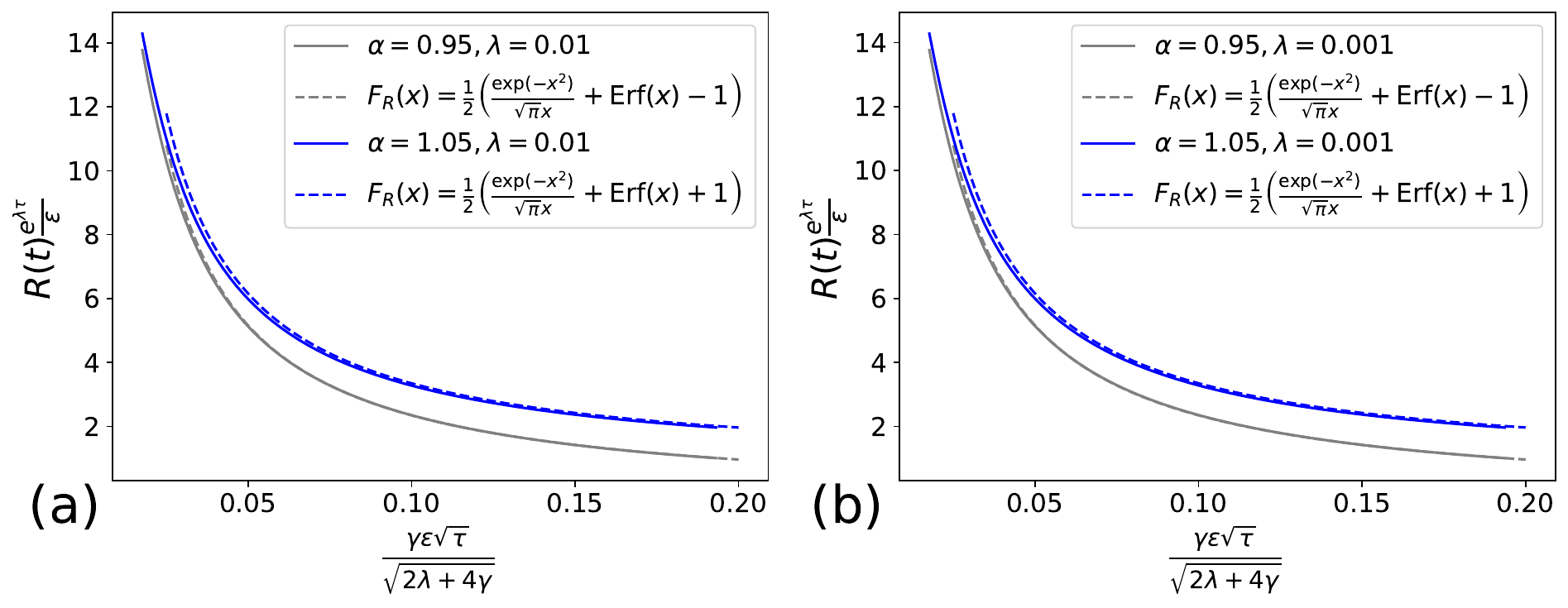}
	\caption{ Data collapse of the response function, demonstrating agreement between the scaling formula (dashed lines) and dynamical mean-field theory (DMFT) simulations (solid lines). 
(a) Results for $\lambda = 0.01$. 
(b) Results for $\lambda = 0.001$.}
	\label{fig3}
\end{figure}

The scaling form of the response function exhibits exponential decay in time, with the decay rate governed by the regularization strength $\lambda$. This behavior highlights the role of $\lambda$ as a temporal cutoff: stronger regularization ($\lambda \gg 0$) suppresses long-time memory by accelerating the decay of the response, whereas weaker regularization $(\lambda \to 0$) allows persistent memory effects and critical slowing down near the transition point. The scaling formula corresponds well with the dynamical mean-field theory (DMFT) simulations as shown in Figure~\ref{fig3}.

\subsubsection{Data Collapse for the Generalization error}
\label{sec:reguscaling}
Given the Laplace-domain expression for the extended generalization error in Eq.~\eqref{eq:gerrorlaplace}, we now analyze its scaling form in the asymptotic regime: small Laplace variables $s_1, s_2 \to 0$, small deviation from criticality $\epsilon = |1 - \alpha| \ll 1$, and weak regularization $\lambda \to 0$. Under these limits, we retain only the leading contributions in both the numerator and denominator of the extended generalization error:
\begin{equation}
\begin{aligned}
\varepsilon_{g}(s_1, s_2) 
&= \frac{ \displaystyle \frac{D}{s_1 s_2} + C(0,0)\, \tilde{R}(s_1)\, \tilde{R}(s_2) + \frac{r (s_1+\lambda)(s_2+\lambda)}{s_1 s_2}\, \tilde{R}(s_1)\, \tilde{R}(s_2) }
{ \displaystyle 1 - \alpha + \alpha \left[ (s_1 + \lambda)\tilde{R}(s_1) + (s_2 + \lambda)\tilde{R}(s_2) - (s_1 + \lambda)(s_2 + \lambda)\tilde{R}(s_1)\tilde{R}(s_2) \right] } \\
&\approx \frac{ \displaystyle \frac{D}{s_1 s_2} }{ \displaystyle 1 - \alpha + \alpha \left[ (s_1 + \lambda)\tilde{R}(s_1) + (s_2 + \lambda)\tilde{R}(s_2) \right] },
\end{aligned}
\end{equation}
where we have dropped subleading contributions involving $\tilde{R}(s_1) \tilde{R}(s_2)$ and $\frac{r (s_1+\lambda)(s_2+\lambda)}{s_1 s_2}\, \tilde{R}(s_1)\, \tilde{R}(s_2)$ since they are suppressed relative to the divergence in the $D / (s_1 s_2)$ term.

To proceed, we approximate the quantity $\alpha(s+\lambda)\tilde{R}(s)$ in the small-$s$, small-$\lambda$ limit. From Eq.~\eqref{eq:Rlaplace2}, we obtain:
\begin{equation}
\begin{aligned}
\alpha (s+\lambda) \tilde{R}(s) 
&= \frac{ -\left[ \alpha(s + \lambda) + \gamma(1 - \alpha) \right] + \sqrt{ \left[ \alpha(s + \lambda) + \gamma(1 - \alpha) \right]^2 + 4\alpha^2 \gamma(s + \lambda) } }{ 2\gamma } \\
&\approx \frac{ -\gamma(1 - \alpha) + \sqrt{A s + B} }{ 2\gamma },
\end{aligned}
\end{equation}
where we have dropped all the sub-leading terms, with the following definitions:
\begin{equation}
\begin{aligned}
A &= 2\lambda\alpha^2 + 2\gamma\alpha + 2\alpha^2 \gamma, \\
B &= \gamma^2(1 - \alpha)^2 + \lambda \left( \alpha^2 \lambda + 2\gamma\alpha + 2\alpha^2 \gamma \right)\approx  \gamma^2(1 - \alpha)^2 + \lambda \left( 2\gamma\alpha + 2\alpha^2 \gamma \right).
\end{aligned}
\end{equation}
Substituting this into the generalization error expression, we obtain the compact form:
\begin{equation}
\varepsilon_{g}(s_1, s_2) 
\approx \frac{ \displaystyle \frac{D}{s_1 s_2} }{ \displaystyle \sqrt{A s_1 + B} + \sqrt{A s_2 + B} }.
\end{equation}

To compute the inverse Laplace transform, we use the known identity for real $a, b > 0$:
\begin{equation}
\mathcal{L}^{-1}\left(\frac{1}{\sqrt{a s+b}+\sqrt{a s^{\prime}+b}}\right)=\frac{e^{-\frac{b(t_1+t_2)}{a}}\left(\frac{a}{t 1+t 2}\right)^{3 / 2}}{2 a^2 \sqrt{\pi}}.
\end{equation}

Applying this result to our expression for $\varepsilon_{g}(s_1, s_2)$, we obtain the time-domain generalization error:
\begin{equation}
\begin{aligned}
\varepsilon_{g}(t) & = \lim_{t_1\to t, t_2\to t} \mathcal{L}^{-1}\left( \varepsilon_{g}(s_1, s_2) \right)(t_1, t_2),\\
&\approx \mathcal{L}^{-1} \left[ \frac{2\gamma D}{s_1 s_2 \left( \sqrt{A s_1 + B} + \sqrt{A s_2 + B} \right) } \right] (t) \\
&=2\gamma D \int_0^t \!\! \int_0^t dt_1 dt_2\, \mathcal{L}^{-1} \left[ \frac{1}{\sqrt{A s_1 + B} + \sqrt{A s_2 + B}} \right](t_1, t_2) \\
& = D\frac{\gamma}{\sqrt{B}}\left(\frac{1}{\sqrt{ \pi}} \left[ 4 e^{-x} \sqrt{x} - 2 \sqrt{2x} e^{-2x} \right] +  \left[ 2(1 + 2x)\, \text{Erf}(\sqrt{x}) - (1 + 4x)\, \text{Erf}(\sqrt{2x}) \right]\right)|_{x =\frac{B t}{A} },\\
& = D\frac{\gamma}{\sqrt{B}} F(x),\\
& = \frac{ D F\left({\left(\frac{\gamma (1-\alpha)^2}{2 \alpha(1+\alpha)}+\lambda\right) t}\right)}{|1-\alpha|\sqrt{1+\frac{2 \alpha(1+\alpha)  \lambda}{\gamma(1-\alpha)^2}}} .
\end{aligned}
\end{equation}
We define the dimensionless scaling variable
\begin{equation}
x ={ \frac{B t}{A}} = {\left(\frac{\gamma (1-\alpha)^2}{2 \alpha(1+\alpha)}+\lambda\right) t},
\end{equation}
so that the generalization error at equal times, $\varepsilon_{g}(t, t)$, can be expressed in terms of a universal scaling function:
\begin{equation}
\begin{aligned}
F(x) & =\frac{1}{\sqrt{\pi}}\left(4 e^{-x} \sqrt{x}-2 \sqrt{2 x} e^{-2 x}\right) \\
& +(2(1+2 x) \operatorname{Erf}[\sqrt{x}]-(1+4 x) \operatorname{Erf}[\sqrt{2 x}])
\label{eq:escaling}
\end{aligned}
\end{equation}

The scaling function $F(x)$ governs the behavior of both the critical and regularized regimes. Its asymptotic behavior is:
\begin{equation}
\begin{aligned}
F(x \to 0) &\approx \frac{4 \sqrt{x}}{\sqrt{\pi}}(2-\sqrt{2}), \\
F(x \to \infty) &\approx 1.
\end{aligned}
\end{equation}

\subsubsection{The Finite-Size Scaling in Statistical Physics}
\label{sec:fss}

Near a second-order phase transition, physical observables such as the specific heat, susceptibility, or order parameter exhibit singular behavior that diverges in the thermodynamic limit. However, in a finite system, these divergences are rounded and shifted—a phenomenon described by finite-size scaling theory\cite{goldenfeld2018lectures}.  

As an illustrative example, consider the specific heat $c(t, L^{-1})$, where $t = (T - T_c)/T_c$ is the reduced temperature and $L$ is the linear system size. In the infinite-size limit, the specific heat follows the power-law singularity
\begin{equation}
c(t) \sim |t|^{-\alpha},
\end{equation}
where $\alpha$ is the specific-heat critical exponent.  
For a finite system of size $L$, the divergence is suppressed, and the scaling form becomes
\begin{equation}
c(t, L^{-1}) = |t|^{-\alpha} F_f^{\pm}(L^{-1} t^{-\nu}),
\label{eq:fss1}
\end{equation}
where $F_f^{\pm}$ is a universal finite-size scaling function (with $\pm$ corresponding to $t>0$ and $t<0$), and $\nu$ is the correlation-length exponent defined by $\xi \sim |t|^{-\nu}$.  

Rewriting Eq.~\eqref{eq:fss1} by rescaling $t$ in terms of $L$, we obtain
\begin{equation}
c(t, L^{-1}) = L^{\alpha/\nu} D^{\pm}(t L^{1/\nu}),
\label{eq:fss2}
\end{equation}
where $D(x)$ is a new dimensionless scaling function, with its argument $x = t L^{1/\nu}$ representing the ratio of the system size to the correlation length ($L / \xi$).  

The maximum of the specific heat occurs when $x = x_0$, corresponding to a finite-size–shifted critical temperature:
\begin{equation}
t_L = x_0 / L^{1/\nu} \propto L^{-1/\nu}.
\label{eq:fss3}
\end{equation}
Thus, the apparent critical temperature in a finite system is displaced from the true critical point by an amount that scales as $L^{-1/\nu}$.  However, $x_0$ can sometimes be zero, like the generalization error in the regularized linear networks.

Similarly, the height of the specific-heat peak is finite and scales with system size as
\begin{equation}
c(t_L, L^{-1}) = L^{\alpha/\nu} D(x_0) \propto L^{\alpha/\nu}.
\label{eq:fss4}
\end{equation}
Equations~\eqref{eq:fss3} and \eqref{eq:fss4} capture the two key features of finite-size scaling:  
(i) the rounding of the singularity, which replaces the true divergence by a finite-size–dependent maximum, and  
(ii) the shift of the apparent critical point away from its thermodynamic limit.  

Physically, this behavior arises because the correlation length $\xi$ cannot exceed the system size $L$. As a result, the singular part of any observable must be a function of the dimensionless ratio $\xi/L$, giving rise to universal scaling forms such as Eq.~\eqref{eq:fss2}.  

In our learning framework, the regularization strength $\lambda$ plays a role analogous to $L^{-1/\nu}$ in statistical mechanics: it cuts off the divergence of the generalization error, shifts the apparent critical point $\alpha_c$, and replaces the sharp transition by a rounded crossover. This analogy provides the conceptual basis for interpreting $\lambda$ as a finite-size parameter in the phase transition of learning dynamics.

\subsubsection{Long Time Behavior}
In this section, we will analyze the long time behavior of the generalization error and training error. In time domain, the limit is $t\to\infty$, and in frequency domain, the limit is $s_1, s_2\to 0$.
\paragraph{Generalization error in $t\to\infty$}
There are two ways that we can check the long time behavior of generalization error in the regularized case. From the original definition of the extended generalization error in frequency domain as shown in Equation~\ref{eq:gerrorlaplace}, if we take $s_1, s_2\to 0$, dropping all the sub-leading terms, and take $\lambda$ as a constant:
\begin{equation}
\begin{aligned}
\lim_{s_1,s_2\to 0} \varepsilon_{g}(s_1, s_2) &\approx  \frac{ \displaystyle  \frac{D}{s_1 s_2} +\frac{r\lambda^2}{s_1s_2} \tilde{R}(0) \tilde{R}\left(0\right) }{\displaystyle 1-\alpha(\lambda \tilde{R}(0)-1)^2},\\
\lim_{t\to \infty}\varepsilon_{g}(t)&\to \frac{ \displaystyle  D +r\lambda^2 \chi^2_{\lambda} }{\displaystyle 1-\alpha(\lambda \chi_{\lambda}-1)^2},
\end{aligned}
\end{equation}
where we defined the susceptibility when $\lambda\neq 0$ as $\chi_{\lambda} = \tilde{R}(0) =\int_0^{\infty} R(t)d t $. The above equation shows that the generalization error in the long time limit is a function of susceptibility, where:
\begin{equation}
\begin{aligned}
\lambda \chi_{\lambda}=\frac{-(\alpha \lambda+\gamma(1-\alpha)) + \sqrt{(\alpha \lambda+\gamma(1-\alpha))^2+4 \alpha^2 \gamma \lambda}}{2 \alpha \gamma}.
\end{aligned}
\end{equation}

To gain further insight into the asymptotic behavior of the generalization error, we revisit the scaling form derived in Section~\ref{sec:reguscaling}. In the long-time limit $t \to \infty$, the dimensionless scaling variable $x = \frac{B t}{A} \to \infty$, and the scaling function approaches a constant. Consequently, the generalization error saturates to:
\begin{equation}
\begin{aligned}
\varepsilon_{g}(t \to \infty)
&= D \frac{\gamma}{\sqrt{B}} F(x \to \infty), \\
&= D \frac{\gamma}{\sqrt{B}}, \\
&= D \frac{\gamma}{\sqrt{ \gamma^2 (1 - \alpha)^2 + \lambda \left( 2 \gamma \alpha + 2 \alpha^2 \gamma \right) }}, \\
&= \frac{D}{\sqrt{\lambda}} \cdot \frac{\gamma}{ \sqrt{ \frac{ \gamma^2 (1 - \alpha)^2 }{ \lambda } + 2 \gamma ( \alpha + \alpha^2 ) } }.
\end{aligned}
\end{equation}
The long-time generalization error remains finite rather than diverging, as the regularization parameter $\lambda$ effectively smooths the critical singularity. A rounded peak emerges near a shifted critical point $\alpha_c < 1$. To analyze its scaling structure, we introduce the small deviation from criticality $\epsilon = 1 - \alpha$ (valid for $\alpha < 1$ near the interpolation threshold) and define the dimensionless scaling variable
\begin{equation}
x = \frac{\gamma^2 \epsilon^2}{\lambda}.
\end{equation}

In the $t \to \infty$ limit, the generalization error can be expressed as
\begin{equation}
\varepsilon_{g}(t \!\to\! \infty) =
\frac{D}{\sqrt{\lambda}}
\cdot
\frac{\gamma}{\sqrt{\frac{\gamma^2 \epsilon^2}{\lambda} + 2\gamma(\epsilon^2 - 3\epsilon + 2)}},
\end{equation}
where $D$ is the noise variance and $\gamma$ is the stochasticity parameter controlling the fraction of data used per update.
Retaining only the dominant terms in $\epsilon$ and $\lambda$ yields
\begin{equation}
\varepsilon_{g}(t \!\to\! \infty)
\approx
\frac{D}{\sqrt{\lambda}}
\cdot
\frac{\gamma}{\sqrt{ x + 4\gamma - 6 \sqrt{\lambda x}}}.
\end{equation}

To leading order, the subleading term $6\sqrt{\lambda x}$ can be neglected, giving the simplified finite-size scaling form
\begin{equation}
\varepsilon_{g}(t \!\to\! \infty)
\simeq
\frac{D}{\sqrt{\lambda}}
\cdot
\frac{\gamma}{\sqrt{\frac{\gamma^2 \epsilon^2}{\lambda} + 4\gamma}}
=
\frac{D}{\sqrt{\lambda}}\,
D^{\pm}\!\left(\frac{\gamma^2 \epsilon^2}{\lambda}\right),
\end{equation}
where $D^{\pm}(x) = \gamma / \sqrt{x + 4\gamma}$ is a dimensionless finite-size scaling function.
This expression highlights the role of $\lambda$ as an effective inverse system size: it cuts off the divergence of $\varepsilon_g$ and introduces a rounded crossover instead of a sharp transition.

Following the finite-size scaling analysis in Section~\ref{sec:fss}, if the leading-order scaling function attains its maximum at $x = x_0$, the apparent shift of the critical capacity obeys
\begin{equation}
\Delta\alpha_c = 1 - \alpha_c = \frac{\sqrt{\lambda x_0}}{\gamma} \propto \sqrt{\lambda}.
\end{equation}
For the present scaling function $D^{\pm}(x)$, the maximum occurs at $x_0 = 0$, indicating that the leading-order contribution produces no shift in $\alpha_c$.
Consequently, the displacement of the critical point arises only from subleading corrections beyond the dominant scaling order.

Including these corrections (e.g., the $6\gamma\sqrt{\lambda x}$ term) slightly shifts the peak below $\alpha = 1$, giving
\begin{equation}
\Delta \alpha_c \sim \lambda.
\end{equation}
Hence, the steady-state generalization error exhibits
\begin{equation}
\varepsilon_{g}^{\text{max}} \sim \lambda^{-1/2}, 
\qquad 
\Delta\alpha_c \sim \lambda,
\end{equation}
mirroring the finite-size scaling behavior observed in continuous phase transitions.
Here, $\lambda$ plays the role of an inverse system size: it limits the effective correlation length and smooths the singularity, just as finite $L$ rounds and shifts the specific-heat peak in critical systems~\cite{goldenfeld2018lectures}.

\subsection{Unregularized Dynamics: $\lambda = 0$}
\label{sec:unregularized}

In this section, we analyze the unregularized dynamics corresponding to $\lambda = 0$, which is the regime where the hallmark features of double descent emerge: the generalization error diverges at the interpolation threshold, while the training error vanishes. These dynamics can be viewed as a limiting case of the regularized setting ($\lambda \neq 0$), and the scaling results derived here will serve as a special case of the more general data collapse formulas presented in Section~\ref{sec:regularized}.

In the long-time limit, the generalization error can be expressed as a function of the susceptibility $\chi = \tilde{R}(0)$, which diverges both as $\lambda \to 0$ and as the model approaches the interpolation threshold $\alpha = 1$ from below. This divergence signals the critical nature of the transition and corresponds to the breakdown of generalization.

Because the generalization error in linear regression scales with the norm of the ground truth target vector $\beta$, its absolute value does not directly reflect learning performance. To address this, we will also introduce a \emph{relative generalization error} that captures the alignment between the learned estimator and the target, independent of scale.

Finally, we show that the role of stochasticity in gradient descent becomes beneficial only in the presence of regularization ($\lambda \neq 0$). In that case, the injected noise helps escape sharp minima and promotes convergence to flatter regions of the loss landscape, which, although not minimizing training error, can generalize significantly better. In contrast, in the unregularized limit, noise provides no such benefit and merely changes the training dynamics.

\subsubsection{Data Collapse for the Response Function}
Given the results from Section~\ref{sec:unResponse}, we can obtain the data collapse formula by taking the limit $\lambda \to 0$. In this unregularized regime, the response function simplifies to:
\begin{equation}
\begin{aligned}
\lim_{\lambda \to 0} R(\tau > 0) 
&\approx \lim_{\lambda \to 0} e^{-\lambda \tau} \, \epsilon \, 
F \left( \frac{\gamma \epsilon \sqrt{\tau}}{\sqrt{2\lambda + 4\gamma}} \right) \\
&= \epsilon \, F \left( \frac{\epsilon \sqrt{\gamma \tau}}{2} \right),
\end{aligned}
\end{equation}
where the scaling function is shown in Equation~\ref{eq:Rscaling}, which is given by
\begin{equation}
F(x) = \frac{1}{2} \left( \frac{e^{-x^2}}{\sqrt{\pi}\, x} + \operatorname{Erf}(x) - \operatorname{Sign}(1 - \alpha) \right).
\end{equation}
This expression reveals the universal scaling form of the response function in the unregularized limit, with the deviation from criticality governed by the small parameter $\epsilon=1-\alpha$, which controls the amplitude of the response function as the system approaches the interpolation threshold.
\begin{figure}[htbp] 
	\centering
	\includegraphics[scale=0.5]{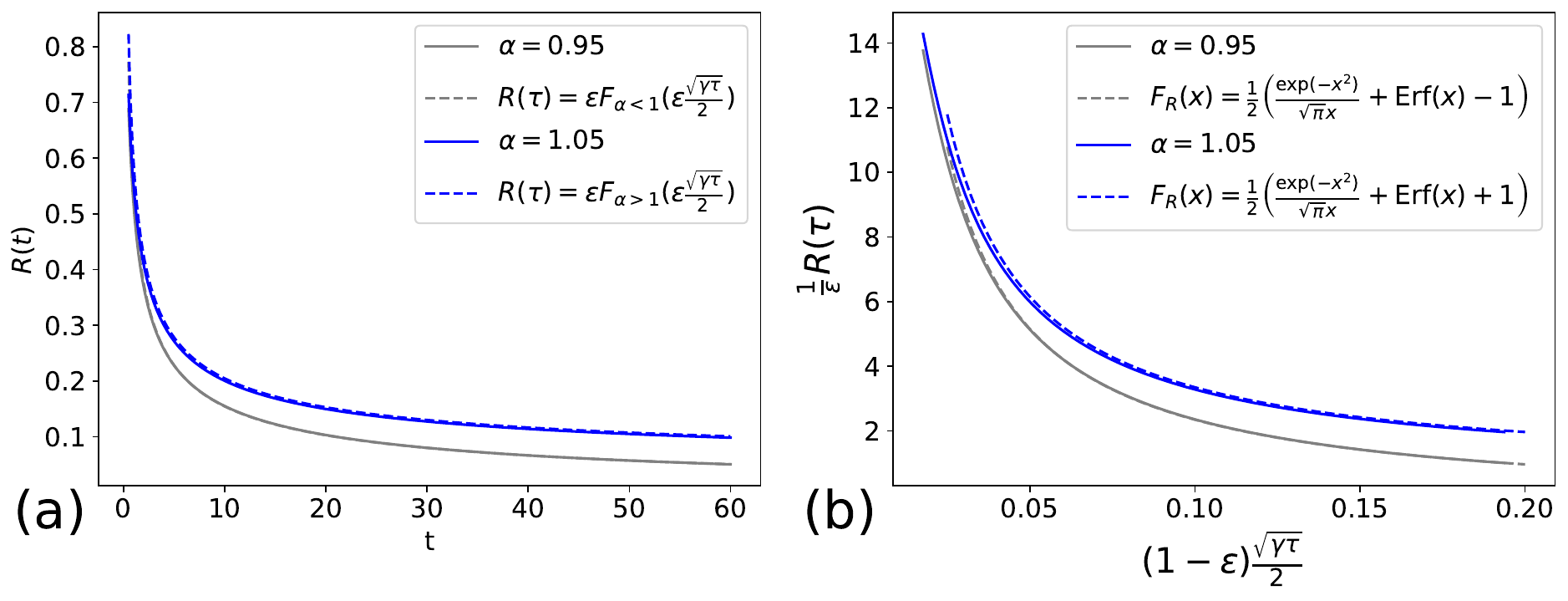}
	\caption{
	(a) Response function $R(\tau)$ in the time domain obtained from dynamical mean-field theory (DMFT) simulations (dashed lines) and the analytical scaling formula (solid lines).\\
	(b) Data collapse of the response function in the unregularized case ($\lambda = 0$), showing excellent agreement between DMFT simulations (dashed lines) and the scaling prediction (solid lines).
	}
	\label{fig4}
\end{figure}

\subsubsection{Data Collapse for the Generalization Error}

By taking the unregularized limit $\lambda \to 0$ of the results derived in Section~\ref{sec:reguscaling}, we obtain the data collapse formula for the generalization error in closed form. In this limit, both $\lambda$ and $\epsilon = |1 - \alpha|$ (near criticality) are small, and the generalization error simplifies to:

\begin{equation}
\begin{aligned}
\lim_{\lambda \to 0,\, \epsilon \to 0} \varepsilon_{g}(t > 0) 
&\approx 
\lim_{\lambda \to 0,\, \epsilon \to 0} 
D \frac{\gamma}{\sqrt{ \gamma^2(1 - \alpha)^2 + \lambda(2\gamma\alpha + 2\alpha^2\gamma) }}
\, F\left({ \frac{ [\gamma^2(1 - \alpha)^2 + \lambda(2\gamma\alpha + 2\alpha^2\gamma)]\, t }{2\lambda \alpha^2 + 2\gamma \alpha + 2\alpha^2 \gamma} }\right), \\
&= D \frac{\gamma}{\gamma |1 - \alpha|} \, F\left( {\frac{\gamma (1 - \alpha)^2 t}{4}} \right), \\
&= \frac{D}{|1 - \alpha|} \, F\left(\frac{|1-\alpha|^2}{4}{ {\gamma  t}} \right).
\label{eq:unscaling}
\end{aligned}
\end{equation}

Here, the universal scaling function $ F(x) $ is identical to that in Equation~\ref{eq:escaling} and is given explicitly by:

\begin{equation}
\begin{aligned}
F(x) & =\frac{1}{\sqrt{\pi}}\left(4 e^{-x} \sqrt{x}-2 \sqrt{2 x} e^{-2 x}\right) \\
& +(2(1+2 x) \operatorname{Erf}[\sqrt{x}]-(1+4 x) \operatorname{Erf}[\sqrt{2 x}])
\end{aligned}
\end{equation}

This data collapse illustrates that, in the absence of regularization, the generalization error exhibits a diverging amplitude as $\alpha \to 1$, with time-dependent dynamics governed by the scaling variable $x =\frac{|1-\alpha|^2}{4}{ {\gamma  t}}$.

\subsubsection{Long-Time Behavior of the Generalization Error and Training Error}
\label{sec:longtimel0}

In this section, we analyze the long-time behavior of both the generalization and training errors in the unregularized case ($\lambda = 0$). 
\newline
\paragraph{Long-time behavior of the generalization error: }In the Laplace frequency domain, the generalization error takes the form:
\begin{equation}
\begin{aligned}
\mathcal{L}\{\varepsilon_{g}(t_1, t_2)\}(s_1, s_2) 
= \frac{ \displaystyle \frac{D}{s_1 s_2} + \left(\hat{C}(0, 0) + r\right)\, \tilde{R}(s_1)\, \tilde{R}(s_2) }
{ \displaystyle 1 - \alpha + \alpha\left(s_1 \tilde{R}(s_1) + s_2 \tilde{R}(s_2) - s_1 s_2 \tilde{R}(s_1)\tilde{R}(s_2)\right) },
\end{aligned}
\end{equation}
where the Laplace-transformed response function when $\lambda=0$ is given by:
\begin{equation}
\tilde{R}(s) = \frac{ -(\alpha s + \gamma(1 - \alpha)) + \sqrt{(\alpha s + \gamma(1 - \alpha))^2 + 4 \alpha^2 \gamma s} }{2 \alpha \gamma s}.
\end{equation}

To evaluate the long-time limit, we examine the behavior as $s_1, s_2 \to 0$. The key quantity controlling the asymptotics is the limit:
\begin{equation}
\lim_{s \to 0} s \tilde{R}(s) =
\begin{cases}
0 & \text{for } \alpha < 1, \\
\frac{\alpha - 1}{\alpha} & \text{for } \alpha > 1.
\end{cases}
\end{equation}
We now analyze the two regimes separately.

\paragraph{$\alpha < 1$:}  
In this regime, the term $s \tilde{R}(s) \to 0$, so the generalization error simplifies to:
\begin{equation}
\begin{aligned}
\lim_{s_1, s_2 \to 0} \mathcal{L}\{\varepsilon_{g}(t_1, t_2)\}(s_1, s_2) 
&= \frac{ \frac{D}{s_1 s_2} }{1 - \alpha}, \\
\Rightarrow \quad \varepsilon_{g}(t \to \infty) 
&= \frac{D}{1 - \alpha}.
\end{aligned}
\end{equation}

This result holds for all $\alpha < 1$, and clearly diverges as $\alpha \to 1^{-}$, indicating a critical point at the interpolation threshold. The same result can also be derived from the scaling form in Equation~\ref{eq:unscaling}, with the scaling variable $x = \frac{|1-\alpha|^2}{4}{ {\gamma  t}}$. In the long-time limit $t \to \infty$, we have $x \to \infty$ and $F(x) \to 1$, reproducing the same divergence.

\paragraph{$\alpha > 1$:}  
In contrast, when $\alpha > 1$, the response function contributes a non-zero limit:
\begin{equation}
\lim_{s \to 0} s \tilde{R}(s) = \frac{\alpha - 1}{\alpha}.
\end{equation}
Plugging this result into the generalization error expression, we obtain:
\begin{equation}
\begin{aligned}
\lim_{s_1, s_2 \to 0} \mathcal{L}\{\varepsilon_{g}(t_1, t_2)\}(s_1, s_2) 
&= \frac{ D+ \left(\hat{C}(0, 0) + r\right)\, \left( \frac{\alpha - 1}{\alpha} \right)^2 }
{ s_1 s_2 \left[ (1 - \alpha) + 2(\alpha - 1) - \frac{\alpha - 1)^2}{\alpha} \right] }, \\
\Rightarrow \quad \varepsilon_{g}(t \to \infty) 
&= \left(\hat{C}(0, 0) + r\right)\, \frac{\alpha - 1}{\alpha} + D\, \frac{\alpha}{\alpha - 1},\\
& = C(0,0)\, \frac{\alpha - 1}{\alpha} + D\, \frac{\alpha}{\alpha - 1},
\end{aligned}
\end{equation}
where $C(0,0) = \langle \hat{\beta}^2(0)\rangle + r$, with $\langle \hat{\beta}^2(0)\rangle$ denoting the variance of the estimator’s initial condition and $r$ the mean squared norm of the ground-truth parameter vector per dimension.

Near criticality from the overparameterized side ($\alpha > 1$), the generalization error asymptotically approaches $\varepsilon_{g}(t \to \infty) \sim \frac{D}{\alpha - 1}$, consistent with the prediction from the scaling function in the limit $x \to \infty$.

The first term, ${C}(0, 0) \frac{\alpha - 1}{\alpha}$, captures the influence of initial conditions and reflects a memory of the early-time dynamics. Notably, the prefactor $\frac{\alpha - 1}{\alpha}$ quantifies the degree of \textit{ergodicity breaking} in the learning dynamics. In the overparameterized regime, the system fails to fully explore its weight-space due to the proliferation of flat directions and isolated minima, leading to broken ergodicity and persistent memory of initialization.

This ergodicity breaking, which becomes stronger as $\alpha$ increases, plays a constructive role in generalization: rather than overfitting to noisy fluctuations, the system implicitly biases toward broader, flatter minima that generalize better. As we will discuss further in Section~\ref{sec:transition}, this asymptotic structure highlights the non-equilibrium nature of learning dynamics in the overparameterized phase, and reveals how broken ergodicity serves as a mechanism for regularization in the absence of explicit constraints.
\paragraph{$\alpha = 1$:}
At the interpolation threshold, the dynamics of the generalization error become singular, and the long-time behavior is best captured using the scaling form. When $\alpha = 1$, the scaling variable reduces to $x = \frac{|1-\alpha|^2}{4}{ {\gamma  t}} \to 0$, meaning that even as $t \to \infty$, the argument of the scaling function remains small in the critical regime. In this limit, the scaling function admits the asymptotic form $F(x \to 0) \approx \frac{4 \sqrt{x}}{\sqrt{\pi}}(2 - \sqrt{2})$, and the generalization error becomes:
\begin{equation}
\begin{aligned}
\lim_{\alpha \rightarrow 1} \varepsilon_{g}(t \to \infty)
&= \frac{D}{|1 - \alpha|} F(x \to 0) \\
&= D \frac{2 \sqrt{\gamma} }{ \sqrt{\pi} } (2 - \sqrt{2}) \sqrt{t}.
\end{aligned}
\end{equation}
Thus, as the system approaches criticality ($\alpha = 1$), the generalization error exhibits a $\sqrt{t}$ divergence, signaling critical slowing down at the interpolation threshold. Near this transition point, convergence becomes progressively slower, and the characteristic timescale for relaxation diverges. 
\paragraph{Generalization error as a function of susceptibility: }
To better understand the divergence of the generalization error at the interpolation threshold, we can reinterpret the long-time generalization error $\varepsilon_{g}(t \to \infty)$ as a function of the \textit{susceptibility}:
\begin{equation}
\chi = \tilde{R}(0) = \int_0^\infty R(\tau)\, d\tau.
\end{equation}
This susceptibility is well-defined only in the equilibrium regime ($\alpha < 1$). Using the identity
\begin{equation}
\frac{\gamma}{\alpha} \cdot \frac{\tilde{R}(s)}{\gamma \tilde{R}(s)+1} = 1 - s\tilde{R}(s),
\end{equation}
we can rewrite the generalization error in Laplace space as:
\begin{equation}
\begin{aligned}
\mathcal{L}\{\varepsilon_{g}(t_1, t_2)\}(s_1, s_2) 
&= \frac{1}{s_1 s_2} \cdot 
\frac{ D + \left(\hat{C}(0, 0) + r\right)s_1 \tilde{R}(s_1) \cdot s_2 \tilde{R}(s_2) }
{ 1 - \frac{\gamma^2}{\alpha} \cdot \frac{\tilde{R}(s_1)}{\gamma \tilde{R}(s_1) + 1} \cdot \frac{\tilde{R}(s_2)}{\gamma \tilde{R}(s_2) + 1} }.
\end{aligned}
\end{equation}

In the long-time limit $s_1, s_2 \to 0$, this becomes:
\begin{equation}
\begin{aligned}
\lim_{s_1, s_2 \to 0} \mathcal{L}\{\varepsilon_{g}(t_1, t_2)\}(s_1, s_2)
&= \frac{1}{s_1 s_2} \cdot 
\frac{ D }
{ 1 - \frac{\gamma^2}{\alpha} \cdot \left( \frac{\chi}{\gamma \chi + 1} \right)^2 }.
\end{aligned}
\end{equation}

The susceptibility satisfies the following condition in equilibrium ($\alpha<1$):
\begin{equation}
\frac{\gamma \chi}{\gamma \chi + 1} = \alpha.
\end{equation}

Solving this expression yields the long-time generalization error:
\begin{equation}
\begin{aligned}
\varepsilon_{g}(t \to \infty) 
&= \frac{D}{1 - \left( \frac{\gamma \chi}{\gamma \chi + 1} \right) }
=D\left( \gamma \chi + 1\right).
\end{aligned}
\end{equation}

This result shows that the generalization error is directly governed by the susceptibility $\chi$, which diverges as $\alpha \to 1^{-}$. Therefore, the divergence of $\varepsilon_{g}$ near the interpolation threshold is a direct consequence of critical behavior in the susceptibility, linking the learning dynamics to equilibrium phase transition phenomena. 

Bringing together the analyses above, we can now interpret the generalization error through the lens of static susceptibility $\chi$. In the regime $\alpha < 1$, the learning dynamics eventually relax to an equilibrium steady state. In the long-time limit, the generalization error takes the asymptotic form
\begin{equation}
\lim_{t \rightarrow \infty} \varepsilon_{g}(t) = D \left( \gamma \chi + 1 \right),
\end{equation}
where $\chi = \int_0^\infty R(\tau)\, d\tau$ is the susceptibility of the learning dynamics and $R(t)$ is the linear response function. \new{This relation follows from linear response theory within the DMFT framework: the generalization error can be expressed in terms of correlation functions of the learning dynamics, which are related to response functions. In the mean-field equations, this leads to a simple linear dependence of the long-time generalization error on the integrated response $\chi$. Beyond the mean-field approximation, more general relations between the generalization error and response functions may arise, reflecting additional dynamical correlations that are not captured at the saddle-point level}.

This relation admits a clear physical interpretation: the generalization error quantifies how sensitive the learned model is to perturbations in the data distribution. When we evaluate generalization by testing the trained network on previously unseen samples, we are effectively introducing a perturbation by augmenting the training dataset with new inputs. The generalization error can thus be viewed as the model’s response to this perturbation. A larger deviation between the estimator $\hat{\beta}$ and the ground truth $\beta$—corresponding to a higher generalization error—leads to a greater response $\delta \hat{\beta}$ under the same perturbation to the training data. This implies a steeper gradient in the loss landscape and, consequently, a larger susceptibility. Since susceptibility is defined as the integrated linear response of the model to external perturbations, it is directly and positively correlated with the generalization error.
\newline

\paragraph{Long-Time Behavior of the Training Error}

Given the expression for the training error in Laplace space (Equation~\ref{eq:trainL}), we analyze its long-time behavior by taking the limit $s_1, s_2 \to 0$ in each of the three regimes.

\paragraph{$\alpha < 1$:}  
In the underparameterized regime, the system reaches equilibrium, and the susceptibility $\chi$ is well-defined as
$$
\chi = \int_0^\infty R(\tau)\, d\tau = \tilde{R}(s \to 0) = \frac{\alpha}{\gamma\left(1-\alpha\right)}.
$$
The training error in Laplace space becomes
\begin{equation}
\begin{aligned}
\lim_{s_1, s_2 \to 0} \varepsilon_{\text{train}}(s_1, s_2)
&= \frac{1}{(\gamma \chi + 1)^2} \cdot \lim_{s_1, s_2 \to 0} \varepsilon_{g}(s_1, s_2) \\
&= \frac{1}{(\gamma \chi + 1)^2} \cdot \frac{D}{(1 - \alpha)\, s_1 s_2},
\end{aligned}
\end{equation}
which gives in the time domain:
\begin{equation}
\varepsilon_{\text{train}}(t \to \infty) = D (1 - \alpha).
\end{equation}

\paragraph{$\alpha > 1$:}  
In the overparameterized regime, the susceptibility diverges and is not well-defined. The Laplace-space response becomes:
$$
\gamma \tilde{R}(s \to 0) + 1 = \frac{\gamma (\alpha - 1)}{\alpha s}.
$$
Using the known long-time limit of the generalization error in this regime, we obtain:
\begin{equation}
\begin{aligned}
\lim_{s_1, s_2 \to 0} \varepsilon_{\text{train}}(s_1, s_2)
&= \left( \frac{\alpha s_1}{\gamma (\alpha - 1)} \right) \cdot \frac{D + (C(0, 0) + r) \left( \frac{\alpha - 1}{\alpha} \right)^2}{s_1 s_2 \cdot \left[ 1 - \alpha + 2(\alpha - 1) - \frac{(\alpha - 1)^2}{\alpha} \right]} \cdot \left( \frac{\alpha s_2}{\gamma (\alpha - 1)} \right) \\
&= \frac{\alpha^2}{\gamma^2 (\alpha - 1)^2} \cdot \frac{D + (C(0, 0) + r) \left( \frac{\alpha - 1}{\alpha} \right)^2}{1 - \alpha + 2(\alpha - 1) - \frac{(\alpha - 1)^2}{\alpha}},
\end{aligned}
\end{equation}
which corresponds to:
\begin{equation}
\varepsilon_{\text{train}}(t \to \infty) = 0.
\end{equation}

\paragraph{$\alpha = 1$:}  
At the interpolation threshold, the response function scales nontrivially with $s$, and training error decays slowly. Using the scaling behavior $\tilde{R}(s) \sim \frac{1}{\sqrt{\gamma s}}$, we get:
\begin{equation}
\begin{aligned}
\lim_{s_1, s_2 \to 0} \varepsilon_{\text{train}}(s_1, s_2)
&\approx \frac{1}{\gamma \tilde{R}(s_1) + 1} \cdot \frac{D/s_1 s_2 + (C(0,0) + r) \tilde{R}(s_1)\tilde{R}(s_2)}{1 + s_1 \tilde{R}(s_1) + s_2 \tilde{R}(s_2) - s_1 s_2 \tilde{R}(s_1)\tilde{R}(s_2)} \cdot \frac{1}{\gamma \tilde{R}(s_2) + 1} \\
&\sim \frac{(C(0, 0) + r)}{\gamma^2} \cdot \frac{1}{\sqrt{\frac{s_1}{\gamma}} + \sqrt{\frac{s_2}{\gamma}}},
\end{aligned}
\end{equation}
which implies:
\begin{equation}
\varepsilon_{\text{train}}(t \to \infty) = 0,
\end{equation}
but with much slower convergence than in the $\alpha > 1$ case, due to critical slowing down.

\subsection{Role of Stochasticity in Training}

Stochasticity plays a qualitatively different role depending on whether the system is regularized ($\lambda \neq 0$) or unregularized ($\lambda = 0$). In our framework, stochasticity is controlled by the parameter $\gamma$, the mean value of the stochastic indicator $s(t)$, representing the fraction of training data used to compute gradient updates—mimicking the effect of stochastic gradient descent (SGD).

In the unregularized regime ($\lambda = 0$), the generalization error exhibits the scaling form
\begin{equation}
\varepsilon_{g}(t) = \frac{1}{\epsilon} F\left( \frac{\epsilon \sqrt{\gamma t}}{2} \right),
\end{equation}
as shown in Equation~\ref{eq:unscaling}. Here, $\gamma$ only enters through the scaling variable $x =\frac{|1-\alpha|^2}{4}{ {\gamma  t}}$, effectively rescaling time. Thus, the role of stochasticity is limited to slowing down the training dynamics via $t \to \gamma t$, without affecting the long-time generalization error. This behavior reflects the fact that in the underparameterized regime ($\alpha < 1$), where a unique minimizer exists, both gradient descent and SGD converge to the same solution. In the overparameterized regime ($\alpha > 1$), the loss landscape contains flat directions, and the noise in SGD is too weak to push the dynamics away from these flat minima. As a result, stochasticity has negligible impact on the final outcome when $\lambda = 0$.

By contrast, in the regularized case ($\lambda \neq 0$), stochasticity affects not only the speed of convergence but also the final generalization performance. Regularization renders the loss function strictly convex:
\begin{equation}
\mathcal{L}(\hat{\beta}) = \sum_{\mu=1}^N \mathcal{L}(\sum_i^d \hat{\beta}_i, {x}_i^\mu {y}^\mu) + \lambda \|\hat{\beta}\|_2^2,
\end{equation}
introducing a competition between data fitting and norm minimization. While gradient descent deterministically converges to the unique minimizer
\begin{equation}
\hat{\beta}^* = \left(X^\top X + \lambda I\right)^{-1} X^\top \mathbf{y},
\end{equation}
where $X\in \mathbb{R}^{N\times d}$ indicates the training data matrix and $\mathbf{y}\in\mathbb{R}^{N\times 1}$ denotes the label vector. This solution often lies in a sharp, narrow valley in the loss landscape and can generalize poorly. In contrast, the noise introduced by SGD helps the dynamics explore flatter minima, which have been empirically linked to better generalization.

This can be interpreted as follows: although the optimal solution minimizes the regularized loss, it may be sharply localized. SGD, by injecting stochasticity, perturbs the dynamics just enough to escape such sharp minima and settle into flatter, suboptimal regions that generalize better. This aligns with the broader understanding that flatter minima correspond to solutions that are more robust to perturbations in the input data and model parameters\cite{hochreiter1997flat,baldassi2021unveiling}.

In summary, stochasticity is largely neutral when $\lambda = 0$ but becomes crucial when $\lambda \neq 0$, as it helps the model find flatter minima with better generalization. This perspective provides a dynamical explanation for why SGD-trained models often outperform those trained with full-batch gradient descent, especially when regularization is present.

\section{Double Descent as a Dynamical Phase Transition: A Superconducting Analogy}
\label{sec:transition}
The double descent phenomenon is often interpreted as a non-monotonic feature of model performance with increasing capacity, associated with the interpolation threshold at $\alpha = 1$. In this section, we propose a novel and distinct interpretation: that double descent, in the specific setting of linear regression trained with stochastic gradient descent (SGD), can be understood as a genuine \textit{dynamical phase transition}, bearing a formal analogy to superconducting transitions in condensed matter physics.

This perspective is made precise by analyzing macroscopic observables, such as the response function, susceptibility, and ergodicity, as functions of the control parameter $\alpha$, defined as the ratio between the number of model parameters and the size of the training dataset. In particular, the analogy becomes sharp in the unregularized limit $\lambda \to 0$, where regularization no longer obscures the intrinsic dynamical structure of the learning dynamics. In this limit, we observe hallmark signatures of criticality: the divergence of the susceptibility as $\alpha \to 1$, the emergence of a simple pole in the response function for $\alpha > 1$, which indicates a persistent current and the absence of dissipation reminiscent of the superconducting phase, and the breakdown of ergodicity in the overparameterized regime, manifested as a long-time memory plateau in the response function.

These results reveal a sharp transition in the dynamical character of the system at $\alpha = 1$ when $\lambda=0$: for $\alpha < 1$, the dynamics are well-behaved, exhibiting finite susceptibility, exponential relaxation, and ergodic behavior consistent with equilibrium statistical mechanics. In contrast, for $\alpha > 1$, the system enters a qualitatively distinct non-equilibrium phase characterized by divergent susceptibility, persistent temporal correlations, and a violation of ergodicity. The transition is not merely a shift in performance metrics but reflects a fundamental change in the underlying dynamics of learning. By identifying the origin of this transition in the analytic structure of the response function, we demonstrate that double descent in this setting can be understood as a dynamical phase transition governed by the control parameter $\alpha$, analogous to superconducting transition with control parameter temperature in physics.

\subsection{Analytic Structure of the Response Function}
\label{sec:pole}

In this section, we analyze the pole and branch structure of the response function $\tilde{R}(\omega)$ to elucidate its causal properties. Under the Fourier convention adopted in Equation~\ref{eq:Fourier}, causality in the time domain requires that $\tilde{R}(\omega)$ be analytic in the lower half of the complex $\omega$-plane.

To demonstrate this analyticity condition (no singularities within the lower half-plane), we consider the complex contour integral:
\begin{equation}
\oint_{\Omega_1} \tilde{R}(\omega) e^{i \omega \tau} \, \mathrm{d} \omega - \int_{-r}^{r} \tilde{R}(\omega) e^{i \omega \tau} \, \mathrm{d} \omega = 0,
\end{equation}
where $\Omega_1$ denotes the semicircular arc of radius $r \to \infty$ in the lower half-plane, and the second term represents the integral over the real axis from $-r$ to $r$.

On the semicircular contour $\Omega_1$, we parameterize $\omega = \omega_r + i \omega_i$ and observe that
\begin{equation}
\begin{aligned}
e^{i \omega t} &= e^{i(\omega_r + i \omega_i)t} = e^{i \omega_r t} \cdot e^{- \omega_i t}.
\end{aligned}
\end{equation}
Since $\omega_i < 0$ along the contour and the exponential factor $e^{-\omega_i \tau} \to 0$ for $\tau < 0$, the integrand vanishes in this limit provided that $\tilde{R}(\omega)$ remains bounded or decays at large $|\omega|$.

Therefore, the contribution from the semicircle vanishes:
\begin{equation}
\oint_{\Omega_1} \tilde{R}(\omega) e^{i \omega \tau} \, \mathrm{d} \omega = 0,
\end{equation}
and we obtain
\begin{equation}
\int_{-\infty}^{\infty} \tilde{R}(\omega) e^{i \omega \tau} \, \mathrm{d} \omega = R(\tau<0) = 0,
\end{equation}
which confirms that the response function $ R(\tau) $ vanishes for $ \tau < 0 $, provided that $ \tilde{R}(\omega) $ is analytic in the lower half of the complex $ \omega $-plane\cite{toll1956causality}. This analytic property is a direct manifestation of causality in the time domain.

With this framework in place, we proceed to examine the analytic structure of $\tilde{R}(\omega)$ by analyzing the cases $\alpha < 1$, $\alpha > 1$, and $\alpha = 1$ separately.

\paragraph{Case $\alpha < 1$:}  
In the underparameterized regime, we consider the unregularized case by setting $\lambda = 0$ in Equation~\ref{eq:RF}, and select the positive root as we discussed in previous sections. The resulting response function takes the form:
\begin{equation}
\tilde{R}(\omega) = \frac{ - (i \omega \alpha + \gamma(1 - \alpha)) + \sqrt{(i \omega \alpha + \gamma(1 - \alpha))^2 + 4 i \omega \alpha^2 \gamma} }{2 i \omega \alpha \gamma}.
\label{eq:smalla}
\end{equation}
Although the expression appears to have a singularity at $\omega = 0$, this is a removable singularity rather than a true pole. The singularity is removable, and the response function is in fact regular at the origin.  The only non-analyticity arises from the square-root branch cut in the complex $\omega$-plane, which does not violate causality.

To identify the branch points and corresponding branch cut introduced by the square root, we define the auxiliary quantity:
\begin{equation}
z(\omega) = \left(i \omega \alpha + \gamma(1 - \alpha)\right)^2 + 4 i \omega \alpha^2 \gamma.
\end{equation}
And this variable can be further reorganized into:
\begin{equation}
\begin{aligned}
z &= \left(i \omega \alpha + \gamma(1 - \alpha)\right)^2 + 4 i \omega \alpha^2 \gamma \\
&= -\omega^2 \alpha^2 + 2i\gamma \alpha(1+\alpha)\omega + \gamma^2(1-\alpha)^2 \\
&= - \alpha^2 \left(\omega - \frac{i\gamma (1+\alpha) + 2i \gamma\sqrt{\alpha}}{\alpha}\right)\left(\omega - \frac{i\gamma (1+\alpha) - 2i \gamma\sqrt{\alpha}}{\alpha}\right) \\
&= -\alpha^2 \left(\omega - i \gamma \frac{(1+\sqrt{\alpha})^2}{\alpha}\right)\left(\omega - i \gamma \frac{(1-\sqrt{\alpha})^2}{\alpha}\right).
\end{aligned}
\end{equation}
This shows that the square root $\sqrt{z}$ introduces two branch points at $\omega = i \gamma \frac{(1 \pm \sqrt{\alpha})^2}{\alpha}$, defining a branch cut along the segment:
$$
\left[ i \gamma \frac{(1 - \sqrt{\alpha})^2}{\alpha}, \, i \gamma \frac{(1 + \sqrt{\alpha})^2}{\alpha} \right].
$$
The entire branch cut lies within the upper half-plane, implying that $\tilde{R}(\omega)$ is analytic in the lower half-plane. This guarantees that the time-domain response function $R(\tau)$ vanishes for $\tau < 0$, thus preserving causality. The branch cut, located along the imaginary axis, corresponds to a continuous range of complex frequencies that contribute to the system's response. Its length is given by $i \gamma \cdot \frac{4}{\sqrt{\alpha}}$, which becomes shorter as $\alpha$ increases, and reaches its minimum at the critical point $\alpha_c=1$. This shrinking of the branch cut reflects a narrowing of the active frequency range, indicating that the response becomes more sharply localized in both time and frequency near criticality.

\paragraph{Case $\alpha > 1$:}  
In the regime $\alpha > 1$, the response function retains the same functional form as in Equation~\ref{eq:smalla}, but the square root must now be evaluated on the negative branch due to the sign change in $1 - \alpha < 0$, which alters the analytic structure in the complex $\omega$-plane.
\begin{equation}
\begin{aligned}
\tilde{R}(\omega) = \frac{ - (i \omega \alpha + \gamma(1 - \alpha)) + \sqrt{(i \omega \alpha + \gamma(1 - \alpha))^2 + 4 i \omega \alpha^2 \gamma} }{2 i \omega \alpha \gamma}.
\label{eq:largea}
\end{aligned}
\end{equation}
In this case, $\omega = 0$ is a genuine \textit{simple pole} of $\tilde{R}(\omega)$, which indicates a persistent long-time memory in the time domain. The non-analytic structure of the response function therefore consists of both a simple pole at $\omega = 0$ and a square-root branch cut located along the interval
$$
\left[ i \gamma \frac{(1 - \sqrt{\alpha})^2}{\alpha}, \, i \gamma \frac{(1 + \sqrt{\alpha})^2}{\alpha} \right],
$$
which is the same as in the $\alpha < 1$ regime. The appearance of a pole at the origin in the expression for $\tilde{R}(\omega)$ when $\alpha > 1$ does not signal a violation of causality, but rather indicates that the derived response function is incomplete, because it omits the contribution at $\omega = 0$. To complete the response function and ensure consistency with its causal and analytic structure, one must incorporate the missing term using the Kramers--Kronig relations, which prescribe the appropriate distributional correction at $\omega = 0$.

\paragraph{Case $\alpha = 1$:}  
The analytic structure of the response function at the critical point $\alpha = 1$ exhibits distinct behavior compared to the $\alpha \neq 1$ cases. The response function simplifies to:
\begin{equation}
\tilde{R}(\omega) = \frac{ - i\omega + \sqrt{-\omega^2 + 4 i \omega \gamma} }{2 i \omega \gamma},
\label{eq: aone}
\end{equation}
where the branch points occur at $\omega = 0$ and $\omega = 4 i \gamma$. Consequently, the branch cut lies along the vertical segment
$$
\left[ 0, \, 4 i \gamma \right].
$$
Although $\omega = 0$ leads to a divergence of the form $\tilde{R}(\omega) \sim \omega^{-1/2}$, this singularity corresponds to a branch point rather than a true pole. This square-root non-analyticity reflects critical slowing down and marks the onset of a non-exponential but power-law decay in the time domain. However, we do not focus on the behavior exactly at the critical point, since the theoretical framework breaks down there. At $\alpha = 1$, key assumptions such as linear response, ergodicity, and time-scale separation no longer hold, and quantities like the susceptibility diverge. As a result, standard perturbative or mean-field approaches become invalid, and a more refined treatment would be required to capture the system's dynamics at criticality\cite{goldenfeld2018lectures}.
\subsection{Kramers–Kronig Relations and Causality}
\label{sec:KK}
A fundamental requirement for any physical response function is \textit{causality}—the principle that a system cannot respond before a perturbation is applied. Mathematically, this implies that the time-domain response function $R(\tau)$ must vanish for $\tau < 0$. When the response function is transformed into the frequency domain via a Fourier transform, causality imposes strong constraints on the analytic structure of the complex response function $\tilde{R}(\omega)$. Specifically, $\tilde{R}(\omega)$ must be analytic in the lower half of the complex $\omega$-plane, as dictated by the Fourier convention adopted in Equation~\ref{eq:Fourier}, which ensures that the time-domain response function $R(\tau)$ is causal.  This analyticity gives rise to the celebrated \textit{Kramers–Kronig relations}\cite{toll1956causality,jackson2021classical}, which link the real part $R^{\prime}(\omega) $ and imaginary part $R^{\prime \prime}(\omega)$ of $\tilde{R}(\omega)$ through Hilbert transforms:
 \begin{equation}
\begin{aligned}
R^{\prime}(\omega) & =\mathcal{P} \int_{-\infty}^{\infty} \frac{d z}{\pi} \frac{R^{\prime \prime}(z)}{\omega-z}, \\
R^{\prime \prime}(\omega) & =-\mathcal{P} \int_{-\infty}^{\infty} \frac{d z}{\pi} \frac{R^{\prime}(z)}{\omega-z}.
\end{aligned}
 \end{equation}
These relations encode the entire function's behavior from its imaginary or real component alone, and serve as a powerful consistency check for physical theories. In our analysis, we will verify that the derived response function satisfies the Kramers–Kronig relations, thereby confirming that our theoretical framework respects causality.  

We now derive the Kramers–Kronig relations for the complex response function $\tilde{R}(\omega)$, which, under the Fourier convention specified in Equation~\ref{eq:Fourier}, is analytic in the lower half of the complex $\omega$-plane when $\alpha < 1$. Consider the complex function
\begin{equation}
\tilde{F}(\omega) = \frac{\tilde{R}(\omega)}{\omega - z - i\epsilon},
\end{equation}
where $z \in \mathbb{R}$ and $\epsilon \to 0$. We integrate $\tilde{F}(\omega)$ along a closed contour $\mathcal{C}$ that consists of the real axis and a semicircle in the lower half-plane, as introduced in Section~\ref{sec:pole}:
\begin{equation}
\oint_{\mathcal{C}} \frac{d \omega}{2 \pi i} \, \tilde{F}(\omega) = \oint_{\mathcal{C}} \frac{\tilde{R}(\omega)}{\omega - z - i \epsilon} \, d\omega = 0,
\end{equation}
by Cauchy’s theorem, since $\tilde{F}(\omega)$ is analytic within $\mathcal{C}$. We now decompose the contour integral into two parts:
\begin{equation}
\oint_{\Omega_1} \frac{\tilde{R}(\omega)}{\omega - z - i \epsilon} \, d\omega - \int_{-r}^{r} \frac{\tilde{R}(\omega)}{\omega - z - i \epsilon} \, d\omega = 0,
\end{equation}
where $\Omega_1$ denotes the semicircular arc with radius $r \to \infty$ in the low-half plane. The first term vanishes in this limit, yielding
\begin{equation}
\int_{-r}^{r} \frac{\tilde{R}(\omega)}{\omega - z - i \epsilon} \, d\omega = 0.
\end{equation}
Next, we decompose the response function into real and imaginary parts,
\begin{equation}
\tilde{R}(\omega) = \tilde{R}^\prime(\omega) + i \tilde{R}^{\prime\prime}(\omega),
\end{equation}
and apply the Sokhotski–Plemelj identity,
\begin{equation}
\frac{1}{x \pm i \epsilon} = \mathcal{P}\left( \frac{1}{x} \right) \mp i \pi \delta(x),
\end{equation}
to rewrite the integral as
\begin{equation}
\int_{-r}^{r} \left[ \tilde{R}^\prime(\omega) + i \tilde{R}^{\prime\prime}(\omega) \right] \left[ \mathcal{P} \left( \frac{1}{\omega - z} \right) + i \pi \delta(\omega - z) \right] \, d\omega = 0.
\end{equation}
Here, $\mathcal{P}$ denotes the Cauchy principal value, defined by
\begin{equation}
\mathcal{P} \int_{-\infty}^{+\infty} \frac{\eta(\omega')}{\omega' - \omega} \, d\omega' = \lim_{\varepsilon \to 0} \left[ \int_{-\infty}^{\omega - \varepsilon} \frac{\eta(\omega')}{\omega' - \omega} \, d\omega' + \int_{\omega + \varepsilon}^{+\infty} \frac{\eta(\omega')}{\omega' - \omega} \, d\omega' \right].
\end{equation}
Taking the limit $r \to \infty$, we obtain
\begin{equation}
\int_{-\infty}^{\infty} \left[ \tilde{R}^\prime(\omega) + i \tilde{R}^{\prime\prime}(\omega) \right] \left[ \mathcal{P} \left( \frac{1}{\omega - z} \right) + i \pi \delta(\omega - z) \right] \, d\omega = 0,
\end{equation}
which implies
\begin{equation}
i \pi \left[ \tilde{R}^\prime(z) + i \tilde{R}^{\prime\prime}(z) \right] = - \int_{-\infty}^{\infty} \left[ \tilde{R}^\prime(\omega) + i \tilde{R}^{\prime\prime}(\omega) \right] \mathcal{P} \left( \frac{1}{\omega - z} \right) \, d\omega.
\end{equation}
Separating the real and imaginary parts, we arrive at the Kramers–Kronig relations:
\begin{equation}
\tilde{R}^\prime(z) = \frac{1}{\pi} \mathcal{P} \int_{-\infty}^{\infty} \frac{\tilde{R}^{\prime\prime}(\omega)}{z - \omega } \, d\omega, \qquad
\tilde{R}^{\prime\prime}(z) =- \frac{1}{\pi} \mathcal{P} \int_{-\infty}^{\infty} \frac{\tilde{R}^\prime(\omega)}{z - \omega} \, d\omega.
\label{eq:KK}
\end{equation}

Next, we examine the response function $\tilde{R}(\omega)$ in the regime $\alpha > 1$, where its expression is given by Equation~\ref{eq:largea}. In this regime, $\tilde{R}(\omega)$ appears to exhibit a simple pole at $\omega = 0$, which may seem inconsistent with the Kramers--Kronig relations and the principle of causality. However, this apparent inconsistency arises because the analytical expression for $\tilde{R}(\omega)$ was derived under the assumption $\omega \neq 0$, and the contribution from the zero-frequency component has not been properly included. To complete the response function and ensure consistency with analyticity in the lower-half complex $\omega$-plane, the missing $\omega = 0$ contribution must be incorporated explicitly via the Kramers--Kronig relations, which prescribe the appropriate distributional correction.

When the imaginary part of the response function has a principal–value pole at the origin (here for $\alpha>1$), the Kramers--Kronig relations indicates the existence of a distributional contribution in the real part at $\omega=0$. Around $\omega=0$, we can expand the response function in frequency domain:
\begin{equation}
\tilde R(\omega)\approx \frac{f(0)}{i\omega}+ \tilde R_{\mathrm{reg}}(\omega),
\qquad f(0)=\frac{\alpha-1}{\alpha},
\end{equation}
where $\tilde R_{\mathrm{reg}}$ is regular at $\omega=0$ and $f$ is smooth. Since $1/(i\omega)=-\,i\,\mathrm{PV}(1/\omega)$, the singular piece contributes
\begin{equation}
\tilde R''(\omega) = \operatorname{Im} \tilde R(\omega) = -\,f(0)\,\mathrm{PV}\!\Big(\frac{1}{\omega}\Big) + \text{regular}.
\end{equation}
By the Kramers–Kronig relations shown in Equation~\ref{eq:KK}, the real part of the response function $\tilde R'(z)$ can be obtained by:
\begin{equation}
\tilde R'(z)= -\,\frac{f(0)}{\pi}\,\mathcal{P}\!\!\int_{-\infty}^{\infty}\frac{\mathrm{PV}(1/\omega)}{z-\omega}\,d\omega \;+\; \frac{1}{\pi}\,\mathcal{P}\!\!\int_{-\infty}^{\infty}\frac{\text{regular}}{z-\omega}\,d\omega.
\end{equation}
We now evaluate the principal–value convolution
\begin{equation}
\label{eq:I-def}
I(z):=\frac{1}{\pi}\,\mathcal{P}\!\!\int_{-\infty}^{\infty}\frac{\mathrm{PV}(1/\omega)}{z-\omega}\,d\omega.
\end{equation}

\paragraph{Step 1: $I(z)=0$ for $z\neq 0$.}
For $z\neq0$, the partial fractions identity
\begin{equation}
\frac{1}{\omega(z-\omega)}=\frac{1}{z}\left(\frac{1}{\omega}+\frac{1}{z-\omega}\right),
\end{equation}
where
\begin{equation}
\mathcal{P}\!\!\int_{-\infty}^{\infty}\frac{d\omega}{\omega}=0,
\qquad
\mathcal{P}\!\!\int_{-\infty}^{\infty}\frac{d\omega}{z-\omega}=0,
\end{equation}
one obtains $I(z)=0$ for all $z\neq0$. Therefore $I(z)=C\,\delta(z)$ for some constant $C$.

\paragraph{Step 2: Determine $C$ by integrating across $z=0$.}
For any $a>0$,
\begin{equation}
\begin{aligned}
J(a)&:=\int_{-a}^{a} I(z)\,dz
= \frac{1}{\pi}\,\mathcal{P}\!\!\int_{-\infty}^{\infty}\frac{d\omega}{\omega}
\left(\mathcal{P}\!\!\int_{-a}^{a}\frac{dz}{z-\omega}\right) \\
&= \frac{1}{\pi}\,\mathcal{P}\!\!\int_{-\infty}^{\infty}\frac{d\omega}{\omega}
\log\left|\frac{a-\omega}{a+\omega}\right|.
\end{aligned}
\end{equation}
Rescale $\omega=at$ to get
\begin{equation}
J(a)=\frac{1}{\pi}\,\mathcal{P}\!\!\int_{-\infty}^{\infty}\frac{dt}{t}
\log\left|\frac{1-t}{1+t}\right|
= \frac{4}{\pi}\int_{0}^{1}\frac{dt}{t}\log\left(\frac{1-t}{1+t}\right).
\end{equation}
With $t=\tanh y$ (so $dt=\mathrm{sech}^{2}y\,dy$ and $\log\frac{1-t}{1+t}=-2y$), this becomes
\begin{equation}
J(a)= -\,\frac{8}{\pi}\int_{0}^{\infty}\frac{y}{\sinh(2y)}\,dy
= -\,\frac{4}{\pi}\int_{0}^{\infty}\frac{u}{\sinh u}\,du \quad (u=2y).
\end{equation}
Using the convergent series $1/\sinh u = 2\sum_{n=0}^{\infty}e^{-(2n+1)u}$,
\begin{equation}
\int_{0}^{\infty}\frac{u}{\sinh u}\,du
=2\sum_{n=0}^{\infty}\int_{0}^{\infty}u\,e^{-(2n+1)u}\,du
=2\sum_{n=0}^{\infty}\frac{1}{(2n+1)^{2}}
=\frac{\pi^{2}}{4}.
\end{equation}
Hence $J(a)=-\pi$ for every $a>0$. Therefore $C=-\pi$ and
\begin{equation}
\label{eq:hilbert-delta}
I(z) = -\,\pi\,\delta(z).
\end{equation}

\paragraph{Consequence for the real part.}
Combining \eqref{eq:I-def} and \eqref{eq:hilbert-delta},
\begin{equation}
\tilde R'(z) = -\,f(0)\,I(z) + \text{(regular Hilbert transform)}
= \pi\,f(0)\,\delta(z) + \text{regular}.
\end{equation}
With $f(0)=(\alpha-1)/\alpha$, the $\delta$-mass is $\pi(\alpha-1)/\alpha$. In complete form, the response function for $\alpha>1$ is
\begin{equation}
\tilde R(\omega) =  \frac{ - (i \omega \alpha + \gamma(1 - \alpha)) + \sqrt{(i \omega \alpha + \gamma(1 - \alpha))^2 + 4 i \omega \alpha^2 \gamma} }{2 i \omega \alpha \gamma}+\;\pi\,\frac{\alpha-1}{\alpha}\,\delta(\omega),
\label{eq:realdelta}
\end{equation}
which obeys the Kramers--Kronig relations \eqref{eq:KK} and ensures causality in the time domain. The Dirac mass precisely compensates the principal–value pole in $\tilde R''(\omega)$ at $\omega=0$; higher–order terms in the expansion of $\tilde{F}(\omega)$ are regular and contribute only smooth Hilbert transforms to $\tilde R'(\omega)$.

\subsection{Dynamical Sum Rules}

We now present a unified expression for the response function $\tilde{R}(\omega)$ across the regimes $\alpha < 1$ and $\alpha > 1$, highlighting the structural symmetry and the origin of a dynamical sum rule. Specifically, the response functions can be written as:
\begin{equation}
\begin{aligned}
\tilde{R}(\omega)_{\alpha<1} &= \frac{ - (i \omega \alpha + \gamma(1 - \alpha)) + \sqrt{(i \omega \alpha + \gamma(1 - \alpha))^2 + 4 i \omega \alpha^2 \gamma} }{2 i \omega \alpha \gamma},\\
\tilde{R}(\omega)_{\alpha>1} &= \frac{ - (i \omega \alpha + \gamma(1 - \alpha)) +\sqrt{(i \omega \alpha + \gamma(1 - \alpha))^2 + 4 i \omega \alpha^2 \gamma} }{2 i \omega \alpha \gamma} + \pi \frac{\alpha - 1}{\alpha} \delta(\omega).
\label{eq:causalR}
\end{aligned}
\end{equation}
Based on the formula $\sqrt{a+b \mathbf{i}}= c+d \mathbf{i}$, then we can solve $c$ and $d$ as: $c=\sqrt{\frac{a+\sqrt{a^2+b^2}}{2}}$ and $d=\operatorname{sgn}(b) \cdot \sqrt{\frac{-a+\sqrt{a^2+b^2}}{2}}$. In this case, we see the real and imaginary part of the response function here:
\begin{equation}
\begin{aligned}
\operatorname{Re}(\tilde{R}(\omega)_{\alpha<1}) & =-\frac{1}{2 \gamma} + \frac{\operatorname{sgn}(\omega)}{2 \omega \alpha \gamma} \sqrt{\frac{-a+\sqrt{a^2+b^2}}{2}}, \\
& =-\frac{1}{2 \gamma} + \frac{1}{2|\omega| \alpha \gamma} \sqrt{\frac{-a+\sqrt{a^2+b^2}}{2}}, \\
\operatorname{Re}(\tilde{R}(\omega)_{\alpha>1}) & =-\frac{1}{2 \gamma} + \frac{1}{2|\omega| \alpha \gamma} \sqrt{\frac{-a+\sqrt{a^2+b^2}}{2}}+ \pi \frac{\alpha - 1}{\alpha} \delta(\omega), \\
\operatorname{Im}(\tilde{R}(\omega)_{\alpha<1}) & =\operatorname{Im}(\tilde{R}(\omega)_{\alpha>1})  = \frac{1-\alpha}{2 \omega \alpha} + \frac{-\sqrt{\frac{a+\sqrt{a^2+b^2}}{2}}}{2 \omega \alpha \gamma} \\
a & =-\omega^2 \alpha^2+\gamma^2(1-\alpha)^2 \\
b & =2 \omega \alpha(1+\alpha) \gamma.
\end{aligned}
\end{equation}
Interestingly, we find that the real part of the response function satisfies a nontrivial sum rule \new{over the full range of $\omega$}:
\begin{equation}
\frac{1}{\pi} \int_{-\infty}^{+\infty} \operatorname{Re}\bigl[\tilde{R}(\omega)\bigr]_{\alpha<1}\, d\omega 
= \frac{1}{\pi} \int_{-\infty}^{+\infty} \operatorname{Re}\bigl[\tilde{R}(\omega)\bigr]_{\alpha>1}\, d\omega 
= 1.
\end{equation}
This identity indicates that the total spectral weight of the response function is conserved across the interpolation threshold, even though its distribution changes qualitatively. 

% The analogy extends to a sum rule. In superconductors, the Ferrell–Glover–Tinkham sum rule states that finite-frequency spectral weight is suppressed by the energy gap and transferred into a zero-frequency $\delta$-function \cite{tinkham2004introduction}, reflecting the formation of a condensate that carries current without dissipation. Likewise, across the double descent transition we obtain (see sec.~VI.C of the Supplemental Material \cite{SM})
% \begin{equation}
% \frac{1}{\pi}\!\int_{0^{+}}^{\infty}\!\!\!\mathrm{Re}[\tilde{R}(\omega)]_{\alpha<1}\,d\omega
% = \frac{1}{\pi}\!\int_{0^{+}}^{\infty}\!\!\!\mathrm{Re}[\tilde{R}(\omega)]_{\alpha>1}\,d\omega
% + \frac{\alpha-1}{2\alpha}.
% \end{equation}
% This identity captures the redistribution of spectral weight in $\tilde{R}(\omega)$ across $\alpha=1$, with the extra term $\tfrac{\alpha-1}{2\alpha}$ representing the emergent $\delta$-function contribution (see sec.~VI.B of the Supplemental Material \cite{SM}). Just as the superconducting condensate reflects the fraction of electrons locked into a dissipationless state, this contribution reflects the fraction of learning dynamics condensed into a persistent memory of the initial condition, which becomes dominant deep in the over-parameterized regime.

A close analogy can be drawn with the celebrated Ferrell--Glover--Tinkham sum rule in superconductivity \cite{ferrell1958conductivity,tinkham1959determination,tinkham2004introduction}, which states that finite-frequency spectral weight is suppressed by the opening of the superconducting gap and reappears as a zero-frequency $\delta$-function. This $\delta$-function encodes the condensate response that supports dissipationless supercurrents. \new{Explicitly, focusing only on the regime of positive $\omega$}:
\begin{equation}
\frac{1}{\pi} \int_{0^+}^{\infty} \! d\omega \,\sigma_N(\omega)
= \frac{1}{\pi} \int_{0^+}^{\infty} \! d\omega \,\sigma_S(\omega) 
+ \frac{n_s e^2}{2m},
\end{equation}
where $\sigma_N$ and $\sigma_S$ denote the normal and superconducting conductivities, respectively, and $n_s$ is the superfluid density. 

In direct analogy, the learning dynamics exhibit a redistribution of spectral weight across the interpolation threshold $\alpha=1$. The sum rule takes the form
\begin{equation}
\frac{1}{\pi} \int_{0^+}^{\infty} \! d\omega \,\operatorname{Re}\bigl[\tilde{R}(\omega)\bigr]_{\alpha<1}
= \frac{1}{\pi} \int_{0^+}^{\infty} \! d\omega \,\operatorname{Re}\bigl[\tilde{R}(\omega)\bigr]_{\alpha>1}
+ \frac{\alpha-1}{2\alpha}.
\end{equation}
The additional $\delta$-function contribution at $\omega=0$ in the $\alpha>1$ regime reflects the onset of ergodicity breaking: part of the spectral weight condenses into a persistent response, just as part of the optical spectral weight condenses into the superfluid component in a superconductor. \new{Just as the superconducting condensate reflects the fraction of electrons locked into a dissipationless state, this contribution reflects the fraction of learning dynamics condensed into a persistent memory of the initial condition, which becomes dominant deep in the over-parameterized regime. } This provides a spectral characterization of the transition, where ergodicity breaking in learning plays the role of superfluid density in superconductivity.  \new{We will say a bit more about this in Sec.~\ref{sec:superconducting}.}

\subsection{Fluctuation–Dissipation Theorem (FDT)}
\label{sec:FDT}
We now present a detailed analysis grounded in the fluctuation–dissipation theorem (FDT), demonstrating that the dynamics of stochastic gradient descent (SGD) exhibit a transition from equilibrium to non-equilibrium behavior as the model complexity parameter $\alpha$ crosses the interpolation threshold at $\alpha = 1$ in the unregularized ($\lambda=0$) linear regression setting.

The fluctuation-dissipation theorem (FDT) is a cornerstone of nonequilibrium statistical mechanics, relating a system's linear response to external perturbations with its spontaneous fluctuations in thermal equilibrium. It holds under the assumptions of time-translational invariance and microscopic reversibility, and captures the idea that dissipation (response) and noise (fluctuations) stem from the same underlying thermal motion\cite{kubo1957statistical,goldenfeld2018lectures,marconi2008fluctuation,kubo1966fluctuation}. In equilibrium, this leads to universal relations between correlation and response functions, both in time and frequency domains.

Under time-translational invariance (TTI), the dynamics of the evolving variable $x(t)$, for instance, $x(t) = \hat{\beta}(t) - \beta$, can be characterized by the centered two-time correlation function
\begin{equation}
C_{xx}(t_1, t_2) = \left\langle \left(x(t_1) - \langle x(t_1) \rangle\right) \left(x(t_2) - \langle x(t_2) \rangle\right) \right\rangle,
\end{equation}
and the retarded response function
\begin{equation}
R_{xx}(t_1, t_2) = \left. \frac{\delta \langle x(t_1) \rangle}{\delta j(t_2)} \right|_{j=0},
\end{equation}
where $j(t)$ is a small external perturbation to the dynamics of $x(t)$.

Under TTI, both $C_{xx}(t_1, t_2)$ and $R_{xx}(t_1, t_2)$ depend only on the time difference $\tau \equiv t_1 - t_2$, and we can write:
\begin{equation}
\text{TTI:} \quad 
\left\{
\begin{aligned}
C_{xx}(t_1, t_2) &= C_{xx}(\tau), \\
R_{xx}(t_1, t_2) &= R_{xx}(\tau).
\end{aligned}
\right.
\end{equation}

In equilibrium, these functions satisfy the fluctuation-dissipation theorem (FDT), which relates spontaneous fluctuations to linear response\cite{castellani2005spin,zamponi2005fluctuation}:
\begin{equation}
R_{xx}(\tau) = -\frac{\theta(\tau)}{T} \frac{d C_{xx}(\tau)}{d\tau},
\quad
\frac{d C_{xx}(\tau)}{d \tau} = T \left[ R_{xx}(-\tau) - R_{xx}(\tau) \right],
\end{equation}
where $T$ denotes an effective temperature, and $\theta(\tau)$ is the Heaviside step function enforcing causality.

Taking the Fourier transform of the second identity yields:
\begin{equation}
\begin{aligned}
i\omega\, \tilde{C}_{xx}(\omega) &= T \left[ \tilde{R}_{xx}(-\omega) - \tilde{R}_{xx}(\omega) \right], \\
&= -2iT\, \operatorname{Im} \left( \tilde{R}_{xx}(\omega) \right), \\
\Rightarrow \quad \tilde{C}_{xx}(\omega) &= -\frac{2T}{\omega} \operatorname{Im} \left( \tilde{R}_{xx}(\omega) \right),
\end{aligned}
\end{equation}
where $\tilde{C}_{xx}(\omega)$ and $\tilde{R}_{xx}(\omega)$ are the Fourier transforms of the correlation and response functions, respectively. 

This frequency-domain form of the FDT shows that the imaginary part of the response function encodes the spectral density of fluctuations, ensuring detailed balance and relaxation toward a stationary equilibrium state.  
To illustrate this relation for a different dynamical variable, consider the case where the evolving quantity is the velocity $v(t)$ rather than the displacement $x(t)$.  
In Fourier space, the two are related by $\tilde{v}(\omega) = i\omega \tilde{x}(\omega)$, leading to the corresponding relations between their response and correlation functions:
\begin{equation}
\tilde{C}_{vv}(\omega) = \omega^2 \tilde{C}_{xx}(\omega), 
\qquad
\tilde{R}_{vv}(\omega) = i\omega \tilde{R}_{xx}(\omega).
\end{equation}
Under this transformation, the fluctuation–dissipation theorem becomes
\begin{equation}
\tilde{C}_{vv}(\omega) = 2T\,\operatorname{Re}\tilde{R}_{vv}(\omega).
\label{eq:vFDT}
\end{equation}

To characterize the dynamics of the variable $x(t) = \hat{\beta}(t) - \beta$ in the unregularized stochastic gradient descent (SGD) learning of a linear regression task, we begin by rewriting its evolution equation, derived from Equation~\ref{eq:dynamics}:
\begin{equation}
\partial_t {x}(t) = -\frac{\gamma}{\alpha} \int dt^{\prime} \, (\gamma R + \delta)^{-1}(t, t^{\prime})\, x(t^\prime) + \eta(t),
\end{equation}
where $\eta(t)$ is a noise term with zero mean, $\langle \eta(t) \rangle = 0$, and two-time correlation 
\begin{equation}
\langle \eta(t_1) \eta(t_2) \rangle = \hat{D}(t_1, t_2).
\end{equation}

Following the Green's function approach introduced in Section~\ref{sec:solveC}, the solution of the linear stochastic equation above can be expressed as:
\begin{equation}
x(t) = \int_0^t dt^\prime \, R(t, t^\prime)\, \eta(t^\prime) + x(0) R(t),
\end{equation}
where $x(0) = \hat{\beta}(0) - \beta$ encodes the initial condition, and $R(t, t^\prime)$ is the retarded Green's function (or response function).

From this solution, we can compute the centered two-time correlation function:
\begin{equation}
\begin{aligned}
C_{xx}(t_1, t_2) &= \langle x(t_1)\, x(t_2) \rangle - \langle x(t_1) \rangle \langle x(t_2) \rangle \\
&= C(t_1, t_2) - \langle \beta \rangle^2 R(t_1)\, R(t_2),\\
\end{aligned}
\end{equation}
where $C(t_1, t_2)$ is the full correlation function of $\hat{\beta}(t)$, and we have used the fact that $\langle x(0) \rangle = -\langle \beta \rangle$ under the assumption $\langle \hat{\beta}(0) \rangle = 0$.

The retarded response function $R_{xx}(t_1,t_2)$ for $v(t)$ obeys the same dynamics as Eq.~\ref{eq:response}, reflecting both linearity and causality. 
Assuming time-translation invariance (TTI), we write $R_{xx}(t_1,t_2)=R_{xx}(t_1-t_2)$ and $C_{xx}(t_1,t_2)=C_{xx}(t_1-t_2)$. 
To compute $C_{xx}(t_1-t_2)$, we first evaluate $C(t_1-t_2)$ in the Fourier domain. 

We extend the temporal domain to $(-\infty,\infty)$ and perform a Fourier transform with respect to $t_1$ and $t_2$, the corresponding Fourier variables are $\omega_1, \omega_2$. 
Unlike the Laplace formulation, which retains transient dynamics, the Fourier approach isolates the stationary component. 
The Fourier representation of the correlation function is
\begin{equation}
\mathcal{F}[C(t_1,t_2)](\omega_1,\omega_2)
= \int_{-\infty}^\infty dt_1 \int_{-\infty}^\infty dt_2\, C(t_1,t_2)\, e^{-i\omega_1 t_1-i\omega_2 t_2}.
\end{equation}

Next, we analyze the unregularized dynamics $\lambda=0$ and introduce the kernel
\begin{equation}
K(t_1-t_2) = \int_{t_2}^{t_1} dt_0\, \gamma R(t_1-t_0)\,(\gamma R+\delta)^{-1}(t_0-t_2).
\end{equation}
Since the zero-frequency component of $C(t_1,t_2)$ does not affect the fluctuation–dissipation theorem (FDT), we can redefine $C(t_1,t_2)+D \to C(t_1,t_2)$ for simplicity. 
Equation~\ref{eq:correlation} for the correlation function then becomes
\begin{equation}
\begin{aligned}
C(t_1,t_2) &=\int_{-\infty}^{t_1} dt^{\prime\prime} \int_{-\infty}^{t_2} dt_0\, 
K(t_1-t^{\prime\prime})\, C(t^{\prime\prime},t_0)\, K(t_2-t_0) \\
&\quad + C(0,0) R(t_1) R(t_2) + D,\\
\mathcal{F}[C(t_1,t_2)] (\omega_1,\omega_2) 
&= \frac{(2 \pi)^2 D \delta(\omega_1) \delta(\omega_2) + C(0,0)\tilde{R}(\omega_1)\tilde{R}(\omega_2)}{1-\tfrac{1}{\alpha}\tilde{K}(\omega_1)\tilde{K}(\omega_2)},
\end{aligned}
\end{equation}
where the Fourier transform of the kernel is
\begin{equation}
\tilde{K}(\omega)=\int_0^\infty K(t)\, e^{-i\omega t}\, dt
= \frac{\gamma \tilde{R}(\omega)}{\gamma\tilde{R}(\omega)+1}.
\end{equation}
At zero frequency, $\tilde{K}(0)=\alpha$ for $\alpha<1$ and $\tilde{K}(0)=1$ for $\alpha>1$. 
Hence the $\delta$-function term in $C(\omega_1, \omega_2)$ contributes only at zero frequency and does not enter the FDT. 
The effective part for correlation function $C(t_1, t_2)$ in Fourier domain is therefore
\begin{equation}
\begin{aligned}
\mathcal{F}[C^{\mathrm{eff}}(t_1,t_2)] (\omega_1,\omega_2) 
&= \frac{C(0,0)\tilde{R}(\omega_1)\tilde{R}(\omega_2)}{1-\tfrac{1}{\alpha}\tilde{K}(\omega_1)\tilde{K}(\omega_2)}.
\end{aligned}
\end{equation}
Next, we define the operator
\begin{equation}
\tilde{K}^{\prime}(\omega_1, \omega_2) =  \frac{1}{1-\tfrac{1}{\alpha}\tilde{K}(\omega_1)\tilde{K}(\omega_2)},
\end{equation}
so that the effective part of the correlation function in the time domain can be written as
\begin{equation}
C^{\mathrm{eff}}(t_1,t_2) = C(0,0)\int_{-\infty}^{t_1}ds \int_{-\infty}^{t_2}dt_0 \, R(t_1- s)\, K^{\prime}(s, t_0)\, R(t_2- t_0).
\end{equation}
If the time-translational invariance condition holds in the long-time limit, the kernel must reduce to a function of the time difference,
\begin{equation}
K^{\prime}(s,t_0) = K^{\prime}(s-t_0),
\end{equation}
which implies
\begin{equation}
C^{\mathrm{eff}}(t_1-t_2) = C(0,0)\int_{-\infty}^{t_1}ds \int_{-\infty}^{t_2}dt_0 \, R(t_1- s)\, K^{\prime}(s-t_0)\, R(t_2- t_0).
\end{equation}
Fourier transforming with respect to $\tau = t_1-t_2$ gives
\begin{equation}
\tilde{C}^{\mathrm{eff}}(\omega)=C(0,0)\,\tilde{R}(\omega)\,\tilde{K}^{\prime}(\omega)\,\tilde{R}(-\omega).
\end{equation}
Since the correlation function is symmetric, $C^{\mathrm{eff}}(\tau) = C^{\mathrm{eff}}(-\tau)$, its Fourier transform is real and even, leading to
\begin{equation}
\begin{aligned}
\tilde{C}^{\mathrm{eff}}(-\omega)&=\tilde{C}^{\mathrm{eff}}(\omega),\\
\tilde{K}^{\prime}(\omega)&=\tilde{K}^{\prime}(-\omega).
\end{aligned}
\end{equation}
Therefore,
\begin{equation}
\tilde{K}^{\prime}(\omega) = \frac{1}{1-\tfrac{1}{\alpha}\tilde{K}(\omega)\tilde{K}(-\omega)}.
\end{equation}
The correlation function in the Fourier domain, with $\omega$ the conjugate variable to $\tau$, is then
\begin{equation}
\tilde{C}^{\mathrm{eff}}(\omega)=C(0,0)\,\tilde{R}(\omega)\,\frac{1}{1-\tfrac{1}{\alpha}\tilde{K}(\omega)\tilde{K}(-\omega)}\,\tilde{R}(-\omega).
\end{equation}
In the long-time limit $t_1,t_2\to\infty$, the extra term $\langle \beta\rangle^2 R(t_1)R(t_2)$ contributes only a constant term and therefore does not affect the fluctuation--dissipation relation (FDT).

\paragraph{Under-parameterized regime.}
For $\alpha<1$, the response function takes the form $\tilde{R}(\omega)_{\alpha<1}$ given in Eq.~\ref{eq:causalR}. In this regime the FDT is exactly satisfied with effective temperature $C(0,0)/2$:
\begin{equation}
\begin{aligned}
\tilde{C}_{xx}(\omega)
&= \frac{C(0,0)\,\tilde{R}(\omega)\tilde{R}(-\omega)}{1-\tfrac{1}{\alpha}\tilde{K}(\omega)\tilde{K}(-\omega)} \\[4pt]
&=C(0,0) \frac{2(-1+\alpha)\gamma+\sqrt{4 i \alpha^2 \gamma \omega+\bigl((-1+\alpha)\gamma-i \alpha \omega\bigr)^2}+\sqrt{-4 i \alpha^2 \gamma \omega+\bigl((-1+\alpha)\gamma+i \alpha \omega\bigr)^2}}{4 \alpha \gamma \omega^2} \\[4pt]
&=C(0,0) \frac{\tilde{R}(\omega)-\tilde{R}(-\omega)}{2i} \\[4pt]
&= -\frac{C(0,0)}{\omega}\,\operatorname{Im}\tilde{R}(\omega).
\end{aligned}
\end{equation}

\paragraph{Over-parameterized regime.}
For $\alpha > 1$, the response function takes the form $\tilde{R}(\omega)_{\alpha>1}$ given in Eq.~\ref{eq:causalR}. In this regime, an analogous expression for $\tilde{C}_{xx}(\omega)$ can be derived; however, the FDT is violated due to the pole structure of the response function, with the resulting additional $\delta(\omega)$ term as derived in Eq.~\ref{eq:realdelta} explicitly breaking the fluctuation–dissipation relation.
\begin{equation}
\begin{aligned}
\tilde{C}_{xx}(\omega)&= \frac{C(0,0)\,\tilde{R}(\omega)\tilde{R}(-\omega)}{1-\tfrac{1}{\alpha}\tilde{K}(\omega)\tilde{K}(-\omega)} \\& \not \propto
\operatorname{Im}\tilde{R}(\omega).
\label{eq:nFDT}
\end{aligned}
\end{equation}

\new{\subsection{Connections to Ergodicity and Ergodicity Breaking}}

\new{To help the reader better understand the significance of this analogy, we clarify that the phenomena discussed here, including ergodicity breaking and the associated dynamical phase transition, are not merely heuristic analogies, but arise directly from explicit calculations within our model when it is viewed as a physical system. In statistical physics, ergodicity refers to the ability of a dynamical systems to explore every part of its phase space over infinitely long times in equilibrium.  This property enables the replacement of time averages by ensemble averages, and is the foundation of Boltzmann and Gibbs's formulation of statistical mechanics (see, e.g. Section 2.10 of \cite{goldenfeld2018lectures}).}

\new{Ergodicity breaking refers to the failure of a system to explore its full phase space over long times, and is connected to the existence of a non-zero order parameter and other emergent phenomena below a critical temperature~\cite{goldenfeld2018lectures}  Without ergodicity breaking, a statistical mechanical system, such as a magnet, cannot exhibit a non-zero spontanous magnetization in the absence of an external field.  Below a critical temperature, the system breaks ergodicity and thus is constrained to remain in the part of its phase space that is in the vicinity of the initial condition, never venturing into regions of phase space that could cancel out emergent order.   In the learning dynamics studied here, ergodicity breaking manifests as a persistent dependence on initial conditions.  Once training enters the over-parameterized regime, the dynamics become confined to a restricted subspace of weight space determined by the initialization.  This confinement limits the network’s ability to fit noise while preserving its capacity to generalize, providing a concrete mechanism for good generalization in the unregularized, over-parameterized regime. As a result, different initial conditions lead to distinct long-time states that are not dynamically connected. This ergodicity breaking is reflected in a response function that decays at long times to a non-zero value.  As shown in the main paper, this has the same mathematical structure as the persistent current in superconductivity (see also below), and thus means that the system cannot be in an equilibrium state.  An equilibrium state would be a network that is ergodic, and because it fully explores its phase space, loses memory of the initial conditions at long times.  If the system is in the analogue of a stationary, current carrying state, then it cannot be in an equilibrium, ergodic state, and has a persistent memory of its initial condition.}

\new{More generally, the emergence of ergodicity breaking at the interpolation threshold reflects the presence of a genuine dynamical phase transition. In the vicinity of such a transition, macroscopic properties of the system become insensitive to microscopic details, a hallmark of universality in statistical physics. As the critical point is approached, the relevant dynamics are dominated by long-wavelength fluctuations\cite{tinkham2004introduction}, rendering local features of the learning rule or architecture largely irrelevant. Consequently, qualitatively different learning models can exhibit the same critical behavior near the transition, characterized by identical scaling laws and response properties. This universality explains why we are interested to view the phenomenology of double descent as a phase transition: when approaching criticality, the associated scaling behavior persists across a broad class of models, despite differences in their microscopic implementation.
}

\subsection{Connections to Superconducting Transition}
\label{sec:superconducting}

To further elucidate the dynamical phase transition underlying double descent, we draw an analogy to the superconducting transition in condensed matter physics\cite{tinkham2004introduction}. In both systems, qualitatively distinct dynamical regimes emerge on either side of a critical point: for superconductors, the transition to dissipationless transport occurs below a critical temperature; in our case, a transition from equilibrium to non-equilibrium dynamics arises as the parameter $\alpha$ crosses the interpolation threshold at $\alpha = 1$. In the overparameterized regime ($\alpha > 1$), the system exhibits persistent memory, breakdown of ergodicity, and vanishing dissipation, mirroring the superconducting phase.

The analogy becomes precise when comparing the response function in our model to the conductivity in Drude theory. In a conventional metal, the motion of conduction electrons can be described within the Drude model\cite{tinkham2004introduction}. The electron velocity $\mathbf{v}(t)$ evolves under a time-dependent electric field $\mathbf{E}(t)$ according to
\begin{equation}
m\,\frac{d\mathbf{v}}{dt}
= e\,\mathbf{E}(t) - \frac{m}{\tau}\,\mathbf{v},
\label{eq:drude_eq}
\end{equation}
where $m$ is the effective electron mass, $e$ the electron charge, and $\tau$ the momentum relaxation time characterizing scattering from impurities, phonons, or other electrons. The first term on the right-hand side represents acceleration due to the external electric field, while the second term introduces a phenomenological damping that drives the system back toward equilibrium.

For a monochromatic field $\mathbf{E}(t) = \mathbf{E}_0 e^{i \omega t}$, the steady-state solution of Eq.~\eqref{eq:drude_eq} yields
the following relation in the frequency domain,
\begin{equation}
\tilde{\mathbf{v}}(\omega)
= \frac{(e \tau / m)}{1 + i \omega \tau}\,\tilde{\mathbf{E}}(\omega).
\label{eq:v_freq}
\end{equation}
Defining the current density $\mathbf{J} = ne\,\mathbf{v}$, where $n$ is the carrier density, one obtains Ohm’s law in frequency space,
\begin{equation}
\mathbf{J}(\omega) = \tilde{\sigma}(\omega)\,\tilde{\mathbf{E}}(\omega),
\qquad
\tilde{\sigma}(\omega)
= \frac{n e^2 \tau / m}{1 + i \omega \tau},
\label{eq:drude_sigma}
\end{equation}
where $\mathrm{Re}\,\tilde{\sigma}(\omega)$ describes dissipative (Ohmic) transport, while $\mathrm{Im}\,\tilde{\sigma}(\omega)$ represents the reactive, inductive response of the electron fluid.

From a microscopic standpoint, Eq.~\eqref{eq:drude_eq} can be viewed as the deterministic limit of a Langevin equation that includes a stochastic force term $\boldsymbol{\xi}(t)$ representing thermal fluctuations:
\begin{equation}
m\,\frac{d\mathbf{v}}{dt}
= e\,\mathbf{E}(t)
- \frac{m}{\tau}\,\mathbf{v}
+ \boldsymbol{\xi}(t),
\qquad
\langle \xi_\alpha(t)\,\xi_\beta(t') \rangle
= 2\,\frac{m}{\tau}\,k_B T\,\delta_{\alpha\beta}\,\delta(t-t').
\label{eq:langevin}
\end{equation}
The noise correlation is fixed by the fluctuation--dissipation theorem, which ensures that the random force compensates the energy loss due to damping, maintaining thermal equilibrium. 

 The model of a superconductor used here is a simple London model \cite{london1935electromagnetic}. In the limit $\tau \to \infty$, scattering processes are completely suppressed and electrons accelerate indefinitely under the applied field, corresponding to a perfectly dissipationless state. In this regime,
\begin{equation}
\begin{aligned}
m\,\frac{d\mathbf{v}}{dt}
&= e\,\mathbf{E}(t).
\end{aligned}
\end{equation}
Physically, this limit describes the onset of the \emph{superconducting phase}, in which electrons move coherently without resistance and the fluctuation--dissipation relation breaks down due to the absence of relaxation and ergodicity.

We now show that the linear response function in our model plays an analogous role as the conductivity in Drude model. In the double descent dynamics, the average deviation $\Delta \hat{\beta}(t) \equiv \langle \hat{\beta}(t) - \beta\rangle$ under an external force $F(t)$ satisfies:
\begin{equation}
\partial_t \Delta {\hat{\beta}}(t) = - \frac{\gamma}{\alpha} \int_0^\infty dt_1 \left[ \gamma R + \delta \right]^{-1}(t - t_1) \Delta \hat{\beta}(t_1) +F(t),
\end{equation}
which becomes in Fourier space:
\begin{equation}
\Delta \hat{\beta}(\omega) = \frac{1}{i \omega + \frac{\gamma}{\alpha \left(\gamma \tilde{R}(\omega) + 1\right)}} \tilde{F}(\omega).
\end{equation}
Using the self-consistent equation for the response function,
\begin{equation}
i \omega \tilde{R}(\omega) = - \frac{\gamma \tilde{R}(\omega)}{\alpha\left( \gamma \tilde{R}(\omega) + 1\right)} + 1,
\end{equation}
we can simplify the prefactor:
\begin{equation}
\frac{1}{i \omega + \frac{\gamma}{\alpha \left(\gamma \tilde{R}(\omega) + 1\right)}} = \tilde{R}(\omega),
\end{equation}
which yields the linear response relation:
\begin{equation}
\Delta \hat{\beta}(\omega) = \tilde{R}(\omega) \tilde{F}(\omega).
\end{equation}
Thus, the response function $\tilde{R}(\omega)$ plays the role of an effective conductivity, directly analogous to $\tilde{\sigma}(\omega)$ in condensed matter systems: it quantifies how an external perturbation $\tilde{F}(\omega)$ induces a dynamical response $\Delta \hat{\beta}(\omega)$. In this sense, the interpolation threshold marks a dynamical transition similar to the normal-to-superconducting transition, with $\tilde{R}(\omega)$ replacing $\tilde{\sigma}(\omega)$ as the central response function. In what follows, we compare the double descent transition with the superconducting transition in detail.

\paragraph{Normal metal / Under-parameterized regime.}
In the under-parameterized regime ($\alpha<1$), the system obeys the fluctuation--dissipation theorem (FDT) at steady state, as derived in Sec.~\ref{sec:FDT}. This guarantees that the stationary dynamics are fully characterized by equilibrium fluctuations and linear response, placing the system in thermal equilibrium. Equivalently, the system follows an Ohm’s law: the induced response is proportional to the applied perturbation, with $\tilde{R}(\omega)$ acting as the effective conductivity. This behavior is directly analogous to an ordinary metallic state above the superconducting critical temperature $T_c$, where the conductivity is given by the Drude formula,
\begin{equation}
\tilde{\sigma}(\omega) = \frac{n e^2 \tau / m}{1 + i \omega \tau}.
\label{Conductivity}
\end{equation}
Here $n$ is the carrier density, $e$ the electron charge, $m$ the effective mass, and $\tau$ the scattering time. The key feature is that dissipation is finite, and transport is governed by scattering processes that maintain equilibrium. The correlation function for the electron velocity is defined as $C_{\text{e}}(t,t^\prime) = \langle \left(v(t)-\langle v(t)\rangle\right) \left(v(t^\prime) -\langle v(t^\prime)\rangle \right)\rangle$. To calculate this quantity, we consider the system with stochastic noise described in Eq.~\ref{eq:langevin}. The solution ${v}_i(t)$ for a single electron $i$ can be written as $v_i(t)=v_i(0) e^{-t / \tau}+\frac{e}{m} \int_0^t d s e^{-(t-s) / \tau} E(s)+\frac{1}{m} \int_0^t d s e^{-(t-s) / \tau} \xi_i(s)$. The correlation function takes the following form:
\begin{equation}
\begin{aligned}
    C_{\text{e}}(t,t^\prime) &= \langle \delta v(t) \delta v(t^\prime) \rangle,
\end{aligned}
\end{equation}
where $\delta v(t) = v(t) - \langle v(t)\rangle$ and $\langle \delta v(t) \delta v(t^\prime) \rangle = \frac{1}{n^2} \sum_{i, j}\left\langle \delta v_i(t) \delta v_j\left(t^{\prime}\right)\right\rangle$. Because the motion of electrons is independent, we can further simplify this quantity as $\left(t, t^{\prime} \geq 0\right)$:
\begin{equation}
\begin{aligned}
    C_{\text{e}}(t,t^\prime) &= \frac{1}{n^2} \sum_{i, j}\left\langle \delta v_i(t) \delta v_j\left(t^{\prime}\right)\right\rangle\\
    & = \frac{1}{n} \sum_{i}\left\langle \delta v_i(t) \delta v_i\left(t^{\prime}\right)\right\rangle\\
    & = \frac{1}{n} \frac{k_B T}{m} e^{-\left|t-t^{\prime}\right| / \tau}.
\end{aligned}
\end{equation}

Take Fourier transform to the variable $\Delta t = t-t^\prime$, and in Fourier domain, we can get:
\begin{equation}
    \begin{aligned}
\tilde{C}_{\text{e}}(\omega) & =\frac{1}{n} \frac{k_B T}{m} \int_{-\infty}^{\infty} d \Delta t e^{i \omega \Delta t} e^{-|\Delta t| / \tau} \\
& =\frac{1}{n} \frac{k_B T}{m} 2 \int_0^{\infty} d t e^{-t / \tau} \cos (\omega t) \\
& =\frac{1}{n} \frac{2 k_B T}{m} \frac{1 / \tau}{(1 / \tau)^2+\omega^2}=\frac{1}{n}\frac{2 k_B T}{m} \frac{\tau}{1+\omega^2 \tau^2}.
\end{aligned}
\end{equation}
The correlation function for the current density can be obtained from $C_{\text{e}}(t,t^\prime)$ by $C_{\text{J}}(t,t^\prime) = \langle J(t) J(t^\prime) \rangle - \langle J(t)\rangle \langle J(t^\prime)\rangle = n^2 e^2 C_{\text{e}}(t,t^\prime)$.
\begin{equation}
\begin{aligned}
    \tilde{C}_{\mathrm{J}}(\omega) &= n e^2 \frac{2 k_B T}{m} \frac{\tau}{1+\omega^2 \tau^2}.\\
\end{aligned}
\end{equation}

It is important to emphasize that the relevant dynamical variable in this context is the electron velocity $v(t)$ rather than the displacement $x(t)$. 
Accordingly, the fluctuation--dissipation theorem (FDT) takes the velocity form given in Eq.~\eqref{eq:vFDT}, where the current--current correlation spectrum is related to the real part of the complex conductivity:
\begin{equation}
\tilde{C}_{\mathrm{J}}(\omega)
= 2 k_B T\,\mathrm{Re}\,\tilde{\sigma}(\omega).
\label{eq:CjFDT}
\end{equation}
Above the superconducting transition temperature $T>T_c$, the system remains in thermal equilibrium and satisfies the FDT. 
In this regime, dissipative scattering processes ensure that fluctuations and dissipation are balanced, and the current fluctuations $\tilde{C}_{\mathrm{J}}(\omega)$ are directly proportional to the Ohmic response $\mathrm{Re}\,\tilde{\sigma}(\omega)$.

Similarly, in the $\alpha<1$ regime of double descent, fluctuations and dissipation balance each other through the FDT, ensuring that the dynamics remain reversible on average.
\newline
\paragraph{Superconductor / Over-parameterized regime.}
In the superconducting phase, the scattering time $\tau$ diverges. Taking the limit $\tau \to \infty$ while assuming $\omega \neq 0$ gives
\begin{equation}
\tilde{\sigma}(\omega) = \frac{n e^2/m}{i \omega}.
\end{equation}
This purely imaginary conductivity reflects an inductive response with no resistive dissipation. By analyticity, however, the conductivity also possesses a real part that follows from the Kramers--Kronig relations. The missing contribution at $\omega=0$ appears as a Dirac $\delta$-function,
\begin{equation}
\operatorname{Re}\,\tilde{\sigma}(\omega) = \frac{\pi n_s e^2}{m}\,\delta(\omega),
\end{equation}
where $n_s$ denotes the superfluid density. Physically, this means that a supercurrent induced by a transient field persists indefinitely, with zero dissipation.

A direct analogue arises in the learning dynamics. In the over-parameterized regime ($\alpha>1$), the response function has a $\delta$-function contribution at $\omega=0$, \new{given by $(\alpha-1)/\alpha \, \delta (\omega)$}
where the prefactor is defined as \textit{ergodicity-breaking degree}, directly analogous to the superfluid density $n_s$. As $\alpha$ increases, learning moves deeper into the over-parameterized phase: the ergodicity breaking grows stronger, and the system retains increasingly persistent memory of its initial conditions.

The generalization error in this regime is given by  
\begin{equation}
\begin{aligned}
\varepsilon_{\mathrm{g}}(t \!\to\! \infty, \alpha>1)
&= (C(0,0)+r)\,\frac{\alpha-1}{\alpha}
+ D\,\frac{\alpha}{\alpha-1}, \\
&= (C(0,0)+r)\,\frac{\alpha-1}{\alpha}
+ D + \bigl|\varepsilon_{\mathrm{g}}(t \!\to\! \infty,\;\alpha<1)\bigr|.
\end{aligned}
\end{equation}
The first term reflects the \textit{memory component} arising from ergodicity breaking, which dominates for large $\alpha$, analogous to the growth of the superfluid density $n_s$ as the temperature decreases below $T_c$.  
The second term shows that the generalization errors in the under- and over-parameterized regimes are continuously connected, revealing their shared underlying structure.

The analogy can be sharpened by comparing sum rules. In superconductors, the Ferrell--Glover--Tinkham sum rule relates the redistribution of optical spectral weight across the transition \cite{ferrell1958conductivity,tinkham1959determination}:
\begin{equation}
\frac{1}{\pi} \int_{0^+}^{\infty} d\omega\,\sigma_N(\omega) 
= \frac{1}{\pi} \int_{0^+}^{\infty} d\omega\,\sigma_S(\omega) 
+ \frac{n_s e^2}{2m}.
\end{equation}
The term on the right-hand side shows that the spectral weight missing from the finite-frequency conductivity $\sigma_S(\omega)$ reappears as a $\delta$-function at $\omega=0$, corresponding to the formation of a dissipationless supercurrent with superfluid density $n_s$. In the learning problem, an analogous sum rule holds for the response function across the interpolation threshold:
\begin{equation}
\frac{1}{\pi} \int_{0^+}^{\infty} \! d\omega \,\operatorname{Re}\bigl[\tilde{R}(\omega)\bigr]_{\alpha<1}
= \frac{1}{\pi} \int_{0^+}^{\infty} \! d\omega \,\operatorname{Re}\bigl[\tilde{R}(\omega)\bigr]_{\alpha>1}
+ \frac{\alpha-1}{2\alpha}.
\end{equation}
This relation captures the redistribution of spectral weight across $\alpha=1$, reflecting the emergence of the $\delta$-function contribution in the ergodicity-broken phase.

Both systems become fundamentally out of equilibrium in their ordered phases, as signaled by the breakdown of the fluctuation–dissipation theorem (FDT).  
In the overparameterized learning regime, this violation appears explicitly in Eq.~\ref{eq:nFDT}, where the correlation and response functions no longer obey the equilibrium relation.  
A similar breakdown occurs in superconductors, where
\begin{equation}
    \tilde{C}_{\mathrm{J}}(\omega) \not\propto \operatorname{Re}\tilde{\sigma}(\omega),
\end{equation}
because the real part of the optical conductivity takes the singular form
\begin{equation}
    \operatorname{Re}\tilde{\sigma}(\omega) = \frac{\pi n_s e^2}{m}\,\delta(\omega),
\end{equation}
reflecting a dissipationless channel associated with the superfluid density $n_s$.  

In superconductors, this manifests as a persistent current that flows indefinitely,
\begin{equation}
    {\partial_t {j}(t)} \propto n_s E(t),
\end{equation}
with zero energy dissipation.  
An analogous phenomenon arises in overparameterized learning: the pole of the response function at $\omega = 0$ produces a persistent current in parameter space,
\begin{equation}
    \partial_t{\langle \hat{\beta} \rangle(t)}
    \propto \frac{\alpha - 1}{\alpha}\,F(t),
\end{equation}
indicating that part of the system’s degrees of freedom remain dynamically active even after training converges.  
This non-equilibrium steady state retains long-term memory of the training trajectory, in direct analogy to the supercurrent that preserves phase coherence below $T_c$.  

Thus, the violation of FDT provides a unifying physical picture: both superconductivity and overparameterized learning represent ordered phases sustained by non-dissipative collective dynamics that break ergodicity and detailed balance.

%\bibliographystyle{apsrev4-2}

% References for supplemental material

\end{document}